# 北京航空航天大学

# 硕 士 学 位 论 文

# 基于时间敏感网络的确定性系统性能分析

作者姓名　　哈 山

学科专业　　信 息 与 通 信 工 程

指导教师　　何 锋

培养院系　　电 子 信 息 工 程 学 院

# Performance Analysis for Deterministic System using Time Sensitive Network

A Dissertation Submitted for the Degree of Master

**Candidate: Md Mehedi Hasan**

**Supervisor: Professor He Feng**

School of Electronic and Information Engineering

Beijing University, Beijing, China

硕 士 学 位 论 文

# 基于时间敏感网络的确定性系统性能分析

| | | | |
|---|---|---|---|
| 作者姓名 | **哈山** | 申请学位级别 | 工学硕士 |
| 指导教师姓名 | **何 锋** | 职 称 | 副教授 |
| 学科专业 | **信 息 与 通 信 工 程** | 研究方向 | **时间敏感网络** |
| 学习时间自 | 2019 年 9 月 3 日 | 起至 | 2022 年 6 月 30 日止 |
| 论文提交日期 | 2022 年 5 月 1 日 | 论文答辩日期 | 2022 年 5 月 27 日 |
| 学位授予单位 | 北京航空航天大学 | 学位授予日期 | 年 月 日 |

# 关于学位论文的独创性声明

    本人郑重声明：所呈交的论文是本人在指导教师指导下独立进行研究工作所取得的成果，论文中有关资料和数据是实事求是的。尽我所知，除文中已经加以标注和致谢外，本论文不包含其他人已经发表或撰写的研究成果，也不包含本人或他人为获得北京航空航天大学或其它教育机构的学位或学历证书而使用过的材料。与我一同工作的同志对研究所做的任何贡献均已在论文中作出了明确的说明。

    若有不实之处，本人愿意承担相关法律责任。

学位论文作者签名：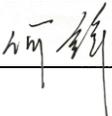      日期： 2022 年 05 月 27 日

# 学位论文使用授权书

    本人完全同意北京航空航天大学有权使用本学位论文（包括但不限于其印刷版和电子版），使用方式包括但不限于：保留学位论文，按规定向国家有关部门（机构）送交学位论文，以学术交流为目的赠送和交换学位论文，允许学位论文被查阅、借阅和复印，将学位论文的全部或部分内容编入有关数据库进行检索，采用影印、缩印或其他复制手段保存学位论文。

    保密学位论文在解密后的使用授权同上。

学位论文作者签名：      日期： 2022 年 05 月 27 日

指导教师签名：      日期： 2022 年 05 月 27 日

# 摘 要

现代工业技术要求使用可靠、快速且廉价的网络作为数据传输的主干。与时间敏感网络（Time-Sensitive Network，TSN）标准相结合的交换式以太网技术以其低成本、高带宽以及具备实时性的优势成为备选，本文对此技术进行了性能研究，用以帮助工业领域进行组网选择。此外，本文还研究了 TSN 标准最近对 IEEE 802.1Qbv 和 IEEE 802.1Qbu 二者做出的修改及其带来的性能变化。

本文在标准以太网的基础上建立了一个时间敏感调度的网络模型，主要研究了以下几个方面的关键技术：

（1）基于标准以太网的确定性通信机制。目前，时间触发以太网（Time-Triggered Ethernet，TTE）和时间敏感网络 TSN 在工业自动化领域得到了广泛的关注，特别是比较关注软实时控制和硬实时控制的组网应用，但 TSN 和 TTE 在设计复杂性和维护成本方面不如人意。本文研究了在标准以太网交换机上使用精确时间协议（Precise Time Protocol，PTP）进行时间触发通信和音视频流（Audio Video Bridger，AVB）业务调度的可行性。

（2）基于 Yen 算法改进的容错拓扑。无论小型网络还是大型网络，故障节点的数据都有可能丢失，严重时可影响系统的确定性通信规则，故障容忍拓扑可以是解决数据丢包的一种办法。本文对基于 Yen 的容错算法进行了改进。

（3）基于数据和帧索引的流量机制。网络流量的增加可能导致时间触发（TT）消息无法满足时延约束，进而破坏 TT 通信的可靠性。我们提出了一种基于数据和帧索引的流量机制，可以最大程度地减少最坏情况下的延迟和抖动。

为了验证并确认上述研究和改善目标，我们使用网络演算来计算最坏延迟理论上界，并使用 Omnet++ 来仿真统计端到端的延迟。为了使得数据具有更好的可对比性，我们还在标准以太网硬件上改造实现了更精细的仿真。结果表明：演算结果、仿真结果和硬件测试结果具有一致性，展示了 TSN 关键机制可以使用标准以太网交换机来实现。

**关键词**：时敏网络, 实时通信, 信用量整形, 时间感知整形, 网络演算, Omnet++, 标准以太网, 故障容忍

# Abstract

Modern technology necessitates the use of dependable, fast, and inexpensive networks as the backbone for data transmission. Switched Ethernet coupled with the Time Sensitive Networking (TSN) standard outperforms the competition because it offers high bandwidth and real-time features while using low-cost hardware. In order for the industry to recognize this technology, comprehensive performance tests must be carried out, and this thesis offers one such study. In particular, the thesis investigates the performance of two IEEE 802.1Qbv and IEEE 802.1Qbu modifications that have recently been added to the TSN standard.

This thesis establishes a time-sensitive network communication model. Based on this, the novel key research technologies in this paper are as follows, (i) Deterministic Communication over a standard Ethernet. At present, TTE switch and TSN switch are getting popular in the industrial automation. But, TSN and TTE switch are not cost effective for design complexity as well as maintenance. This thesis investigates the feasibility of time-triggered communication and AVB traffic over a standard Ethernet switch by using precise time protocol (PTP). (ii) Either small scale network or large-scale network, data can be lost for faulty node as a consequence the rule of deterministic system is severely affected. Fault-resilience topology is one of the possible solutions. To reduce this problem, in this thesis, we have modified Yen's algorithm for fault-resilience, and Yen's Algorithm update adjacent hops to ensure that the path become free and reliable to send frame as well as it can choose shortest path if more than one reliable path is found (iii) Due to increase number of traffic, time triggered can miss the deadline as a result the reliability of the TT communication are getting worst. We have demonstrated a unique traffic mechanism which is known as Data and frame indexing. This mechanism minimizes worst case delay and minimal jitter. To enhance quality of service in the context of deterministic system, time critical traffic should be low latency as well as Audio video traffic also concern about time and transfer before deadline, which is already ensured by TTE and TSN switch except standard ethernet switch by using our main objectives and the relationship between three objectives also established in TSN and TTE network.

To verification and validation in above objectives, we have network calculus for knowing theoretical bound as well as omnet++ simulation have used for end-to-end delay calculation.

For comparison of the simulations value, we have implemented hardware for more precision. The result shows that three type of analogue is more or less similar. TSN network can also be implemented by using standard ethernet switch.

# Table of Contents

# List of Tables

# List of Figures

# List of Abbreviations

| | |
|---|---|
| ECU | Electronic Control Unit |
| CPU | Control Processing Unit |
| IoT | Internet of Things |
| TT | Time Triggered |
| RC | Rate Constrained |
| BE | Best Effort |
| TSN | Time Sensitive Networking |
| GCL | Gate Control List |
| TAS | Time Aware Shaper |
| LAN | Local Access Network |
| WAN | Wide Access Network |
| MAC | Multiple Access Control |
| FCS | Frame Check Sequence |
| QoS | Quality of Service |
| VLAN | Virtual LAN |
| RSTP | Rapid Spanning Tree Protocol |
| MRP | Media Redundancy Protocol |
| STP | Spanning Tree Protocol |
| PTP | Precision Time Protocol |
| OSI | Open System Interconnection |
| gPTP | generalized Precision Time Protocol |
| VL | Virtual Link |
| BAG | Bandwidth Allocation Gap |
| SM | Synchronization Master |
| CM | Compression Master |
| AVB | Audio Video Bridging |
| SRP | Stream Reservation Protocol |
| SR | Stream Reservation |
| CBS | Credit Based Shaping |
| ACL | Access Control List |
| MTU | Maximum Transmission Unit |
| CNC | Centralized Network Configuration |
| LS | List Scheduler |
| MTTF | Mean Time To Failure |
| CCU | Central Computing Unit |
| HMI | Human Machine Interface |
| FTGA | Fault Tolerant-Genetic Algorithm |
| ATS | Asynchronous Traffic Shaper |

# 1. Introduction

## 1.1 Background

Ethernet is a technology that is as extensively utilized in industries as it is in everyday life, even in the home. First, it came out in 1980, then, in 1983, it was standardized by the IEEE, with help from a working group, which would later become a sub-committee. Soon, it superseded all previous wired local area network (LAN) technologies. Initially, it was capable of reaching a maximum speed of 10 Mbit/s at the outset. Since then, many enhancements have been made, and additional standards have been established. Further, data speed of ethernet is boost up which is reached around 400 Gbits/s in 2017. IEEE 802.3, the IEEE standard which defines Ethernet, comprises all of these standards. The physical and data connection layers of Ethernet are defined by the Media Access Control (MAC). The first two levels of Open Systems Interconnection (OSI) model are also known as the two layers of the OSI model. Standards for (LAN) Local Area Networks as well as Metropolitan Area Networks (MAN) fall under the IEEE 802 initiative (MAN). An organizing committee IEEE 802.1 specifies the administration of LAN/MAN bridging. A group in this organization is composed of Time Sensitive Network (TSN) task group [1]. The organization's name changed in 2012 in order to enable its work to continue; before to this, the task group was called Audio Video Bridging (AVB), and it was originally part of IEEE. In the TSN network, Time Triggered (TT) traffic is the most prominent traffic as consider higher priority as well as this traffic are controlled in precise time which is very important for industrial automation.

There are several reasons to support the usage of TSN, one of which is the increasing momentum and popularity of Ethernet-based implementations due to high bandwidth, scalability, and compatibility, but the biggest benefit is that they are predictable in delivering data [2]. Due to data transmission constraints, real-time embedded systems are limited in implementation. As data flow increases in dispersed systems, the time restrictions become more difficult to meet. The industry demands efficient and rapid performance, and this is especially important for the automation process. To provide the best service, the network must be able to handle messages in a very short amount of time. Within this area, applications are time-sensitive, and therefore latency must be met and enforced through strict determinism. The TSN working group came up with a standard set of criteria which helps to use standard Ethernet to communicate in a consistent, timely manner [3].

We look at how TSN networks can be implementable for all industrial automation where industrial companies as well as home network should not concern about high cost financial barrier. In this context, TSN network can be implementable by using standard Ethernet switch along with Precision Time Protocol clock, in addition, fault-tolerance topology is used for deterministic data transmission guarantee. For further verification of the mention network, we have used mathematical analysis of network by using Network Calculus (NC) as well as to compare NC, omnet++ simulation tool is used. Besides, we also demonstrate hardware verification for more precision.

## 1.2 Literature Review

In recognition of the fact that the thesis is on an interdisciplinary subject, the related work has been divided into subsections to represent the close relationship between various research areas. First and foremost, you should keep in mind that TSN is not the only choice available for real-time Ethernet. Detailed descriptions of current solutions, including their most notable characteristics, benefits and drawbacks are provided in Subsection 1.2.1. The TSN and its assessment are the primary topic of many of the articles. Subsections 1.2.2 section gives an overview of the considerable research that has been done recently on time synchronization in time-triggered Ethernet networks and time-synchronization networks (TSNs). There are a few articles that look at the main components of TSN one by one, while others look at TSN as a communication system that is achieving its full potential by including all of the available standards. TSN simulation is one of the many assessment approaches that have been suggested, and it is one of the most popular. Subsections 1.2.3 demonstrates synchronization of the clock for time-sensitive networking where it mentions precision time protocol (PTP) clock synchronization for real time communication. Subsections 1.2.4 presents several types of fault-tolerance topology for standard Ethernet, TTE, TSN switch as well. Sub-Section 1.2.5 illustrate minimum path cost algorithm for fault-tolerance algorithm where in subsection 1.2.6 is mentioning optimization approaches for TSN network as well as subsection 1.2.7 is mentioning jointly solved routing and scheduling algorithm. Sub-section 1.2.8 describes the simulators for TSN that have been created recently, which have- made use of a variety of simulation tools as well as mathematical analysis, such as network calculus. This part is very important to the success of the thesis.

### 1.2.1    Review of Real-time Ethernet Solutions

The many advantages of Ethernet include its wide use, its ability to coexist with other technologies, and its affordability. It also has a disadvantage because of the slow CSMA/CD arbitration mechanism, which leads to a random communication process. Several methods to handle this are highlighted in the following articles.

The non-deterministic applications of Ethernet are the ones constrained by the characteristics defined by P. Doyle in [12]. Using full-duplex connections with switches in a network architecture can help prevent collisions. In order to guarantee determinism, Switched Ethernet may be coupled with real-time protocols. Several current real-time solutions that use the aforementioned principles are discussed in the following issue [13]. The protocols EtherNet/IP, PROFInet, Ether-CAT, and ETHERNET Powerlink are all studied, and the benefits and drawbacks of each method are discussed in details. Every solution's typical network architecture and frame format are shown. Among the earliest research articles to deal with real-time Ethernet, this work served as a foundation for a large number of subsequent research publications. This chapter also includes an introduction to the IEEE 1588 standard, which guarantees that distributed clocks are synchronized to sub-microsecond precision.

Using real-time Ethernet, this study [14] by Kingstar investigates five alternative protocols. EtherCAT, EtherNet/IP, Ethernet Powerlink, PROFINET IRT, and SERCOS III are real-time protocols that have been compared with one another. This research recommends EtherCAT as the most cost-effective and high-performing option. It is not necessary to use the technology of the original vendor in order to use any of the protocols under consideration. With respect to real-time Ethernet, this article offers a fair overview of the current status of the practice. In contrast, the method used in the article is not appropriate for the academic research community, and the experiment design is not described, as a consequence, the findings are not repeatable.

T. Steinbach and colleagues [15] investigate the usefulness of switching Ethernet in in-car communication networks. Many current in-car communication systems, also including Controller Area Network (CAN) as well as FlexRay, are not very scalable, which is a critical network characteristic given the continuous growth in the number of network nodes. Switched Ethernet in conjunction with real-time protocols can meet all of the time restrictions. In this article, researchers investigate TTEthernet as a potential future backbone for in-car networks. TTEthernet supports three types of traffic: time-triggered traffic, rate-constrained traffic, and best-effort traffic. Even if the best-effort class jitter did not satisfy the criteria, the use-case demonstrates that time restrictions may be addressed in certain circumstances. The article, on the other hand, does not disclose the simulation tool or experiment setup that was utilized for this research, making it impossible to replicate the results obtained.

Audio-Video Bridging (AVB) protocol to conduct an analysis of how well car networks with switched Ethernet connections run. AVB, which was established before TSN, was created in response to certain needs expressed by equipment manufacturers and system integrators and contains four standards (IEEE 802.1BA, IEEE 802.1AS, IEEE 802.1Qat, and IEEE 802.1Qav) [16]. It lays out two classes of traffic: A for the higher-priority class and B for the lower-priority class. The Credit-Based Shaper (CBS) algorithm, which is part of the standard, avoids traffic bursts. The BMW Group Research and Technology has been working on a use-case model as described in Paper [17]. The results of the experiment showed that AVB may help keep the job on schedule. Using the same model setup and same use-case, this thesis significantly depends on the work done by M. Ashjaei et al [18].

A comprehensive worst-case study of IEEE 802.1Qbv for TSN is provided by Luxi Zhao et al. [19], who use network calculus to conduct their research. This article focuses only on scheduled traffic and does not include frame preemption since it would complicate the analysis. An IEEE 802.1Qbv-capable switch with three inputs and one output port is described in detail in the first section of this article. Immediately after the revision of the routing table, the switch directs all incoming traffic to the specified output port. Following the PCP, priority filtering is performed, and traffic is routed to the proper queue depending on its priority status. The gates of each priority queue are under the control of GCL (gate control list) [20]. If the gate is closed, the traffic is buffered in the queue; otherwise, it is sent to the output port, as shown. When several gates are opened at the same time, traffic is routed according to the priority of the queue. Assumed in this article is that GCLs are not limited, i.e. that the planned traffic windows are not overlapping, and that the worst-case end-to-end latency of scheduled traffic is calculated. Different planned traffic classes with varying priorities may be accommodated in this manner. The accuracy of the analysis is evaluated in two test cases: a synthetic test case consisting of six nodes and two switches, and a realistic test case consisting of 31 nodes and fifteen switches. It is possible to meet the real-time requirements of planned traffic in both scenarios. TSN's real-time property is guaranteed by the provided analysis, which may also be utilized for the synthesis of GCLs, making TSN a highly scalable architecture.

The use of the IEEE 802.1Qbv amendment raises the issue of establishing a proper schedule for time-triggered traffic, which is a difficult challenge to solve. Silviu S. Craciunas and colleagues [8] solve this issue by identifying the functional characteristics of the amendment and establishing its constraints, as well as deterministic Ethernet constraints, for the amendment. Based on these requirements, they offer an offline scheduler that ensures low and predictable latency for time-triggered traffic while maintaining high availability. Device capabilities and queue configuration are two of the most important factors highlighted in the study. The scheduling of switches and end-systems is referred to as the device capabilities of the device. The emphasis is on deterministic networks, which means that both the switches and the end-systems are planned in advance. When it comes to functional parameters, the number of queues is one that is important since it indicates the number of priorities that may be handled. The gate that is associated with the queue may function in a variety of ways. It may be open at all times and adhere to a tight priority policy, or it can be closed for a certain period of time. The use of gating results in a completely predictable delay for planned traffic when it is implemented. There are many different scheduling restrictions specified. Because of the frame restriction, the offset and frame duration must both be inside the frame period. By virtue of the link restriction, frames that share the same connection cannot overlap. The restriction on frame transmission means that the flow must occur in the same sequence that the frames are sent. End-to-end restriction is defined by the arrival time and the sending time. The amendment imposes restrictions on flow isolation and frame isolation, among other things. The scheduler determines the offset for each frame and the queue assignments based on the restrictions that were previously mentioned. When it comes to addressing the scheduling issue, Satisfiability Modulo Theories (SMT) are used. Based on their tests, they have concluded that as the network's usage increases, the issue gets more difficult to solve. A number of different optimization approaches are suggested.

W. Jia and colleagues [1] have carried out a performance assessment of IEEE 802.1Qbu. Time-critical traffic (TCT) requires predictable end-to-end latency, which is impossible to provide without frame preemption. This may result in missed deadlines, which can be avoided by adding a frame preemption mechanism. TCT would suffer MTU transmission delay in the event of non-time-critical traffic (NTCT) transmission. In accordance with IEEE 802.1Qbu, TCT is permitted to preempt NTCT as soon as it arrives at the frame preemption switch, and once TCT transmission is completed, NTCT transmission is permitted to continue. The priority of a frame is recognized by the MAC layer of a switch via the addition of a 3-bit priority field to the normal Ethernet header. Frame preemption is indicated by the use of the control symbols HOLD and RETRIEVAL, which are used to notify the receiver that it has happened. It is only when the receiver has received the HOLD message and recognized it that it may begin receiving the incoming traffic since the present traffic has been preempted by the higher priority one.

Once the receiver recognizes the RETRIEVAL sign, it returns to the prior traffic receiving mode. There are three distinct scheduling methods that are used in the simulation section: first-in, first-out (FIFO), non-preemptive priority scheduling (NPPS), as well as preemptive priority scheduling (PPS). The results demonstrate that the total jitter and latency are substantially decreased when PPS is used. The article correctly pinpoints the beneficial features of IEEE 802.1Qbu, and it demonstrates that real-time communication across an Ethernet network is substantially enhanced as a result of this modification.

According to Lin Zhao et al. [21], a comparison analysis of TSN and TTEthernet is shown. Between the two technologies, there are numerous parallels to be found. TDMA (Time-Division Multiple Access) is used by both of them, which entails dividing traffic transmission time in order to avoid collisions between them. Following the technical explanations, the authors take the reader on a tour of the network architecture. When it comes to bandwidth allocation, TSN outperforms TTEthernet. When there is a large amount of high-priority traffic, low-priority traffic flows may be suffocated as a result of transmission. TSN additionally makes use of the Stream Reservation Protocol (SRP) for traffic class distinction, which allows lower priority traffic to benefit from higher bandwidth allocations. Following the completion of the study, a delay analysis comparison between TSN and TTEthernet will be conducted. Because of TSN methods like as CBS and frame preemption, high bandwidth is associated with low end-to-end latency. There are three different strategies used by the TTEthernet to blend TT and RC traffic together: a delay strategy, in which TT traffic is simply pushed off the wire until the RC is finished, a preemptive strategy, in which RC traffic is stopped and TT traffic is forwarded before the RC starts again, and finally, a time-conscious method, in which the RC transmission is never started if it would conflict with the broadcast. As an added bonus, TSN offers superior redundancy capabilities than TTEthernet. TSN utilizes several paths for each packet in order to avoid data loss, while TTEthernet does not have a reliable method for dealing with packet loss. In general, TSN is much more flexible to changes in the topology, while TTE requires a whole new schedule to be created in order to accommodate the modifications. For the second time, this research emphasizes the enormous potential of TSN while also providing rationale for the idea itself

## 1.2.2 Synchronization of the timing of time-triggered Ethernet

The AS6802 TTEthernet protocol (SAE 2011) [22] distinguishes three types of nodes, each of which has a specific function. These are the compression masters, the synchronization masters, and the synchronization clients, respectively. These two masters of synchronization and compression work together to jointly create and maintain an internationally synchronized time-base via the exchange of protocol control frames (PCFs). Moreover, the synchronization clients are passively synchronized to this synchronized time-base by only receiving PCFs from the masters. This allows the clients to maintain a consistent state of synchronization. Some of the time-triggered (TT) switches serve as compression masters, while some of the end systems act as synchronization masters, and all other TT-switches and end systems act as synchronization clients in the standard TTEthernet configuration. As we can see, TTEthernet is a mix of master–slave architectures with distributed synchronization mechanisms. For the time being, we will be concentrating on the dispersed synchronization between the masters while we deal with the fundamental issues.

During a synchronization round in TTEthernet, the synchronization masters broadcast periodic packets of data known as PCFs to the switched network, thus initiating the round. These PCFs are delivered with the highest priority in order to establish rapid self-stabilizing startup progress. They are classified into cold-start (CS), cold-start acknowledge (CA), and integration (IN) frames by a type-field defined in the PCF header, which causes specific actions in the state-machines of the masters to be triggered in response to the PCFs. It should be noted that a common COTS Ethernet switch does not distinguish between different frame priorities and types, nor can specific actions in response to different PCF types be supported, which means that the self-stabilizing startup with COTS Ethernet switches could not follow the same idea as the self-stabilizing startup with TTEthernet switches.

Specifically, compression masters (which are present in the synchronization masters) periodically receive PCFs (information about the membership state and configuration) from the synchronization masters and determine whether to compress, re-generate, and deliver them in accordance with the current membership state and off-line configuration. Moreover, in a system where the compression masters employ an identical and redundant synchronization method, these compression masters are also in charge of controlling the timing of when they send the compressed PCFs, which is determined in collaboration with synchronization masters that may be partially in a defective state. Every non-faulty compression master has a consistent sequence of compressed PCFs, all of which are timed almost the same and thus have a common sequence of non-redundant compressed PCFs, because each non-faulty compression master always has its redundant compressed PCFs broadcast nearly simultaneously with the support of transparent clocks. Additionally, the compressed PCFs may be received by at least a part of the network, because the compressed PCFs are only possible as a result of this process. Unavoidable truth to be aware of is that a simple COTS Ethernet switch does not provide transparent clocks or the minimum transmission delay jitters that are common on conventional switches. These limitations could not be ignored even in non-queuing situations.

### 1.2.3 Synchronization of the clock for time-sensitive networking

The IEEE 802.1AS protocol (Garner 2011) [23], which is included in the published IEEE TSN standards, provides extensive precision time protocol capabilities in Layer-2-time synchronization applications (gPTP). When compared to gPTP, which is based mostly on IEEE 1588-2008, the protocol's approach provides more of a background for border application than gPTP. Because the protocol follows the grandmaster-slave synchronization paradigm, which is defined by IEEE Std 1588-2008, mistakes in the grandmaster (GM) cannot be tolerated. As well as delays and ARP attacks, the general-purpose transport protocol (gPTP) is susceptible to a number of additional weaknesses [24].

In the next version of IEEE802.1AS, a large number of instances of gPTP may be used to represent redundant GMs and numerous synchronized spanning trees [25] . At their option, end devices may take use of the time offered by a large number of redundant sources and channels. It is difficult to imagine how these GMs might be coordinated without the use of externally synchronized time references, which are not available without the use of an externally synchronized time reference. In order to do this, the new standard differs from traditional Ethernet in that it allows for a greater number of spanning trees. Because of the numerous Byzantine attacks that occur in reality, connectivity needs may be excessive [26] . [27] In 1982, Dolev published his research on the subject. There are currently no COTS odes available. As a potential resource for a greater degree of network connectivity, we think that COTS switches might be useful.

### 1.2.4 Overview of Fault Tolerance Topology

Fault tolerance topology is significant feature for real-time data communication, especially, where deterministic data transferring within a limited. There are several researchers used several algorithms about how to tolerant the fault in the hops of the network as well as research also find the tolerance level in different point of view by using various algorithm. In the following passages, we have briefly discussed several algorithms in the context of time sensitive network and deterministic network.

[28] Using TSN, Time-Triggered (TT) communication is enabled, which provides reduced latency as well as a predictable timing. For the sake of ensuring that the service is not interrupted,

TSN employ the notion of seamless redundancy. In deterministic systems, the TSN network is well-established in terms of the relationship between fault-tolerant and fault-resilient components. End-to-end latency in each device may be altered in a version of the TSN model using Yen's approach [29]. As an additional point of interest, arriving critical messages, such as audio or video traffic, may often be affected by rate limitation messages. Due to the changing environment, researchers are interested in ensuring zero level jitter for real time communication in a predictable manner. As a consequence, TSN switch configuration should be changeable in tandem with end devices, according to their findings. It is necessary to conduct a comparative study of the TSN architectural model as well as the devices in order to identify the optimal network traffic topology where the transmission of the TT ad RC messages is deterministic and end-to-end latency is reduced while the most important data is not lost.

In the context of transmission reliability, some researchers have focused on ILP formulation which usually used for TT messages transmission. [30], [31]In switch Ethernet networks, transient defects are investigated via a systematic investigation of the transmission reliability of messages. It is necessary to develop a routing algorithm that recognizes the importance of the dependability of the TSN network in order to create a routing algorithm that is reliable. The ILP-based formulation determines the route and redundancy level for each message, and thus ensures that the necessary Mean-Time-to-Detect-Error (MTTDE) is met. Nonetheless, this paper is concerned with a network breakdown in which a large number of end devices may be capable of sending TT messages. If we use deterministic switches in conjunction with TSNs, we may be able to limit RC messages. Alternatives include completing traffic modeling for all TSN variants at the same time. such as CBS+TAS and simply TAS, before moving on the next step. Failure resilience cannot be utilized to cope with permanent connection failures, despite the fact that it is more efficient in temporal redundancy than fault tolerance.

In addition to TT communications, AVB streams with restricted end-to-end latency, and random communication are all possible with TSN technology [30]. The author of this research is interested in controlling TT stream scheduling and decreasing end-to-end latency in a distributed system. In order to address the large delays and multi-hop routing that are inherent in contemporary wireless networks, they proposed a novel technique that has the potential to substantially decrease the routing time as well as the number of hops. Despite the fact that this technique may assist reduce the likelihood of a TT message colliding with another, it is only effective for small networks, which means it is not ideal for industrial applications. At the same time, different models of TSN switches may restrict the transmission of TT and RC signals.

The first step in developing an exact technique for producing an implementation with proper routing and scheduling in one stage is to solve a 0-1 ILP(Integrated Linear Programming) [32], [33] .Futhermore,0-1 ILP may be used to optimize space exploration routing and scheduling, as well as to interface with non-scheduling traffic and the number of specified port slots, among other things. It is import to note that although the ILP method may offer high reliability for optimum capability, it has one major flaw: it is not scalable. For tiny networks, this is only relevant.

When dealing with time-triggered traffic, there are routing and scheduling issues that are NP-hard to solve [34]. This problem has been solved by making extensive use of abstractions wherever it has been possible to do so. The authors provide a novel ILP formulation that can only be utilized for routing and scheduling in TT Ethernet networks, as shown by experiments. According to the findings of the research, the JRaS scheduling and routing tool was shown to be the most beneficial for time-triggered networks. The problem is one of scalability: it is very difficult to deal with. It is only available on a few networks.

The author of the paper presented several well-established network ideas that may be used to solve the issue of enhanced service quality [35]. To determine the reliability of a network, two approaches have been suggested. They are both of a broad nature, the computational tests showed that, despite the fact that the suggested methods reduce computing efforts, they retain the required design quality for the application. Even though it is obvious that thus optimization technique is critical for dependability in networks, deterministic networks are difficult to construct since they are intended to adapt in reaction to failures, making it impossible to run the network in a predictable manner.

TTEthernet networks are used to relay real-time device communications, and they are becoming more popular [36]. To put it simply, Tabu is a search-based metaheuristic that may be used to a variety of tasks, including, but not limited to, optimizing the routing of rate-constrained traffic. The WCDs of TTEthernet systems should be reduced to a minimum as much as feasible. Because of this, the frames may be planned more easily, and the degree of schedule-ability is enhanced as a result of this optimization. In table 1, we can see some several features of TSN redundancy where mentions the alternate approach along with IEEE approach.

Several Deterministic redundancy protocols are given below

- ➢ Spanning Tree Protocol (STP) (2001)
- ➢ Rapid Spanning Protocol (RSP) (2001)
- ➢ Media Redundancy Protocol (MRP) (2012)
- ➢ Parallel Redundancy Protocol (PRP) (2012)
- ➢ High Availability Seamless Redundancy (HSR) (2012)
- ➢ Decoupling Stream Reservation and Redundancy (2012)
- ➢ Harmonized Stream Reservation and Redundancy (2012)

**Table 1 Different Features of TSN Redundancy**

| Features | Alternate Approach | IEEE Standard |
|---|---|---|
| Interdependence of stream reservation and redundancy Protocol | No | Yes |
| Arbitrary redundancy protocols | Yes | No |
| Seamless redundancy | Yes | Yes |
| Stream redundancy | Yes | No |
| Automatic Stream reservation | Yes | No |
| Engineering of streams | (Yes) | Yes |
| Minimal Protocol Overhead | No | Yes |

### 1.2.5    Path Cost Minimization Algorithm

Transmitting information from one end device to another requires the use of switches; in this case, we are referring to a time sensitive switch. Some data may suffer lengthy route lengths to transfer from source to destination as a result of transferring data across long distances. Different factors, including as time, influence the path cost. Several studies have suggested different methods to reduce the weight of the route in the context of a time-sensitive network.

Following is a list of the popular algorithms, which has been arranged in alphabetical order.

> ➤ Yen's Algorithm
> ➤ JTRSS Algorithm
> ➤ Network Level Redundancy
> ➤ Separate Synthesis

### 1.2.6 Optimization Approaches for TSN Network

There are two kinds of methods for TSN network optimization's messages routing optimization and Audio and Video bridging (AVB) message optimization.TT messages routing optimization is the first of these ways. Several algorithms have been created by various researchers on the basic of two optimizations, the most notable of which are listed below. For TT messages, there are a plethora of optimization methods available, with the four most common ones shown below.

> ➤ A heuristic Solution
> ➤ Greedy Randomized Adaptive Search Procedural (GEASP)
> ➤ A Constraints Programming based Model
> ➤ Search-Space Reduction Techniques

For AVB message, among a lot of optimization approaches, two most popular approaches are given below

> ➤ Greedy Randomized Adaptive Search Procedural (GRASP)
> ➤ Search-Space Reduction Techniques

### 1.2.7 Jointly Solved Routing and Scheduling Algorithm

The protocol and method described above are either routing optimization algorithm or scheduling algorithms for data transmission in a deterministic way, depending on the situation. However, some research has developed jointly routing and scheduling algorithms, which can handle two problems with a single algorithm by combing two different techniques. Among the algorithms that may be used to handle routing and scheduling problems are two that are quite popular.

> ➤ Integer Linear Programming Formulation (ILR Formulation)
> ➤ Mixed Integer Program Formulation (MIP Formulation)

### 1.2.8    End to End Delay Evaluation

Model-checking, formal modeling, and some statistical analysis approaches are now accessible as worst-case delay analysis tools, with the first two being the most often used. The model-checking [37] approach is based on a thorough examination of all conceivable scenarios in order to determine the worst-case end-to-end latency between two points. Unfortunately, due to the combinational explosion issue, this method is unable to deal with the complex combinations that are seen in industrial settings. Even while simulation, which tends to overshoot, can identify the worst-case delay in a large number of test scenarios, it cannot predict the longest delay in advance. Network calculus [38] the trajectory method [39]– [41] and the holistic approach [42] are example of techniques that depend on modeling upper-bound delay with a certain degree if pessimism [43] . In particular, network calculus, a well-known approach that has previously shown successful in AFDX network worst-case delay analysis [44], has been demonstrated to be effective [41]. The article [45] that has been cited is concerned with Ethernet-AVB network calculus models. The study (detailed in [46]) makes use of network calculus to investigate how the AVB traffic in TSN networks would be impacted by the time of the AVB transmissions [47]. For close per delay bounds on the Ethernet-AVB network with serialization constraints [48], which uses a more accurate trajectory method than the previous work.

Many network simulation tools have developed in the past couple of years, but not all of them have been well received by the academic community as a result of their widespread use. There are a variety of commercial network simulation programs available, including Ns216, Ns317, J-Sim18, OPNET19, QualNet20, and many more. Overall, the aim of these tools is to provide a simulator platform on which other simulation frameworks may be built. Because omnet++ offers a large number of network topology components, is open-source, and is built on the C++ programming language, it was an obvious option for this kind of study.

## 1.3 Motivation

TSN is a further development of Ethernet and is a collection of IEEE 802.1 Task Group specifications [4]. Switch Ethernet is enabled by TSN, which is compatible since it uses a switch rather than a hub. This supports advanced traffic management. Virtual Local Area Network (VLAN) support is introduced in the standard with the help of IEEE 802.1Q, which uses an 802.1Q tag in an Ethernet frame [5]. In addition, it makes it possible to route and buffer data in the network. The traffic scheduler mentioned above is a method of sorting messages based on priority levels and preventing switch ports from being overloaded. To allow preemption of non-critical communications and the transmission of messages prior to receipt, modifications 802.1Qbv and 802.1Qbu are being proposed. Scheduled traffic comes with the first amendment, while frame preemption is introduced with the second amendment [6]. The implementation of these two adjustments allows for quick, open communication over the TSN network [7]. Based on paper tests, TSN can guarantee minimal latency in comparison to Ethernet standards. the omnet++ simulation model confirms this. To allow planned traffic and time synchronization, the IEEE 802.1Qbv and IEEE 802.1QAS standards were used. The study referenced by this blog post studies the IEEE 802.1Qbv amendment alone. It outlines the operational limits of 802.1Qbv hardware and how they influence the administration of temporal traffic movement [8]. It is the aim of this thesis to analyze how 802.1Qbv and 802.1Qbu function when coupled, and two functions are used in standard ethernet (IEEE 802.3u) along PTP clock [9].

With increase of the electronic device either industrial control or modern electric vehicle, the number of flow and traffic are also significant increased as a consequence the industry or electric vehicles are installed several switches such as TTEthernet switch, AVB switch or TSN switch which are not economical feasible, on the other hand, maintenance and design cost are significantly high. Moreover, as number of switches and hops are increased during the data communication, there are high chance to lose data for sending adjacent hops as a result the deterministic behavior will be affected. Besides, high priority traffic such as Time-Triggered traffic will be arrival same time in the queue which is known as worst-case problem. To enhanced the quality of the deterministic system as well as economic feasibility, our motivation is to use standard switch replaced the specialized switch as well as performance analysis of the deterministic system in context of fault-resilience and improvement worst-case Time-Triggered (TT) traffic.

For mathematical evolution, we have used network calculus for knowing service curve of each network node output and every node arrival curves and service curves are eventually provide end-to-end delay. A simulation tool, based on omnet++, was used to evaluate the performance of time-critical traffic in terms of end-to-end latency and link usage. We have built some small Hardware network implementation to measure network precision, especially for critical time traffic such as Time Triggered (TT).

## 1.4 Problem Statement

According to studies and literature reviews, we can identify a few flaws and problems with our study that need to be addressed. Based on our literature review, we have confined problem statement within three major issues such as (i) High Design and Maintenance Cost (ii) Fault Tolerance for deterministic Network over a standard Ethernet (iii) Worst-case Time Triggered Problem.

### 1.4.1   High Design and Maintenance Cost

Systems such as Fog computing and the Industrial Internet-of-Things (IIoT) rely on technologies such as Distributed Cyber-Physical Systems (CPS) for determinism in cloud-to-machine and machine-to-machine communication [10]. Sensor, actuation, and monitoring devices make up CPS networks, which are comprised of a collection of devices that provide time-critical information using a real-time communication protocol, such as the time-triggered protocol (TTP) [11] . These protocols provide Time-Triggered (TT) communication by having a cooperative schedule that aligns to a global standard for time.TT has been used several protocols, though Ethernet was favored for industrial and automotive networks in the past, Ethernet is only available through a few proprietary industrial Ethernet protocols like PROFINET and EtherCAT. TSN is developing a whole suite of totally deterministic standards for local-area Ethernet networks. These include the IEEE 801.1ASrev standard for time synchronization, as well as IEEE 802.1Qbv, which makes time-triggered data transfer deterministic.

Even while TTEthernet and TSN have emerged as the dominant standards in the field of TT communication, their implementation comes at a significant cost in terms of design complexity and maintenance.

**1.4.2    Fault-Resilience for deterministic Network over a Standard Ethernet**

In a variety of applications ranging from industrial automation to automotive systems, TSN is widely utilized today. When it comes to autonomous driving designs in the automobile industry, fault-resilience TSN networks are being considered because they have the bandwidth needs to combine data from numerous sensors and the dependability necessary for autonomous driving.

Despite the fact that Ethernet is very fast and inexpensive, it is not suitable for real-time or safety-critical applications. Consequently, even in half-duplex systems, collisions cause transmission delays to be indefinitely extended indefinitely. It is discussed in this article how the requirements for a live network may be met and how Ethernet can be improved to fulfill those criteria.

Many protocols have previously been established for TSN and TT communication in the context of fault-tolerance or fault-resilience which is discuss in literature review section, however these protocols are only utilized in either a TSN switch or a TT switch, depending on the application but standard ethernet does not use fault-resilience in the context of TSN and deterministic network. However, maintaining and ensuring high dependability in designs is difficult and time-consuming, and it is not cost-effective either for TSN and TT switch.

**1.4.3    Worst-Case Time Triggered Problem**

Although TT messages are supported by TSN and TT switches, there is a limit of one scheduling queue per switch in each configuration. If more than two TT messages are imposed at the same time, and the TT messages are given equal priority, the scheduling queues are shared. In this case, two or more TT messages are received by the switch which is shown in fig. 1, and there is no way to resolve the situation effectively. Currently, this is kind of scenario(arrived frame to queue exact same time) solved by sending frame to adjacent hops randomly if the traffic considers the highest priority like TT traffic and there is no alternative option that has been discovered via current research efforts.

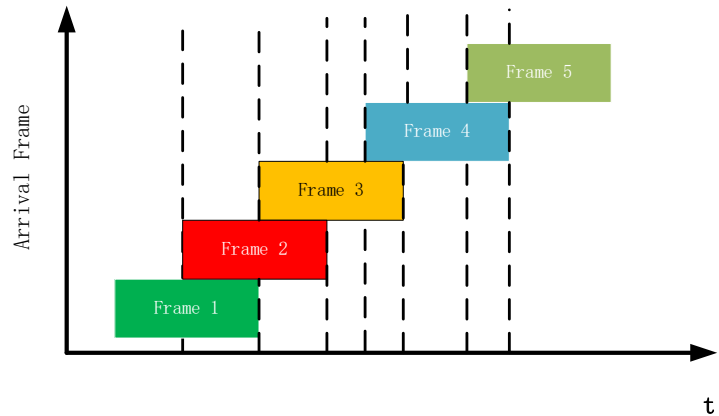

**Figure 1 Worst-Case Arrival Problem**

## 1.5 Objectives

The thesis will examine the TT and AVB communication in deterministic networks using a typical Ethernet switch (standard ethernet switch), which will be based on literature and research findings. For the purpose of performance analysis, three main parameters are used.

(i)    Deterministic Communication over a standard Ethernet

According to literature review, we have found time sensitive network can only implement by using some sophisticated network switch like TSN switch, AVB switch, TT switch and so on. But, maintenance of the switch along with some internal and external complexity are often shown in this type of networking. Moreover, cost of the switches is comparatively high. So, one goal is that we can get similar output by using standard ethernet as the maintenance of the switch, cost and complexity will be significantly reduced.

(ii)    Fault resilience topology over a standard Ethernet for TSN network

Fault resilience topology is one of the significant phenomena in the deterministic network. The main purpose of the time sensitive network is to ensure that data can transmit source to destination within short amount of time and jitter and latency should be zero. Node failure or high transmission is fully responsible for increase end to end delay. Another prominent objective is that fault resilience topology used over a standard ethernet in the context of TSN network. For network reconfiguration and fault-resilience, we have modified Yen's algorithm for update adjacent node for ensuring free and shortest path.

(iii)    Data and Frame Indexing Transmission

Due to increase for the number of traffic, the data transmissions have faced tribble problem because length of the frame are increase as well as higher priority traffic are very frequent to transfer as a consequence the end to end delay are very significant not only higher priority but also lower priority. On the other hand, worst case delay is also massive obstacle for data transmission. So, our last objective is that to shrink the higher priority traffic by using frame indexing method where the end to end delays are significant to improve and worst-case problem will reduce. Shrink highest priority traffic refer that the frame of highest will be compressed before transmission by following unique frame indexing method. But other traffic frame will remain unchanged.

In the basis of the objectives, we have analyzed the performance of the standard ethernet in the context of the time sensitive network and the network should be deterministic. On the basis of the above background, investigate end-to-end delay estimates as well as comparisons between regular ethernet communication, TT communication over standard Ethernet, and TT+AVB communication over standard ethernet.

## 1.6 Thesis Structure

This thesis is organized in the following chapters.

**Chapter 1,** The chapter develops a basic understanding and motivation about the requirement of the thesis work. It provides the primary significance of research work. It also provides use cases and a future vision of TSN network over standard Ethernet. Further, this chapter provides a literature overview and strengths as well as weakness of similar work in literature.

**Chapter 2,** This chapter gives a brief description of the real time communication via Ethernet. In this chapter, we will discuss several types of ethernet switch such standard ethernet switch, time triggered switch, time sensitive network switch and all of the of the are ensuring real time communication. Moreover, in this chapter, discuss some mathematical analysis of network calculus which helps for calculating end-to-end delay latency in theoretical way. Besides introducing network simulation tools for instance omnet++.

**Chapter 3,** This chapter illustrates a concise description of time sensitive networking over standard ethernet. Basically, in this chapter is showing time sensitive network and hardware architecture of TSN network. Some problems are formulating in the context of TSN network over standard ethernet. Besides, improved service curve by using non-overlapped gate problem. is solved.

**Chapter 4,** This chapter demonstrates that fault-resilient topology and traffic configuration have used for time sensitive network. Several type of topology syntheses have been used for data communication.

**Chapter 5,** This chapter shows data and frame indexing transmission. Basically, in this chapter, we will solve worst-case time data for real time communication. Specially for time triggered traffic is using this indexing in real time.

**Chapter 6,** in this chapter, we have shown several case studies on different network topology and analyzing end-to-end by using network calculus as well as omnet++ simulation tools. Besides, hardware implementation is also used in this purpose.

**Chapter 7,** we discuss the conclusion of the mathematical analysis as well as simulation result.

# 2. Real Time Communication via Ethernet

The usage of switched Ethernet in ethernet networks has mostly superseded the use of shared Ethernet since it is a more efficient and easy method of increasing the bandwidth of the network. Using TSN, which is an extension of Ethernet, it is possible to obtain real-time characteristics of the network and to fully use its capabilities. For the purpose of providing performance findings, network calculus for mathematical end-to-end delay latency and omnet++ were chosen as the simulation environment, since they can offer results of upper limit latency. Many topics are covered in more depth later in the text, including real-time systems, switched Ethernet, TSN, and network simulators.

## 2.1 Real-Time Systems

Real-time systems include a variety of components, including human operators, controllable objects, and real-time systems. A time-sensitive system has a direct effect on the internal state of the system since time is a constant factor in the system. A real-time computer system's output timing is just as critical to accuracy as the calculations themselves [49] when it comes to ensuring that the system is accurate.

It is possible to classify real-time systems as either hard real-time systems or soft real-time systems. Breaking the timing constraints of hard real-time systems, such as a train's brake-by-wire system, may have catastrophic consequences. Soft real-time systems, such as an entertainment system on a train, may, on the other hand, continue to function properly even while experiencing a specified degree of delay [50].

A variety of services are attempted by each real-time system. In order to achieve the goal of real-time systems, the exchange of data between controlled objects and computer systems happens via the use of actuators and sensors. When digital information is sent to other systems of computers, data communication is required. Data communication is also required when conventional and continuous information is transferred to other devices. Not only is it possible to scatter the system, but it is also possible to do it in real time. During the operation of a distributed system, computers communicate with one another in real time, allowing information to be shared between them. Nodes' service may be described in messages. They are produced by the network, which responds to inputs, the passage of time, and the state of the node at that moment. As a result, the service provided by the node is time-critical [51].

## 2.2 Ethernet

The idea of local network communication via a variety of protocols, referred to as "Ethernet," is a relatively recent contribution to the world of technology (LAN). The IEEE 802.3 standard was initially made accessible to the public in 1983, and it was formally created the following year. It was originally developed at the XEROX Palo Alto Research Center (PARC) in 1976. It is capable of creating a variety of network topologies, including bus, tree, star, line, ring, and others. It represents the physical and data connection layers, which may be found in the Open Systems Interconnection (OSI) standard model [52]. The physical layer is the most fundamental layer of the OSI model, and it is comprised of the electrical circuit transmission technologies that are used in a network's communication infrastructure. Coaxial cable, twisted pair cable, or even optical fiber may be used for Ethernet connections. The physical layer determines the majority of the speed of the Ethernet network, which may reach up to 400 Gbit/s in certain cases. The data link layer is composed of two sublayers: The Media Access Control (MAC) sublayer and the Logical Link Control (LLC). Data link layer synchronization, flow management, and error checking are all handled by the LLC sub-layer of the network stack. A MAC address is a six-byte number assigned to each node on the network. It is at the MAC sub-layer that network arbitration is carried out. Carrier Sense Multiple Access/Collision Detection (CSMA/CD) is an arbitration mechanism that is the foundation of the Ethernet protocol suite. Every node is continuously monitoring the network state (Carrier Sense), multiple nodes can begin transmission if they detect that the network is quiet (Multiple Access), and in the case of multiple concurrent transmission, each of the nodes must detect a collision, stop its transmission, and try again after some random time interval (Collision Detection) [52]. If a frame collides with the same framework 16 times, that frame is removed and will not be sent [53]. Under high traffic conditions, the primary drawback of this arbitration method is that it is prone to collisions, which results in unbounded end-to-end transmission latency. Because switches operate at the data connection layer, they have the potential to improve the arbitration process.

## 2.3 Switched Ethernet

Full-duplex connections, temporal dependability, limited delays, and minimal jitter are all provided by switched Ethernet, which takes use of Ethernet's high bandwidth, scalability, and cost effectiveness. Fig. 2 depicts the network architecture of traditional Ethernet as well as switched Ethernet systems. In contrast to shared Ethernet, switched Ethernet makes use of a switch rather than a hub as its primary distinguishing characteristic. In contrast to the hub, which can identify the address of the receiver node and send the message solely to that node, a switch may determine the address of the receiver node and convey the message exclusively to that node. If the receivers are different, this allows for numerous simultaneous broadcasts in the network. Due to the fact that the switch maintains its own MAC address table in which it keeps the MAC addresses of all immediately accessible nodes, the cost of utilizing the switch rather than a hub is a little increase in latency when looking up the database. Another distinction between a switch and a hub is the kind of connection that they are capable of establishing and maintaining. Unlike a switch, which can handle a full-duplex connection, the hub only offers a half-duplex link, allowing the node to send and receive messages at the same time[54]. Switched Ethernet is a good platform for real-time traffic solutions such as AVB and TSN because of its inherent properties.

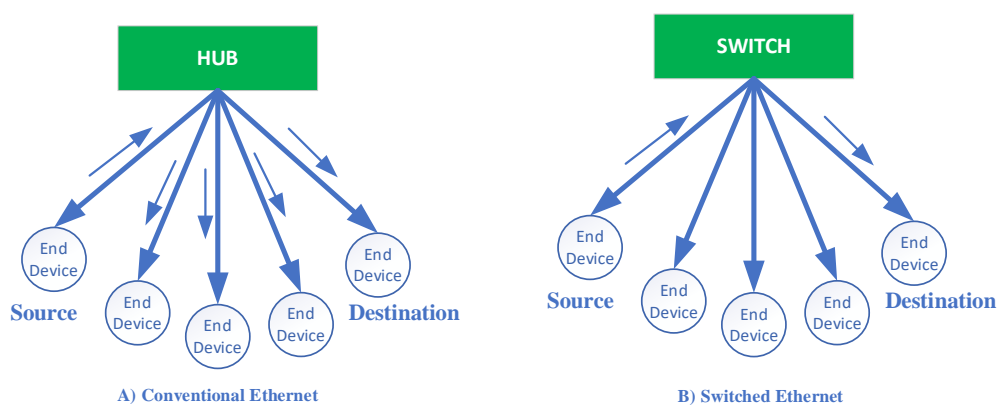

**Figure 2 Conventional and Switched Ethernet Topology**

## 2.4 Real-time Ethernet

Because of its limitations, shared Ethernet is not appropriate for usage in the industrial environment, automation, or process control applications. In order to compensate for the lack of determinism and offer low latency, several standards are used. It is also known as Industrial Ethernet or Realtime Ethernet standards, since they make use of modified MAC sublayers. TCP, UDP, and IP are not appropriate protocols since they do not provide real-time traffic characteristics. However, the hardware utilized by these protocols may still be coupled with certain other real-time protocols. Because of the cost-effectiveness of this technique, many industrial real-time Ethernet systems have been developed, including EtherCAT5, TTEthernet6, EtherNet/IP7, PROFINET8, Ethernet POWERLINK9, SERCOS III 10, and others. Aside from that, there is TSN switched Ethernet as well [21], [55]. In addition to providing secure data transfer and real-time capabilities, the TSN Task Group has produced a number of modifications. TSN also offers high data transmission speeds, up to and including gigabytes, in addition to these capabilities. But only if both the talker and the listener, as well as the Ethernet switches, implement TSN functionalities and standards will TSN be completely successful [41]. To connect to the Internet through a full-duplex Ethernet connection, all of the methods need the usage of a network switch. TSN, as compared to TTEthernet, provides improved transmission context stability, lower low-priority message starvation, and more flexibility, among other benefits. This is shown by [56],as compared to the other real-time options discussed, the TSN switched Ethernet network provides more bandwidth and is more cost-effective, demonstrating that future study and development of this technology is feasible.

## 2.5 Time Sensitive Networking

The IEEE 802 set of standards is one of the most important IEEE standards. Local area networks and metropolitan area networks are covered by this group of IEEE standards. Networking standards are primarily divided into two categories: wired (Ethernet,IEEE 802.3) and wireless (IEEE 802.11 and IEEE 802.16) networks. IEEE 802.3 is a working group that specifies the media access control of wired Ethernet at the physical layer and data link layer. It also contributes to IEEE 802.1 standards. The TSN Task Group, which operates under the auspices of the IEEE 802.1 Working Group, is in charge of developing TSN standards. This set of standards enables the transmission of data across Ethernet networks that must be provided in a timely manner. The vast majority of these are improvements to IEEE 802.1Q that include Virtual Local Area Networks (VLANs) (VLAN). The IEEE 802.1Q protocol is located at the data link layer of the Open Systems Interconnection (OSI) stack.

TSN's primary emphasis is on time sensitivity, as implied by the name of the company. On the basis of IEEE 802.1 and IEEE 802.3 standards, TSN ensures essential real-time communication while also providing deterministic messaging across switched Ethernet networks, among other things. TSN is built on AVB, a set of technological standards that enable streaming services to be delivered over IEEE 802 (Ethernet) networks with the features of time synchronization and low latency. TSN is a joint initiative of the Internet Society and the National Science Foundation. Table 2 lists the standards that have been established by the AVB Task Group.

**Table 2 IEEE Standard for AVB Task Group**

| IEEE Standard | Name |
| --- | --- |
| 802.1 BA | AVB systems |
| 802.1 AS | Timing and Synchronization for Time-Sensitive Applications (gPTP) |
| 802.1 Qat | Stream Reservation Protocol (SRP) |
| 802.1 Qav | Forwarding and Queuing for Time-Sensitive Streams (FQTSS) |

The aim of these standards was to enable users to quickly and easily construct plug-and-play ad hoc networks that provided latency and jitter that were well-contained. A talker and a listener were to communicate in the following ways, according to the designers: Using SRP, a listener would request a route across the network that met specified bandwidth, jitter, and latency criteria. The request would be sent along to the speaker in the relay. The jitter, latency, and bandwidth of the information being sent would be gathered and sent to the other end. After this, the data would be allowed to flow freely and start transmitting. In accordance with the Credit-Based Shaper (CBS) specified in 802.1Qav, the traffic will be shaped. People saw that AVB might be beneficial to the business. This is because it would improve the standards of a better system. The CBS was determined to be not robust enough, therefore the decision was made to create a new group to address its faults and add other capabilities. TSN Task Group became the name of the new team which is shown in Table 3.

**Table 3 IEEE Standard for New AVB Task Group**

| IEEE Standard | Name |
| --- | --- |
| IEEE 802.1 ASrev | Timing & Synchronization for Time-Sensitive Applications |
| IEEE 802.1 Qbu | Forwarding & Queuing – Frame Preemption |
| IEEE 802.1 Qbv | Forwarding & Queuing – Enhancements for Scheduled Traffic |
| IEEE 802.1 Qca | Stream Reservation (SRP) – Path Control and Reservation |
| IEEE 802.1 CB | Stream Reservation (SRP) – Seamless Redundancy |
| IEEE 802.1 Qcc | Stream Reservation (SRP) – Enhancements & Performance Improvements |
| IEEE 802.1 Qci | Forwarding & Queuing – Per-Stream Filtering & Policing |
| IEEE 802.1 Qch | Forwarding & Queuing-Cyclic Queuing & Forwarding |
| IEEE 802.1 CM | Vertical – Time-Sensitive Networking for Fronthaul |
| IEEE 802.1 Qc | Forwarding & Queuing – Asynchronous Traffic Shaping |

Eighth Generation Routing and Switching (802.1Q) is a 32-bit (4-byte) header that is inserted between the source MAC address and the EtherType fields of an initial Ethernet packet to aid in the prioritizing and preemption of transmissions by 802.1Q14 (Fig. 3)

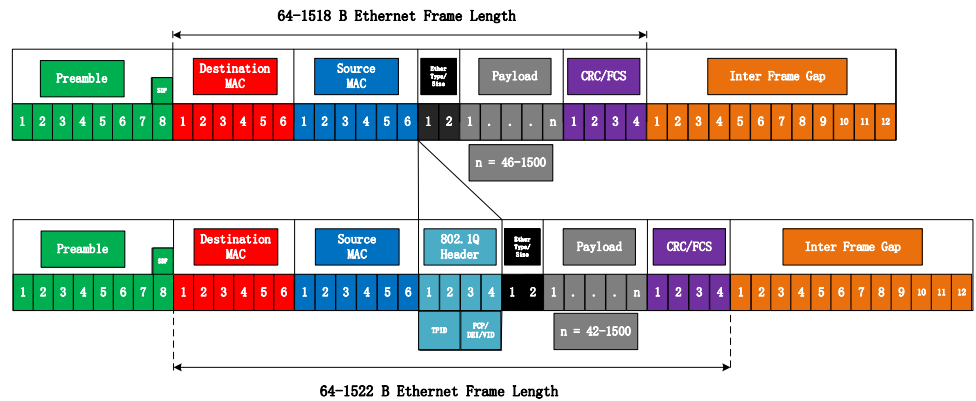

**Figure 3 Ethernet Frame along with 802.1Q**

There are two bytes dedicated to the tag protocol identification (TPID), and two bytes dedicated to tag control information (TCI) (TCI). In order to distinguish the frame from other IEEE 802.1Q-tagged frames, the TPID field must be filled with a constant value (0x8100). A total of three sub-fields are included in the TCI which is shown in table 4:

**Table 4 Tag Control Information (TCI)**

| IEEE Standard | Name |
|---|---|
| Priority code point (PCP) | A 3-bit field which specific the frame priority level |
| Drop eligible indicator | A 1-bit field |
| VLAN identifier (VID) | A 12-bit field which specifies the VLAN to which the frame belongs |

**Table 5 802.1Q Header**

| 16 bits | 3 bits | 1 bit | 12 bits |
|---|---|---|---|
| TPID = 0x8100 | TCI | | |
| | PCP | DEI | VID |

PCP is a 3-bit field that allows for the definition of up to eight distinct traffic classes. Although it is not specified how traffic should be handled after it has been allocated to a particular class, the IEEE has made several suggestions, which are shown in Table 5.

**Table 6 Priority levels**

| PCP value | Priority | Acronym | Traffic Type |
|---|---|---|---|
| 1 | 0 (lowest) | BK | Background |
| 0 | 1(default) | BE | Best effort |
| 2 | 2 | EE | Excellent effort |
| 3 | 3 | CA | Critical applications |
| 4 | 4 | VI | Video (<100 ms latency and jitter) |
| 5 | 5 | VO | Voice (< 10 ms latency and jitter) |
| 6 | 6 | IC | Internetwork control |
| 7 | 7(highest) | NC | Network control |

**Table 7 Mapping traffic classes to queues**

| | | Number of available traffic queues | | | | | | | |
|---|---|---|---|---|---|---|---|---|---|
| | | 1 | 2 | 3 | 4 | 5 | 6 | 7 | 8 |
| Priority | 0 | 0 | 0 | 0 | 0 | 0 | 1 | 1 | 1 |
| | 1 | 0 | 0 | 0 | 0 | 0 | 0 | 0 | 0 |
| | 2 | 0 | 0 | 0 | 1 | 1 | 2 | 2 | 2 |
| | 3 | 0 | 0 | 0 | 1 | 1 | 2 | 3 | 3 |
| | 4 | 0 | 1 | 1 | 2 | 2 | 3 | 4 | 4 |
| | 5 | 0 | 1 | 1 | 2 | 2 | 3 | 4 | 5 |
| | 6 | 0 | 1 | 2 | 3 | 3 | 4 | 5 | 6 |
| | 7 | 0 | 1 | 2 | 3 | 4 | 5 | 6 | 7 |

According to the traffic classification, there may be up to eight lineups. Several distinct kinds of traffic classes may use each queue if there are less than eight queues implemented. In order to map traffic classes depending on their priority to accessible queues, there is a specific method that is shown in table 6. Suppose there are only two queues accessible, and the first includes traffic classes 0-3 while the second comprises traffic classes 4-7 which is shown in table 7.

### 2.5.1   IEEE 802.1Qav – Forwarding and Queuing for Time Sensitive Streams

A modification to IEEE 802.1Qav, produced by the AVB Task Group, is one of the amendments. The primary goal of this amendment is to shape traffic for media streams in order to maximize efficiency. All control messages must be sent as quickly as feasible and must have a minimal end-to-end latency. Media streaming, on the other hand, is not subject to the same restrictions. Providing an uninterrupted stream of audio-video frames and distributing data more equitably are more essential than ever before.

Early in 2009, the IEEE 802.1 working group developed the Credit-Based Shaper for Audio/Video Bridging, the predecessor of TSN. This was the first time a credit-based shaper had been established. (AVB). In accordance with its name, this technology is mainly used for audio/video and similar applications. During an audio/video transmission, the Credit-Based Shaper's objective is to ensure that the maximum amount of bandwidth required is available in a certain amount of time without substantially interfering with other best-effort data traffic on the same network. As data streams travel through the Credit-Based Shaper, credit is assigned to those that have been granted bandwidth. It begins off with a credit of zero.

As long as the transmission credit is more than zero, it is preferred to transmit data packets that have been given a certain amount of allotted bandwidth. However, this is not always possible zero. Transmission credit decreases with time as a consequence of the use of preferred transmission methods and technologies. The reserved bandwidth in a data packet is temporarily held back and not transmitted after the transmission has been finished and while the credit for transmission is negative. When a preferred transmission is interrupted for a brief period of time, best-effort traffic is sent. While they wait for the transmission credit of the AVB packets to revert to zero, the credit of the packets recovers. It is possible for transmission credit of the relevant data stream to exceed the "0" mark if the forwarding of AVB packets is delayed by the transmission of longer best-effort packets, showing in fig. 5b. Since the best-effort transfer has been finished, the AVB packets that were temporarily held back may now be transmitted back-to-back once the best-effort transfer has been completed, for as long as the transmission credit is still valid. Now that the bandwidth has been conserved, it may be statistically possible for it to catch up to time-critical frames [37].

The Credit-Based Shaper is especially well-suited for video surveillance applications because to its prioritizing behavior, which allows for the desired transmission of audio/video data to be sent first. This is particularly essential if the receiving end-ability stations to buffer data is limited in any way. It has been shown [57], [58] that the maximum end-to-end latency of 2 ms (AVB Class A) or 50 ms (AVB Class B) over 7 hops specified in the standard may be achieved in worst-case scenarios, despite the fact that it is not guaranteed. Because of this restriction, the Credit Shaper cannot be used in areas such as process management, where exact guarantees of maximum end-to-end latency are absolutely required, such as in real-time data transmission. Because the end-to-end latencies cannot be controlled by network design or communication patterns, the IEEE has created and is actively developing new traffic shapers to solve this problem, which are now under development.

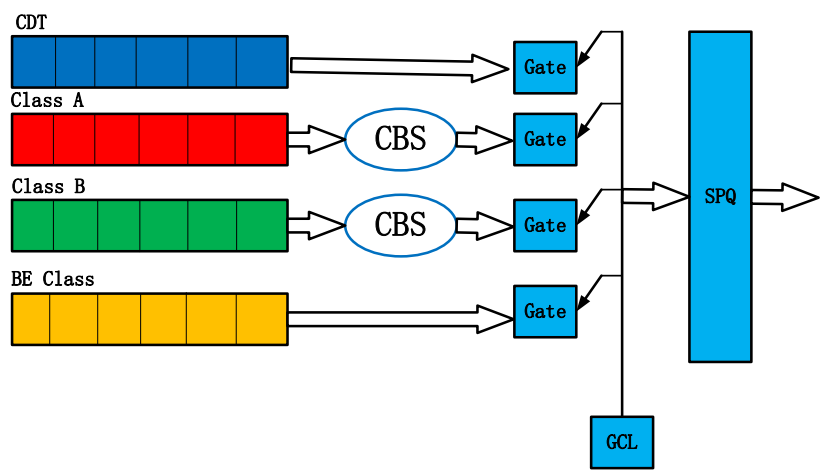

a)

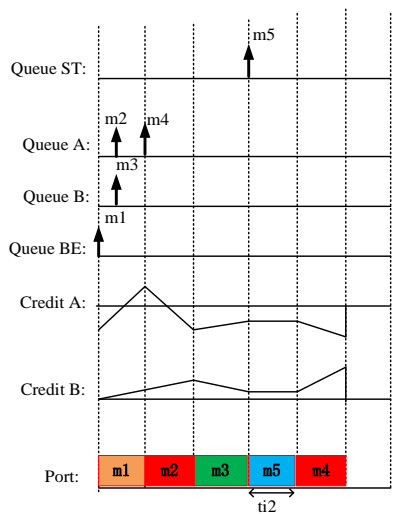

b)

**Figure 4 a) TSN model for CBS IEEE 802.1Qbv b) Credit-Based Shaping IEEE 802.1 Qav**

For a stream, CBS mellows out the traffic by distributing frames equally throughout the time frame spectrum. CBS may be allocated to a queue in a variety of ways. More importantly, the credit is given to those who possess it. If there is credit on the frame, it is feasible to transmit it. Otherwise, it is not. If the credit balance is positive, the balance is reduced to zero by the system. In the case of frames waiting for transmission in the queue, or when there are no messages and the credit is negative, it is raised at a customizable rate known as the idle slope. With a customizable rate termed the transmit slope while frame or frame rate, it is reduced further. Frames are being sent [37], with the number of frames communicated dependent on how much credit has been accumulated before.

Fig. 5 depicts an example of IEEE 802.1Qav and IEEE 802.1Qbv modifications. There are four distinct types of traffic classes: scheduled traffic, Class A traffic, Class B traffic, and best effort traffic. Message $m_3$ should have been sent after messages $m_1$ and $m_2$, but it was not because the gate for scheduled traffic was open and all other gates were closed. Message $m_5$ contains control data, and it is known that it will arrive during time interval $t_{i_2}$, therefore the gate of the ST queue is scheduled to be open at that period. It is clear from this example that gating reduces the latency of a control message from beginning to finish.

### 2.5.2    IEEE 802.1Qbv – Enhancements for scheduled traffic

In accordance with IEEE 802.1Qbv, the Time-Aware Shaper (TAS) is defined. Gate Control Lists (GCLs) are introduced by scheduled traffic [59]. A transmission gate is associated with each queue in the network. When it comes to controlling the gate state, GCLs are utilized. Each gate has two possible states: open and closed. When a gate is in the open state, transmission of frames is permitted. If the gate is closed, no frames are sent [5]. As soon as the gate is opened, frames from the appropriate queue are sent on to the transmission channel. The transmission of frames with a higher priority is prioritized over lower priority traffic when two or more gates are open at the same time [19]. Scheduling is often synchronized with a world clock in order to provide real-time and predictable network operations. IEEE 802.1ASrev may be used for synchronization in time-sensitive networks [60], and it is now in use which is shown in fig. 5.

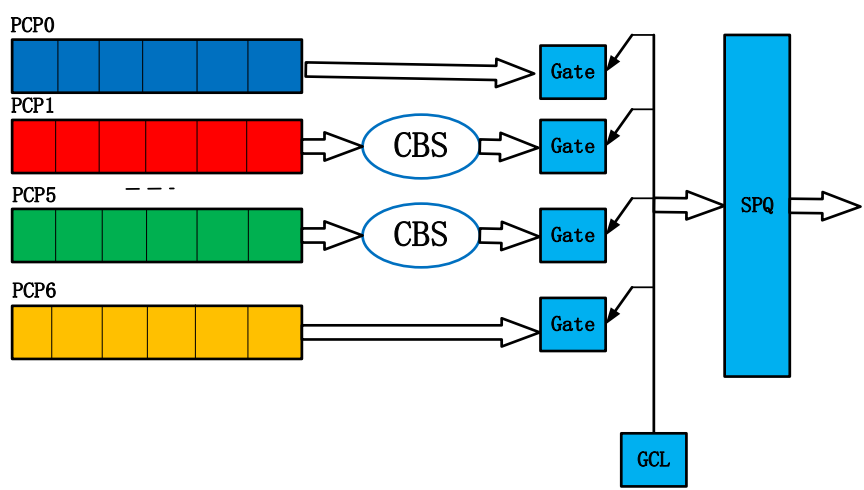

**Figure 5 Enhancement for Scheduled Traffic**

Depending on the kind of traffic, a switch may have up to eight distinct queues. A three-bit marker known as Priority Code Point is included in every frame (PCP). A frame is sent to the appropriate queue, where it is buffered, if this tag is present. Transmission via the port is only possible if the gate associated with the queue is open, which is not always the case. The gate schedule is defined by GCL. The data is often represented as a list of bit vectors, each of which represents a different configuration for a particular gate. It also includes the length of time for each bit vector in the bit vector arrays. A single amount of time must be set aside for the configuration of gates at each entrance point. Unlike the gate control list, which is cyclic and defined by industry, the gate control list must be constructed by hand. The traffic in queues 1 and 5 is also shaped using CBS, as defined by amendment 802.1Qav [61]. Prioritization alone will not ensure that time-critical communications are delivered at the appropriate moment. For example, if a control message comes while a lower priority message is being sent, the control message will be queued and transmitted after the current message. Control messages, on the other hand, are often planned in the industrial sector. GCLs may be established in accordance with their schedules, ensuring that only the gates of the scheduled traffic are opened for a particular amount of time during that time. As a result, planned control messages are not delayed by lower priority traffic since the time period allocated for the scheduled traffic is used exclusively for the scheduled traffic.

### 2.5.3    IEEE 802.1Qbu – Frame preemption

It is possible to use either non-preemptive or preemptive priority queuing techniques for prioritizing tasks. Priority queuing is used in switched Ethernet, and it is non-preemptive. An arriving time-critical frame is not handled immediately if another non-time-critical frame is already in the process of being processed at the time of the arrival. It is placed at the front of the queue and processed as quickly as feasible after the non-time-critical frame is sent. With the quantity of traffic on the network, the time-critical frame suffers from a delay. Preemptive queuing is introduced as part of the IEEE 802.1Qbu revision in order to minimize latency. It reduces delay for time-critical frames, but it also protects non-time-critical frames in order to reduce the impact on them if the delay is too great. If a time-critical frame is received while a non-time-critical frame is being handled, the processing of the non-time-critical frame is halted until the time-critical frame is received. Once the processing of the time-critical frame has been completed, it will proceed. Frames that are not time-critical may be preempted as many times as necessary until the preempted burst limit has been reached. The IEEE 802.1Bbu amendment specifies that this restriction must be adhered to. The goal of IEEE 802.1Q is to avoid the starvation of non-time-critical frames, while the goal of IEEE 802.1Qbu is to decrease the latency of time-critical frames [62].

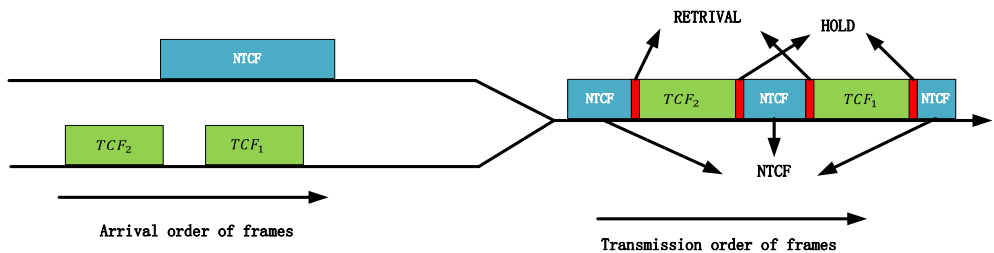

**Figure 6 Priority-based frame preemption**

Fig. 6 depicts the process of priority-based frame preemption. $TCF_1$ and $TCF_2$ preempt a non-time-critical frame NTCF on two separate occasions. They are processed as soon as they arrive. In order to indicate preemptive insertion of a time-critical frame into a non-time-critical frame, the control symbols HOLD (K28.2) and RETRIEVAL (K28.3) are reserved.

## 2.6 Network Calculus (NC)

Network Calculus is a collection of recent discoveries that offer profound insights into the flow issues that arise in networking environments. The mathematical theory of dioids, and in particular the Min-Plus dioid, serves as the basis for network calculus (also called Min-Plus algebra). We can grasp certain basic characteristics of integrated services networks, window flow management, scheduling, and buffer or delay dimensioning, thanks to network calculus.

A variety of NC implementations are available, including deterministic, real-time, and stochastic. This article covers deterministic nonlinearity (NC), which is the most advanced NC theory. [63] is an excellent place to start if we want to understand more about deterministic NC. The deterministic component of the NC real-time method is present. At the time of its development, it was used to evaluate the capabilities of integrated devices. Additionally, rather of depending only on the lower and higher curves as in deterministic NC, it considers the service for each component, both upper and lower. Describes a kind of calculus known as real-time calculus, which is presented in [4]. In the case of non-trivial traffic sources, stochastic NC, as previously outlined in [64], may be more effective in dealing with them when statistical multiplexing and scheduling become inconsistent.

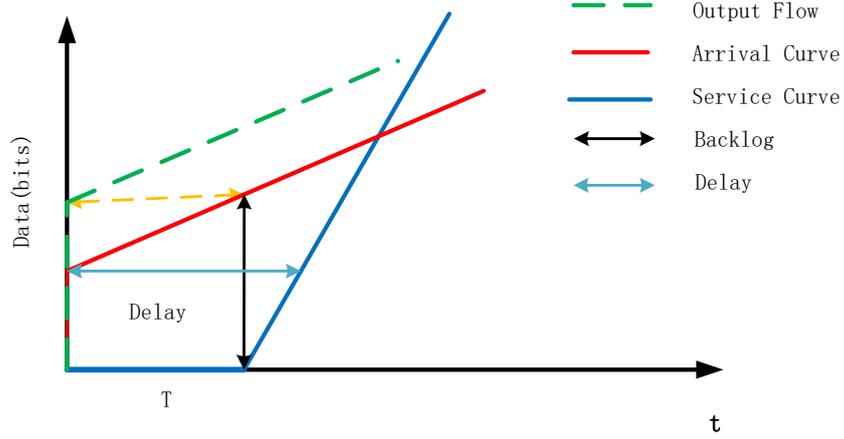

**Figure 7 Basic NC Curve, Arrival curve, Service Curve, Output Curve**

Due to the fact that TSN traffic sources evolve at the end-device level [47], deterministic NC may be advantageous in pure TSN networks. NC and Min-Plus are two types of algebra. When it comes to algorithms, Min-Plus and NC are often used in a variety of situations. Min-plus algebra has a wide variety of symbols, all of which are commutative dioids, which makes it a very versatile algebra. For example, convolution (1) and deconvolution (2) are both described as such in [48].

$$(x \otimes y)(t) = \inf\{x(t-s) + y(s)\} \qquad (2.1)$$
$$0 \le s \le t$$
$$(x \oslash y)(t) = sup\{x(t-s) - y(s)\} \qquad (2.2)$$
$$0 \le s \le t$$

In North Carolina, as will be described in more detail in the next section, these are often used formulae. NC Modeling Concept: NC examines specific queues, whole nodes, and entire networks. In general, we utilize curves to abstract incoming data and each system's scheduling and extract limits per system, such as per-hop or even end-to-end guarantees. On the right is a representation of the fundamental NC curves, while on the left is a representation of the derivation of delay and backlog. Both are explored in detail in the next sections. The actual incoming and outgoing signals are represented by $R(t)$ and $R_0(t)$, respectively a system's traffic flow (t).

Curves may be represented in a number of methods in NC (fig. 7), ranging from discrete to left- or right-continuous, and utilizing either packets or fluid bit-wise models. We exclusively make use of left-continuous, bitwise models since they have been shown to be effective in prior

TSN research [44]. Curves in NC must always be positive and non-decreasing in order to be valid.

### 2.6.1 Arrival Curve

During each given interval of time t, an arrival curve $\alpha(t)$ records the maximum number of bits that arrive to the system. [38]defines it as

$$R(t) \leq (R \otimes \alpha)\ (t) \qquad (2.3)$$

Application code analysis and establishing application requirements are a few examples of how to get a value for $\alpha(t)$. The standard version is termed rate-burst (often also named leaky- or token-bucket). The $\alpha(t)$ function is known as $\alpha(t) = b + rt$, where b represents the initial jolt and r is a constant rate.

### 2.6.2 (Min-Plus) Service Curve

Service curves are used to keep the TSN packet scheduling specifics abstract. The smallest service offered to a flow during a time period is $\beta(t)$. [44] defines $\beta(t)$ as follows:

$$R_0(t) \geq (R \otimes \beta)\ (t) \qquad (2.4)$$

The whole service control of NC is modeled via an examination of queuing, scheduling, and forwarding processes, and is done so using service curves. In addition, the rate-latency function $\beta(t) = R*[t-T]^+$ specification where $[t]^+$ represents the maximum time-invariant serving rate after the worst-case delay of T.

### 2.6.3 Maximum Service Curve

In the definition of maximum service curve $\beta_{max}(t)$ which, as the name suggests, represents the maximum service [65].

$$R_0(t) \leq (R \otimes \beta_{max})\ (t) \qquad (2.5)$$

### 2.6.4 Shaper Curve

Shapers compel a flow's arrival curve to be $\sigma(t)$. It is impossible for them to fully portray the behavior of all systems. The system behaves according to how the inputs are specified [66]

$$R_0(t) = (R \otimes \sigma_{max})\ (t) \qquad (2.6)$$

### 2.6.5 Derived Guarantees

Because just the arrival curve $\alpha(t)$ and service curve $\beta(t)$ have been defined, NC is able to give three assured worst-case limits. The three guarantees are as follows: delay (also known as latency); backlog (also known as memory or buffer size); and maximum output (also known as maximum output).

### 2.6.6 Delay/Latency

The delay is simply the longest horizontal distance between the arrival and service curves of a flow, as shown in Fig. 7. [67] defines it as follows:

$$d(t) \leq h(\alpha, \beta) \tag{2.7}$$

### 2.6.7 Backlog/Memory

The backlog, which is sometimes referred to as the backlog size, is the sum of all bits present in the system at a given time and is denoted by the vertical distance between the service and arrival curves.

$$R(t) - R^*(t) \leq \sup_{s \geq 0}\{\alpha(s) - \beta(s)\} \tag{2.8}$$

### 2.6.8 Maximum Output Flow

In the definition of a constraint on the output in terms of a particular arrival and service curve [68].

$$\alpha^* \ (t) = (\alpha \oslash \beta) \ (t) \tag{2.9}$$

At the start of the system, b bits are present and will stay in the worst-case scenario for T time units. During this period, it is possible that more bits will be received by the system. Assuming that only the arrival and service curves are defined, we have no way of knowing what the maximum output behavior will be, so we must assume that all bits $b_0$ will be released at the same time after the delay T and that all bits $b_0$ will be released at the same output rate as the input rate, which results in a value of (t). If we know what the service's top limit is (the maximum service curve max(t) and/or the shaper curve (t)), we can make improvements to the curve. It is possible to compute the better curve by using the following formula[66].

$$\alpha^* \ (t) = \min(\sigma(t), \ (\alpha \otimes \beta^{max}) \oslash \beta(t) \tag{2.10}$$

## 2.7 Omnet++

Since 1997, omnet++ has been offered as an open-source discrete event simulation environment built on the C++ programming language. For the last few decades, it has been widely recognized by the academic community, mostly because of its broad application. It is compatible with all widely used systems, including Windows, Mac OS, and Linux. Its primary benefit is that it offers fundamental logic and functional components that may be utilized to build more sophisticated network modeling frameworks. omnet++ has great scalability, which means that the software is modular and inter-operable, simulation models can be viewed and debugged, it has its own Integrated Development Environment (IDE), and data interfaces are as open and generic as possible [69]. It is being utilized in this study to build a simulation model for the TSN network, and it is the foundation for all performance analysis.

Compounds and simple modules are two kinds of modules that comprise the model structure of omnet++. They interact via messaging. At the bottom of the omnet++ design, C++ programs are used to code simple modules. Combining many smaller modules is how complex structures are made. Modules that use this module type may either use direct message exchange or message exchange through gates, as shown in Fig. 8. There is a pointer that points to the next and previous modules within the compound module, whereas the simple module has just one pointer that points to the previous or next module. The omnet++ Network Description Language (NED) lets the user construct a model structure. Generally, network settings may vary, and thus initial configurations are not included in the NED since they change often. INI files are used to hold initialization data [60]. omnet++ comes with a graphical editor built into its IDE. The program's graphical editor includes a tool for changing the network topology which uses either the direct route from the NED source view or via a graphical option. omnet++'s primary unit of work is the message, and in their models, messages may represent things like packets, instructions, and other entities [70]. One may set them using MSG files. omnet++, a software tool, is integrated with the Eclipse environment, and includes an analysis tool. Histograms are one of the many ways in which the outcomes of the simulation may be stored. Statistical techniques may be used to glean data that may help in making conclusions, and this is accomplished via automated usage of analysis files (ANF).

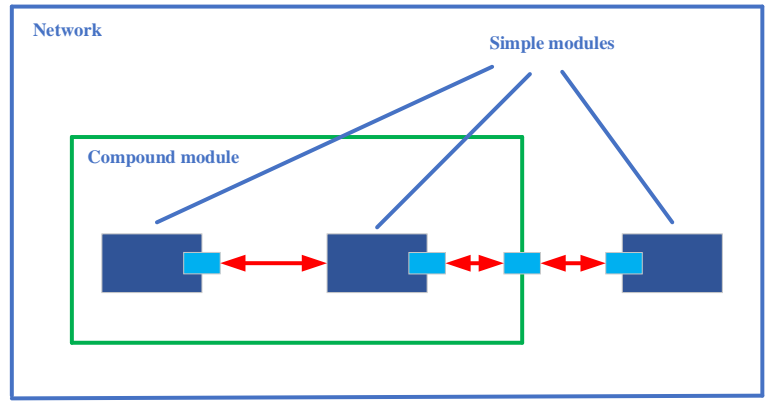

**Figure 8 Simple and Compound module**

# 3. Time Sensitive Network over Standard Ethernet

## 3.1 Time Sensitive Network

Strong, scalable, and time-constrained networks are required to offer safe, reliable, and predictable systems. In response to the widespread use and popularity of Ethernet technology, the IEEE 802.1 Time Sensitive Networking (TSN) task group began working on protocol improvements to the IEEE 802.1 Ethernet standard (Ethernet 43). This is a concise list of standards that are currently in use and that improve real-time capabilities and performance. The mixing of synchronous, asynchronous, and best-effort traffic across a single network is the most significant characteristic of TSN. TSN standards are built on the basis of AVB protocol suites, which were developed for the Audio Video Bridging (AVB) protocol. Using AVB, it can ensure that audio and video transmissions are delivered with a predefined degree of delay, as well as that jitter remains constant no matter how long the network connection is. As a result, it is necessary to ensure that there is a sufficient amount of bandwidth available for the whole transmission. In spite of the fact that AVB has been very successful and is used in a wide range of automotive networks, it is not capable of meeting the demands of highly accurate applications, such as those found in demanding real-time systems [71].

The primary objective of TSN is to concentrate on the regions of AVB sub-standards that have not yet been explored in detail. TSN task group is working on fault-tolerant synchronization methods, a time-sensitive transport protocol, and improvement techniques for the Stream Reservation Protocol in [8]order to accomplish this goal. As part of its efforts, the task group has developed a strong redundancy mechanism to avoid traffic loss in the event of a breakdown at any of the network's many tiers. The time-aware scheduling and enforcement methods included in TSN are also time-sensitive. Because of the characteristics listed above, TSN is a real-time competent, dependable, and interoperable standard that may be used in a variety of industrial automation and control networks applications (e.g. railway, avionics and automotive). Even while TTEthernet and TSN have emerged as the dominant standards in the field of TT communication, their implementation comes at a significant cost in terms of design complexity and maintenance.

## 3.2 Network architecture

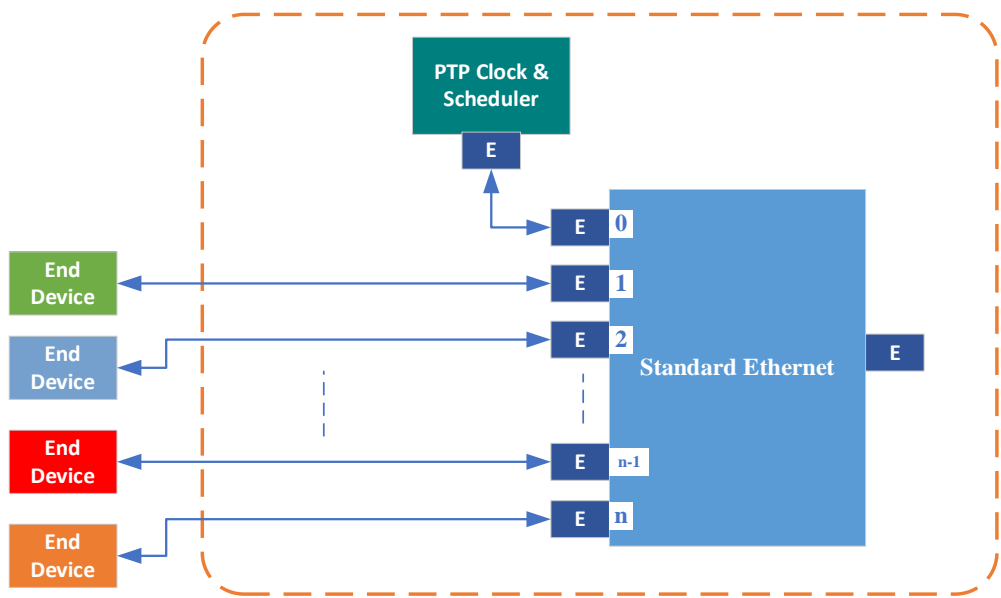

**Figure 9 Modified Standard for Time-Sensitive Network**

In Fig. 9, it is feasible for a standard switch (SS) to interact with a Precision Time Protocol (PTP) clock via an ethernet connection, provided that the ethernet port is configured to operate in full duplex mode and this fig. 9 is motivated from the author [72] though the paper used Walden switch which is applicable for CTOS switch as well as they used decoupled linker but here we have used PTP clock and scheduler. Because the scheduling of the PTP clock takes place in real time, the bandwidth of the communication channels may be controlled and planned in accordance with the demands of the users. Because PTP clocks are maintained on a separate channel, a dedicated port 0 has been designated for these purposes. The 0 port is the sole external port that is not connected to any other device. Overall, the transition is simple to utilize as a result of the PTP clocks and PTP servers that are available.

In fig. 10, Eight end systems are linked to two Ethernet-AVB switches: Switch Front (SF) is responsible for flow forwarding within the front portion of the network, and Switch Rear (SB) is in charge of flow forwarding within the back half of the network and this network is collected from in this paper which is partially modified [73]. Two cameras are placed on each side of the vehicle to provide driver assistance. One is positioned in front of the vehicle, one is installed in the rear, and the remaining two are installed on each side of the vehicle (Left Camera: LC and Right Camera: RC). VS signals (Video Signals) are collected by these cameras and then sent to Top View (T) (TV). Each of these four video streams is combined by the TV end system, which then delivers the combined visual signal to the Head Unit (HU).

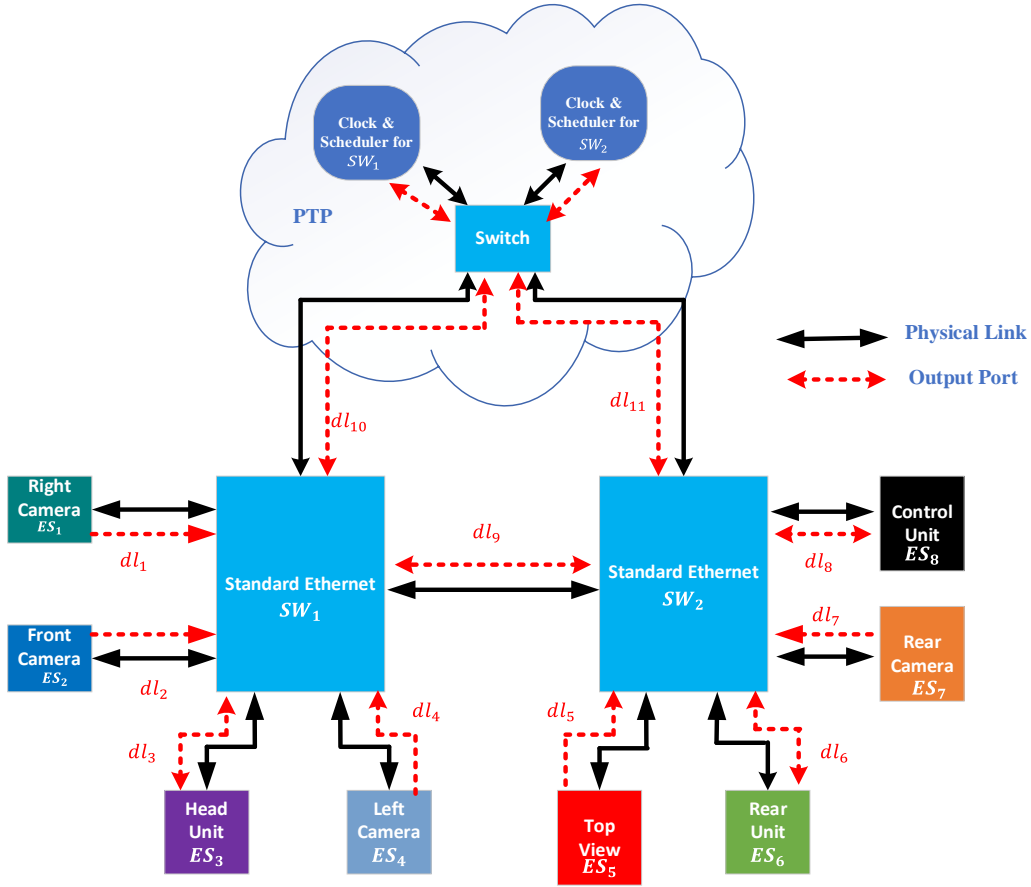

**Figure 10 TSN Network over Standard Ethernet Topology for Industrial Case**

A visual signal from the Rear Camera (RearC) is shown in conjunction with these findings on the HU end system. Additional Multimedia Video Signal (MVS), such as navigation information, is sent from HU to the Rear Unit, which is also controlled by HU (RU). Each end system receives control signals (CS) from the Control Unit (CU), which is in charge of network administration across the whole network.

Besides, PTP clock of each SS is connected with another PTP clock as a consequence, PTP clock can easily communicate between SS behavior. Fig.10 also depicts the idle-slope characteristics for SR-A and SR-B, as well as the actual total load per class. For example, the idle-slope of SR-A class for the Right Camera is 10 Mbps, while the idle-slope of SR-B class is 65 Mbps for the same camera. Total load on SR-A is currently 0.67 Mbps, while total load on SR-B is at 17.2 Mbps. Unless otherwise stated, all of these idle-slope parameters were obtained from [36].

## 3.3 Basic system symbols

Each cluster in a switch network (SW) network contributes to the overall network's performance. The end systems (ESes) in each cluster are linked by connections and switches to form a network (SWs). A buffer for the output port is included in each ES, which connects to a single input port on the SW. Each SW has no input buffers on any of its input ports, but does have an output buffer on each of its output channels. ES or another switch input port is linked to one of the output ports of SW. There is full duplex communication on the connections, which allows for communication in both directions, and the networks may be multi-hop in nature. Fig. 10 depicts an example cluster consisting of four ESes ( $ES_1$ $to$ $ES_8$ ) and three SWs ($SW_1$ $to$ $SW_3$). Table 8 demonstrate network symbol and descriptions

**Table 8 Network Symbol and Descriptions**

| Symbol | Descriptions |
|---|---|
| $G$ $(\boldsymbol{E}, \boldsymbol{V})$ | Undirected Graph |
| $\boldsymbol{V} = \boldsymbol{ES} \cup \boldsymbol{SW}$ | $\boldsymbol{V} = \{ES_1, ES_2, ES_3, ES_4, ES_5, ES_6, ES_7, ES_8\}$ $\cup \{SW_1, SW_2\}$ |
| $EW$ | End Systems |
| $SW$ | Standard Ethernet Switches |
| $dl_i = [v_a, v_b] \in \boldsymbol{L}$ | Dataflow Link, $\boldsymbol{L} = Set\ of\ data\ flow$ |
| $v_a \rightarrow v_b$ | Direction edge $v_a\ to\ v_b$ |
| $v_a \& v_b \in \boldsymbol{V}$ | ESes & SWes |
| $dl_i.C$ | Physical Link Rate |
| $h$ | Output Port |
| $\delta^S = d_{max}^S - d_{min}^S$ | Maximum jitter |
| $d_{max}^V$ | Message process real-time Units |
| $t_0$ | Nonfaulty at real-time |
| $d_{max}^E$ | All messages dispatched |

$$d_{max} = d_{max}^E + d_{min}^V \qquad \text{Maximum Delay}$$

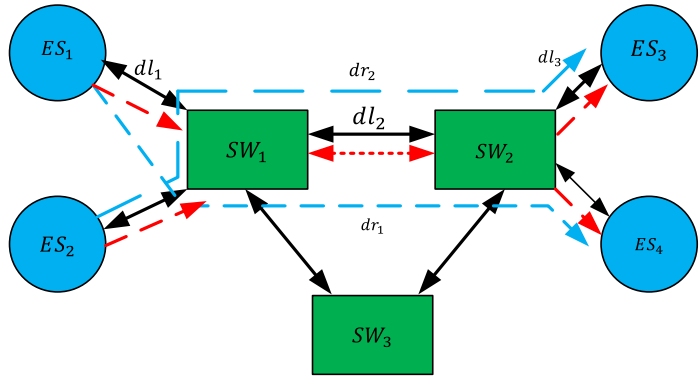

**Figure 11 Topological Example**

When several endpoints of a network are linked at the same time, this is referred to as a Time Sensitive Network. Devices (ES) and switches (SW) are connected to one another via the use of a computer network. A variety of physical cables, such as an Ethernet cable, are used. This device has full duplex and communication connection characteristics that may be used. It is possible to engage with those outside of the network as well as inside it. In real-world network applications, switches are used to link many hops between two or more nodes. A hop is the term used to describe this. Fig. 11 depicts an illustration of system's topology. There are four ESes and three SWes in the grouping. In table 8 contains some objects that are comparable to the ones mentioned above.

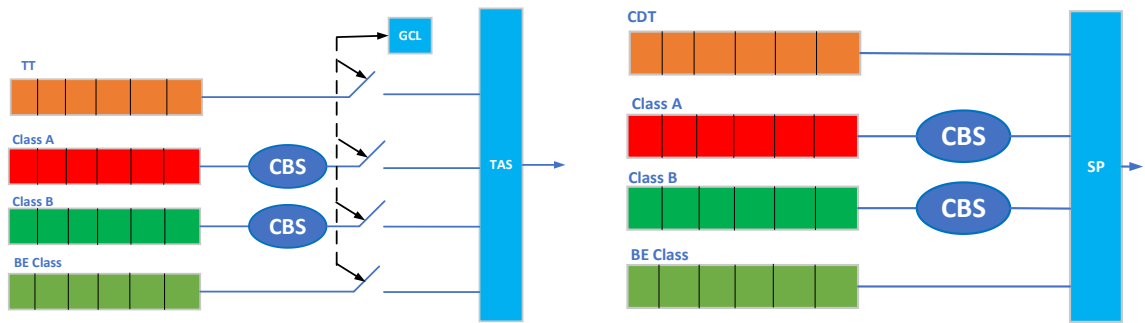

**Figure 12 Output Scheduling of the Network**

The main goals of the application model are the transfer of data from one ES to one or more ESes (destination) via a physical link between the two ESs.TT, AVB class A, AVB class B (some networks also offer AVB class C) and BE traffic types are all supported by TSN networks, as are a number of other kinds of traffic. Example of a TSN model for CBS is shown in Fig. 12. Design engineers, we have believed, are responsible for determining the number of individuals who will utilize a model. In appendix A, we have shown some optimal timing analyses of CBS scheduling. The above notation (table 8) has been developed for our system in order to make it simpler to understand.

## 3.4 Upper bound delay calculation

According to network calculus theory, the upper limit latency of a Class x flow in the output port h is determined by the greatest horizontal deviation between the arrival curve of AVB and the service curve of AVB in the output port h which is shown in fig.13.

$$d_{traffic}^h(t) = h\left(\alpha_{traffic}^h(t), \beta_{traffic}^h(t)\right), traffic \in \{TT, A, B\}$$ 3.1

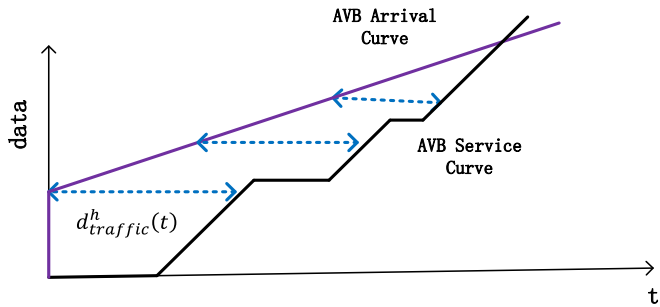

**Figure 13 Example of horizontal delay**

For calculating end to end delay, in fig. 14 demonstrates the various delays of a packets from one end device to another end delay. Consider each node has consisted FIFO queue and TSN scheduling is available and each device is connected by physical Ethernet cable. Here $T_{queue}$ is considered as FIFO queuing time, secondly $d_{traffic}^h(t)$ is considered as scheduling time, thirdly, $\frac{L_{traffic}}{C}$ is physical cable speed time. Equation 1 is total end to end delay for one intermediate node. If node more than one, then intermediate delay will be sum up according to number of nodes.

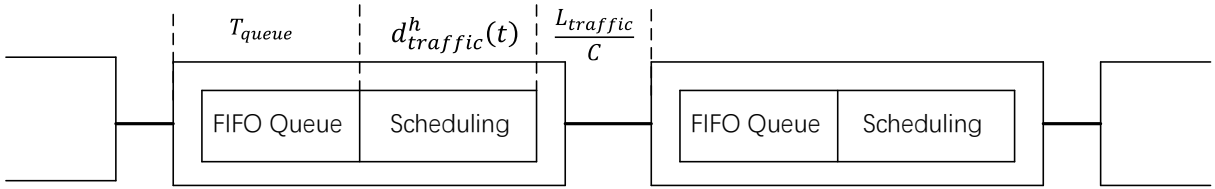

**Figure 14 Timing Model between two nodes**

$$D_{total_{traffic}}^{e2e}(t) = T_{queue} + d_{traffic}^{h}(t) + \frac{L_{traffic}}{C}$$  3.2

In equation 3.2 may be vary according to changing time $d_{traffic}^{h}(t)$ because $T_{queue}$ and $T_{queue}$ remain same.

## 3.5 Fixed Frame Synchronization

Fixed frame synchronization is a frame where consists some basic information of the several traffic for instance, frame size of the traffic, idle slope, send slope, source address, destination address, frame propagation time, ACK time, buffer time information and so on. The frame size will 70 bytes (6 bytes pay load) which is considered as tiny frame all of traffic and the size of the frame remain unchanged. This frame is basically used time synchronization frame (syn frame). This frame only transmits through the syn window.

From the PTP clock, each cycle of fixed frame synchronization is started. The reason for this is because the PTP clock has two modes of operation: one is the transmission of synchronization frame and the other is the rigorous preservation of data scheduling information. Along with the end devices, these pulses are sent and received by the nodes of the Standard switch. The fixed frame is treated in order to reduce implementation challenges and avoid being involved in authentication and masquerading issues.

To understand fixed frame synchronization, at first, we consider data are sending from $ES_8$ to $ES_2$ in fig. 15 and in fig. 16 we have given an example of fixed synchronous frame.

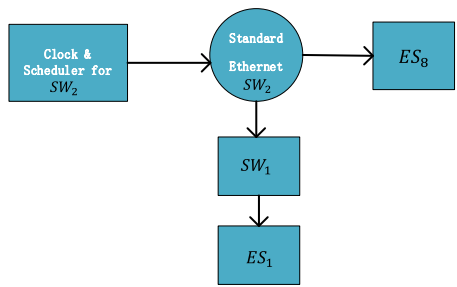

**Figure 15 Example of Sending fixed Synchronous Frame from Source to Destination**

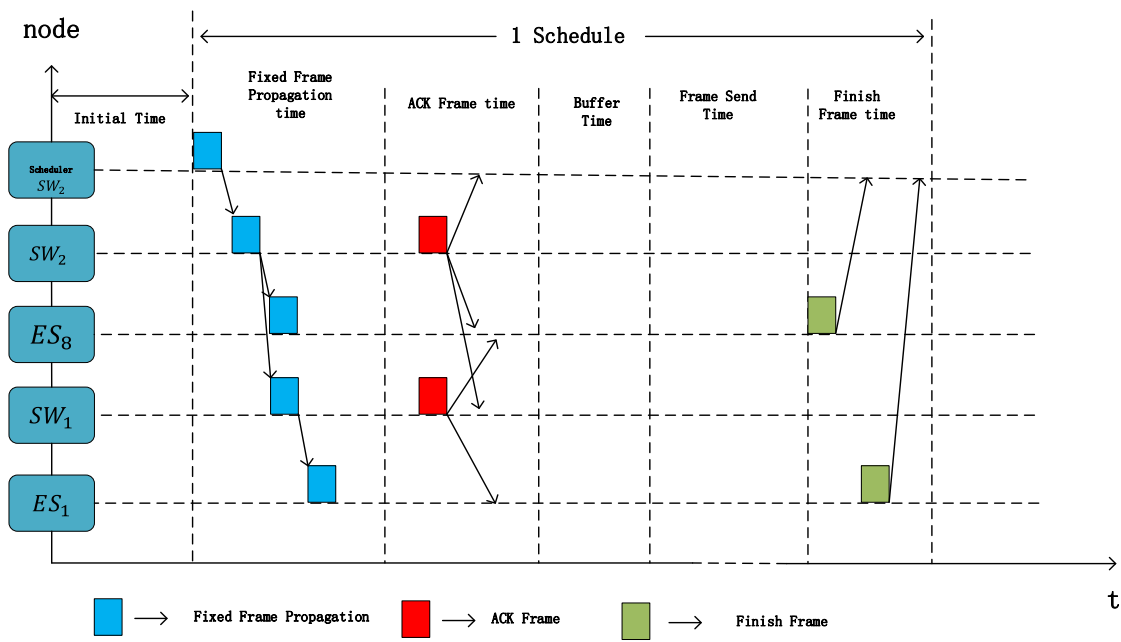

**Figure 16 Scheduling Synchronization from Clock Synchronization and Destination**

## 1) *Initial time State*

In fig. 16, initialization is the stage in which each node gets a fixed frame upon arrival, and it is referred to as the setup stage. Receiving a response to each stage of the startup process has a specified time limit. If no frame is received in the initial state, the initialization time will be halted and reset. After then, everything will start over from the beginning again. Initial time state slot should be 0.20ms.

## 2) *Fixed Frame Propagation Time*

In fig. 16, each node is assigned a predefined frame upon arrival, which is referred to as the starting stage. It is possible to break down the initialization process into several stages, each of which has a specified time restriction for receiving a response. No further initialization time will be allowed, and everything is going to be reset if no frames are received during the initialization step. After then, it will go back to the beginning and repeat the procedure once again. In fig. 17, we have shown fixed ethernet frame for synchronization. Here, in payload size 32 bytes divided into for 4 sub part, 1st 8 bytes allocated for idle slope information of actual frame (AVB),2nd 8 bytes allocated for send slope of actual frame (AVB),3rd 8 bytes allocated for frame length of the actual frame (TT, AVB),4th 8 bytes allocated from frame length of the type of traffic like TT, AVB. Fixed frame propagation time slot should less 0.5 ms.

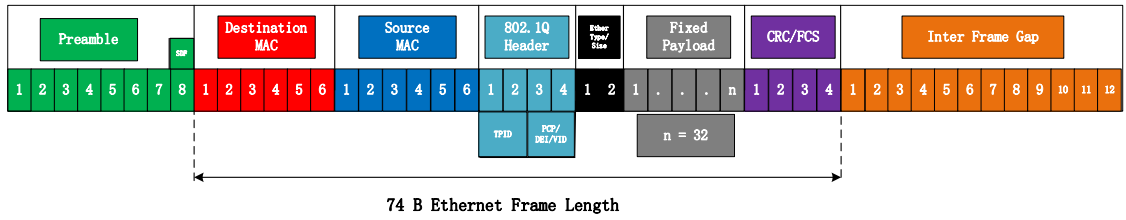

**Figure 17 Fixed Frame Propagation Time frame**

### 3)  *ACK(Acknowledge) Frame time*

In fig. 16, fixed frame propagation timing will be completed when the fixed frame reaches the ACK frame time. This is the condition in which the main is responsible for sending an ACK signal to the sender in order to inform that a fixed frame has been received. SW nodes are responsible for transmitting these ACK signals. In fig. 18, fixed ethernet frame for ack and the frame size is 64 bytes and payload 22 bytes. In the 22 bytes payload, only ack information will be delivered frame time slot should be less 1 ms.

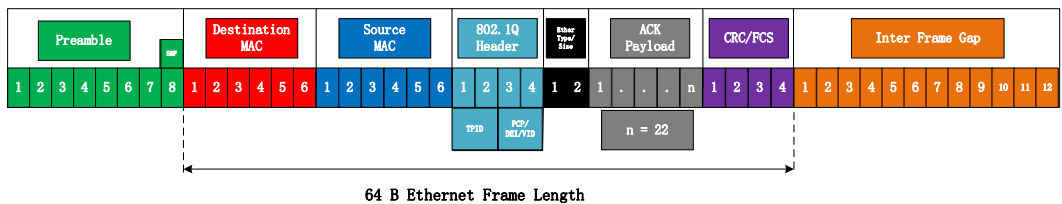

**Figure 18 ACK(Acknowledge) Frame**

### 4)  *Buffer Time*

In fig. 16, each node enters the buffer time state after completing the ACK time. If a node has the buffer time state enabled, it may use it to estimate and then fix the synchronization error of its local state-timer by setting the state-timer with a new value while the state is enabled. If no changes were made, the state-timer would automatically expire after a specified amount of time if no adjustments were made. Maximum buffer time slot should be not more than 0.25 ms.

### 5)  *Frame Send Time*

In fig.16, when a node's timeout in the buffer time state expires, the node is automatically moved into the frame send time state. The condition of frame send time is larger than the sum of all of the phases combined. In accordance with the schedule, frame send times may be either TT frame sends or AVB frame sends. Actual frame send timing are discuss rest of the consecutive chapters in details.

*6) Finish Frame time*

In fig. 16, after completing frame transmit time, the ES node should send a completion signal to the PTP clock via the SW, as a result, the PTP clock may be made aware of the receipt, which is very essential in a deterministic network like the one described above. In fig.19, fixed ethernet frame for finish and the frame size is 64 bytes and payload 22 bytes. In the 22 bytes payload, only finish information will be delivered. Finish Frame time slot should be less than 0.2ms.

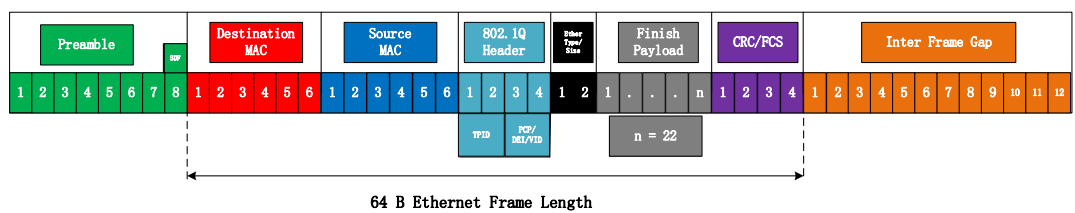

**Figure 19 Finish Frame**

## 3.6 Synchronization Algorithms

The basic synchronization algorithm for the ES and SW nodes are presented in table 9. In the pseudo-codes, we have demonstrated algorithm in the basis of initial time state, fixed frame propagation time, ACK time state, buffer time, frame send time, finish time.

**Table 9 Synchronization Algorithm**

| Synchronization Algorithm |
|---|
| 1    *// Initial time state* |
| 2    *send fixed frame to SW from PTP clock* |
| 3    *if    fixed frame received* |
| 4    │    *Wait until initial time finished* |
| 5    *end* |
| 6    *else* |
| 7    │    *Reset initial time and start from begin* |
| 8    *end* |
| 9    *// Fixed Frame Propagation Time* |
| 10    *if    Fixed frame received* |

| 11 | | *Frame Delivery to Adjacent node according to Schedule* |
| 12 | ***end*** | |
| 13 | ***else*** | |
| 14 | | *Fixed time will not be start until getting any frame from initial state* |
| 15 | ***end*** | |
| 16 | ***//ACK Frame Time*** | |
| 17 | ***if*** | *Fixed frame received* |
| 10 | | *Send ACK signal to source and destination node* |
| 11 | ***end*** | |
| 12 | ***else*** | |
| 13 | | *Fixed time will not be start until getting any frame from previous state* |
| 14 | ***end*** | |
| 15 | ***//Buffer Time*** | |
| 16 | ***if*** | *(Fixed frame size || arrival fixed time) > ACK frame Time* |
| 17 | | *Buffer time slot is used* |
| 18 | ***end*** | |
| 19 | ***else*** | |
| 20 | | *Buffer will be eliminated to minimize synchronization time* |
| 21 | ***end*** | |
| 22 | ***//Frame Send time*** | |
| 23 | ***if*** | *Frame send time is start* |
| 24 | | ***While***    *Source to destination channel is free* |
| 25 | | *Send TT or AVB according to scheduling* |
| 26 | | ***end*** |
| 27 | ***end*** | |
| 28 | ***elseif*** | *Frame send time is finished* |
| 29 | | *Stop to send TT and AVB* |
| 30 | ***end*** | |
| 31 | ***elseif*** | *No synchronization frame arrives in Frame Send Time* |
| 32 | | *No data will not be sent* |
| 33 | ***end*** | |
| 34 | ***// Finish Frame Time*** | |
| 35 | ***if*** | *Finish frame send time* |
| 35 | | *ES send finish signal to PTP clock* |
| 37 | | *Terminate schedule* |
| 38 | ***end*** | |

## 3.7 TSN integration mode

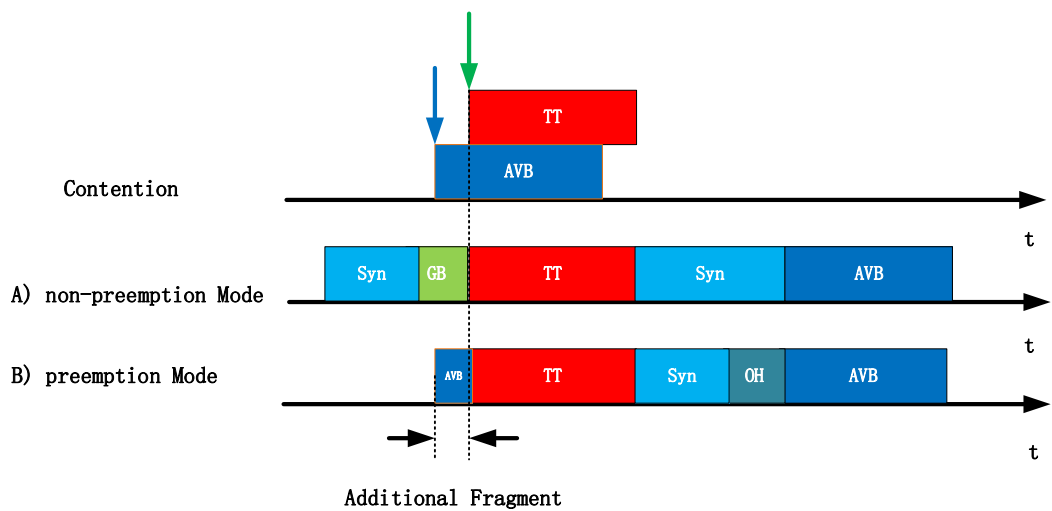

**Figure 20 TSN integration mode**

TSN makes advantage of two different integration modes. In fig. 20A, a non-preemption "syn-window" is presented before each time window of the TT frame [8][33], which prevents preemption from occurring. Syn-window is dedicated window for sending synchronization frame which is already discuss in section 3.5. When a guard band is used, it is the length of the largest possible frame that may interfere with the relevant TT traffic. In the worst-case scenario, this is the Ethernet Maximum Transmission Unit (MTU), which is 1518 bytes. During the syn-band, the gates associated with AVB and BE traffic are closed in advance to ensure that the link is idle when a TT queue is ready to transmit data to the destination network. In addition to wasting bandwidth owing to the guard band, the non-preemption integration option assures that TT traffic is not delayed at all. According to IEEE 802.1Qbu [74][75], the preemption mode, which is shown in Fig. 20b, is the alternate integration option to choose from. When preemption is used, an AVB frame will be interrupted by a TT frame, and its transmission will be restarted from the place where it was stopped while the TT frame was being sent. As soon as the associated gate for the preemptable AVB frame that is now in transmission is available for opening, the related gate for the TT queue is opened for the period necessary to complete the transmission of a fragment before the TT queue may be opened [76]. When transmission is resumed, the remaining AVB frame will contain an overhead that will be used to split and reassemble the frames at the destination ES when the transmission is resumed. Due to the fact that overhead may be minor when compared to the guard band, the use of preemption can reduce the latency of AVB traffic while simultaneously increasing capacity. In the case of TT traffic, it will, however, produce some jitter.

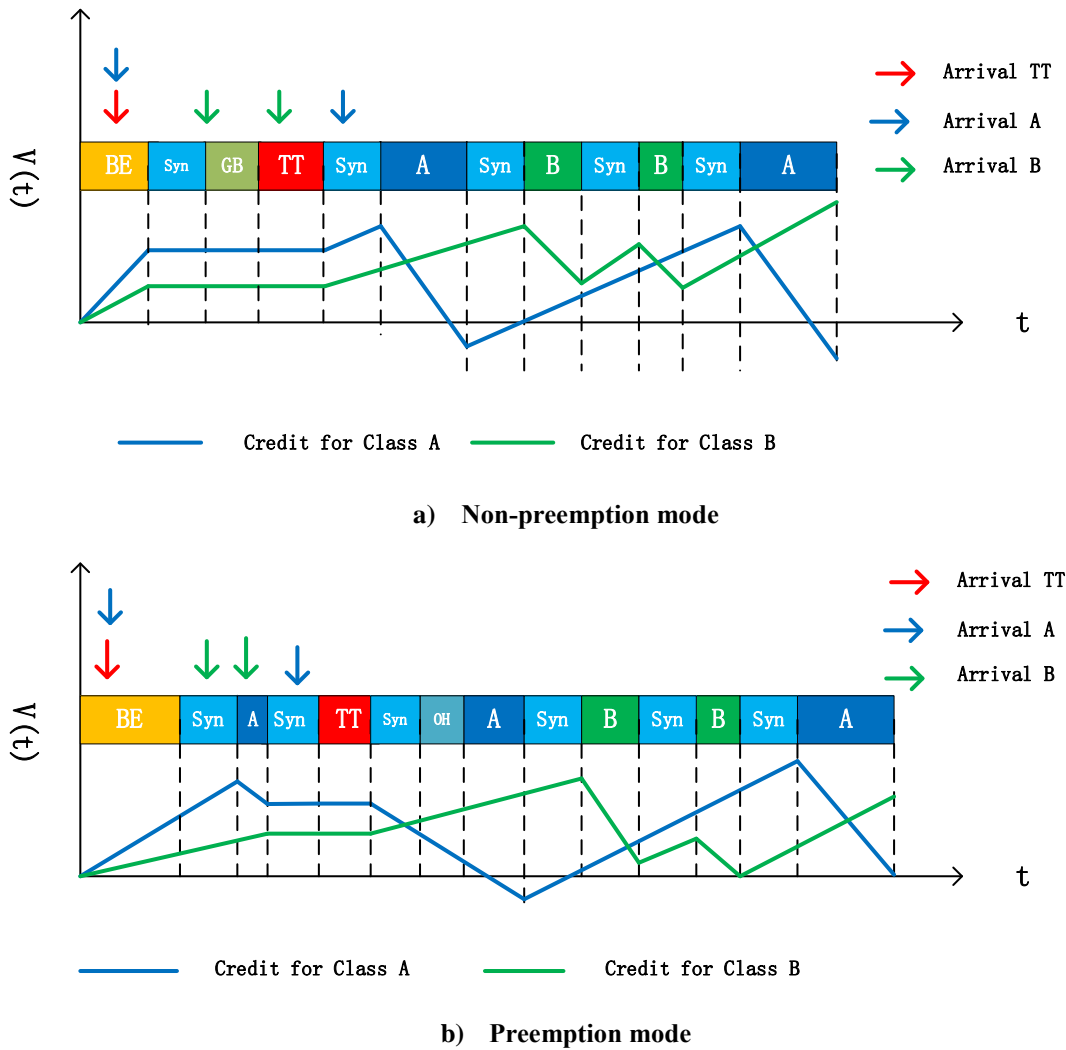

a) **Non-preemption mode**

b) **Preemption mode**

**Figure 21 CBS example for two modes**

Additionally, the availability of an AVB queue is governed by a Credit-Based Shaper (CBS), whose primary function is to prevent lower priority flows from being starved. Accordingly, when (i) the queue gate is open, (ii) the CBS permits it, and (iii) no other higher priority AVB frames are being sent, an enqueued AVB frame is permitted to be broadcast. Begin by outlining the manner in which CBS is implemented in TSN. Several credits are awarded for each AVB class. In order for an AVB frame to be delivered, this credit must not be negative at the time of transmission. Furthermore, the credit is cancelled out as well.

During the transmission of an AVB Class frame, if the AVB queue is not empty, credit may be lowered with a send slope of $S^x$ $Where\ x \in A, B$, and credit can be raised with an idle slope $I^x$ $Where$ $x \in A, B$, while Class M frames are waiting to be communicated. Furthermore, if the AVB queue is empty and its credit is positive, the credit is set to zero; otherwise, the credit is raised with the idle slope until it reaches zero. If the gate for the linked AVB traffic queue is open in the TSN network, the situation is the same as it is in the WAN network. Furthermore, when the time gate is closed in a TSN network, there must be extra attention given to AVB credit, which is currently an unresolved subject [75]. In order to prevent credit overflow, we are particularly interested in the frozen form in this study, which is to say that the credit is frozen when the related AVB gate is shut. In fig. 21a and 21b, we demonstrate how CBS works interfered with TT and AVB frames, respectively, when using two different integration modes, using two different integration modes. Transmission times of frames are shown by rectangles on the first-time line, while arrival times of frames are represented by down arrows at the top.

**Table 10 Summary of Notation**

| Symbols | Description |
|---------|-------------|
| $V^x$ | Credit of AVB class $x \in A, B$ |
| $N^h$ | Number of TT traffic in h port |
| $L_{TT}^h$ | TT windows in h port |
| $L_{syn+GB}^h$ | Syn windows along with GB Band |
| $o^h$ | Offset |
| $L_{syn}^h$ | Syn window |
| $\alpha_{syn+TT}^{h,*}$ | Aggregate Arrival Curve (* different scheduling mechanism) |
| $L_{syn+OH}^h$ | Syn windows along with OH Band |
| $T_{GCL}$ | Hyper-period |
| $\beta_{AVB,x}^{h,*,**}$ | Service Curve of AVB traffic (* different mode, **different scheduling mechanism) |
| $C$ | Physical Link rate |
| $V_{max}^x$ | Maximum credit of AVB |
| $ES$ | End Systems |
| $SW$ | Network Switches |
| $I^x$ | Idle slope of class x |

| $S^x$ | Send slope of class x |
|---|---|
| $\overline{L^x}$ | Maximum frame of class x |
| $t_{syn}$ | Syn time |
| $\Delta t = t_2 - t_1$ | $t_2$= final credit time of a frame, $t_1$ = initial credit time of a frame |
| $\Delta t^+$ | Increasing credit time |
| $\Delta t^-$ | Decreasing credit time |
| $\Delta t^0$ | Frozen time |

In table 11, we have used following non-preemption mode algorithm for time sensitive network over standard ethernet switch.

**Table 11 Non-preemption mode algorithm**

**Non-preemption mode Algorithm**

| | |
|---|---|
| *1* | ***// Beginning of the data transfer*** |
| *2* | $\boldsymbol{V^x \leftarrow 0}$ , |
| *3* | ***if*** $\boldsymbol{t_{BE} > t_{avb}}$   ***//$t_{BE}$ ← BE Arrival time, $t_{avb}$ ← avb Arrival time*** |
| *4* | ***do*** *(BE frame are transferring)* |
| *5* | $V^x \leftarrow V^x + I^x . \dfrac{\overline{L^x}}{C}$ |
| *6* | ***until*** *BE frame finished* |
| *7* | ***end*** |
| *8* | ***if*** *(BE frame transfer finished && TT frame available in Queue)* |
| *9* | ***do*** *(TT frame send through  $L_{TT}$  window)* |
| *10* | $L_{GB} \leftarrow Guard\ Band$ |
| *11* | $V^x \leftarrow V^x + 0$ |
| *12* | ***until*** *TT frame send frame finish* |
| *13* | ***end*** |
| *14* | ***if*** *(BE frame transfer finished && No TT frame available in Queue)* |
| *15* | ***do*** *(avb frame send through  $\boldsymbol{L_{avb}}$  window)* |
| *16* | $V^x \leftarrow V^x + S^x . \dfrac{\overline{L^x}}{C}$ |
| *17* | ***until*** $\boldsymbol{V^x \leftarrow V^x_{min}}$ |
| *18* | ***end*** |

In table 12, we have used following preemption mode algorithm for time sensitive network over standard ethernet switch.

**Table 12 Preemption mode algorithm**

| | **Preemption mode Algorithm** |
|---|---|
| *1* | *// **Beginning of the data transfer*** |
| *2* | $V^x \leftarrow 0$ , |
| *3* | ***if*** $t_{BE} > t_{avb}$   *//$t_{BE} \leftarrow BE\ Arrival\ time, t_{avb} \leftarrow avb\ Arrival\ time$* |
| *4* |     ***do** (BE frame are transferring)* |
| *5* |       $V^x \leftarrow V^x + I^x . \dfrac{\overline{L^x}}{C}$ |
| *6* |     ***until** BE frame finished* |
| *7* | ***end*** |
| *8* | ***if** (BE frame transfer finished && No TT frame available in Queue)* |
| *9* |     ***do** (avb frame send through $L_{avb}$ window)* |
| *10* |       $V^x \leftarrow V^x + S^x . \dfrac{\overline{L^x}}{C}$ |
| *11* |       ***if** (TT arrival to transfer)* |
| *12* |         ***do** (TT frame send through $L_{TT}$ window)* |
| *13* |           $V^x \leftarrow V^x + I^x . \dfrac{\overline{L^x}}{C}$ |
| *14* |         ***Until** TT frame finished* |
| *15* |     *end* |
| *16* |     $L_{OH} \leftarrow Overhead\ Band$ |
| *17* |     $V^x \leftarrow V^x + 0$ |
| *18* |   ***until*** $V^x \leftarrow V^x_{min}$ |
| *19* | ***end*** |

### 3.7.1 Service Curve for TT and AVB traffic with non-preemption in CBS+TAS

As a result of the availability of TT traffic, we address the service curve analysis for AVB Class x where x ∈ A, B accessible in an output port h in this section. The non-preemption and preemption modes are used to analyze the service curves, respectively. Starting with the aggregate arrival curve, we should evaluate the influence of TT traffic on the output port h, since the remaining service for AVB traffic is dependent on it. According to [77], the TT aggregate arrival curve is created by adding the arrival curves of each single intersecting periodic TT flow changing with relative offsets to get the TT aggregate arrival curve. TT queue gate states are controlled by GCLs in TSN, but not by the TT frames themselves. If the end systems are not scheduled [8], the nature of the periodicity in TT flows may change throughout the transmission line as a result of the non-scheduling. By using the TT traffic window, we are able to calculate the aggregate arrival curve with an influence from TT traffic in this study. In fig. 22, aside from that, the constraint of syn-band effect for the non-preemption mode is integrated into the TT aggregate arrival curve, and the constraint curve of overheads for the preemption mode is constructed as well.

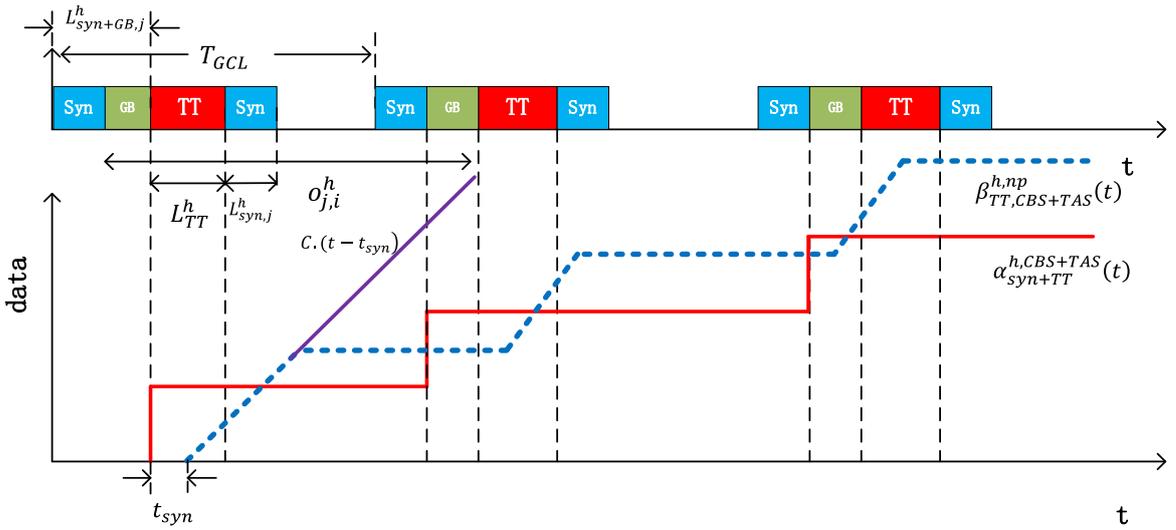

**Figure 22 Example of arrival curve and service curve corresponding windows and guard bands**

**Lemma 3.1** Under non-preemption conditions, the aggregate arrival curve for intersecting TT flows and syn-bands in an output port h is given by the equation

$$\alpha_{syn+TT}^{h,CBS+TAS}(t) = max_{0 \leq i \leq N^h-1} \left\{ \sum_{j=i}^{i+N^h-1} \left( L_{TT,j}^h + L_{syn+GB,j}^h \right).C.\left\lceil \frac{\left(t-o_{j,i}^h+t_{syn,j}^h-L_{syn+GB,j}^h\right)}{T_{GCL}} \right\rceil \right\} \qquad 3.3$$

This is identical to the curves in equation 3.3. This is obtained by multiplying the worst-case number of bits generated by the twice of $L_{syn,j}^h$ syn-window by the maximum number of bits broadcast in the $L_{TT,j}^h$ TT traffic window, both of which cause AVB traffic to be delayed until after the service for AVB traffic has been restored to normal. Aside from that, the distance between new "Syn+TT" window+band may vary as a result of the possibility of a varied length of syn-window.

**Lemma 3.2** Under non-preemption conditions, the Service curve for intersecting TT flows and Syn-band in an output port h is given by the equation

$$\beta_{TT,CBS+TAS}^{h,np}(t) = C.\max\left(\left\lfloor \frac{t-t_{syn}}{T_{GCL}} \right\rfloor L_{TT}^h, t - \left\lceil \frac{t-t_{syn}}{T_{GCL}} \right\rceil \left(T_{GCL} - L_{TT}^h\right)\right)^+ \qquad 3.4$$

In equation 3.4 refers that TT frames are flowing according to opening TT window, on the other hands the seconds term of the equation 3.4 indicates when the TT windows will be finished, the value of the will be frozen. An example of the remaining service curve for AVB flows is given by the blue dotted line in fig. 22, and the aggregate arrival curve of TT flows also shown.

***Theorem 3.1*** *According to non-preemption mode, service curve of AVB class x where* $x \in \{A, B\}$ *in an output port of ethernet switch port h,*

$$\beta_{AVB,x}^{h,np}(t) = I^x\left[\Delta t - \frac{\alpha_{syn+TT}^{h,CBS+TAS}(\Delta t)}{C} - \frac{V_{max}^x}{I^x}\right]^+ \qquad 3.5$$

*Where np indicates the non-preemption integration mode.*

***Proof:*** s is the beginning of the most recent phase of server overload. All backlogs for all flows have been cleared at this point in time such as $O_x^{h*}(s) = O_x^h(s)$, $O_{TT}^{h*}(s) = O_{TT}^h(s)$, $O_{syn}^{h*}(s) = O_{syn}^h(s)$ and $V^x(s) = 0$

With respect to any given time $t \geq s$, the interval between two consecutive times $\Delta t = t - s$ may be decomposed as follows:

$$\Delta t = \Delta t^+ + \Delta t^- + \Delta t^0 \qquad 3.6$$

Because of syn bands and TT traffic windows in non-preemption mode, time zero is created by

$$\Delta t^0 = \Delta t_{TT}^0 + \Delta t_{syn}^0 \qquad 3.7$$

where t denotes the length of time between AVB Class x frames. As a result, we have

$$\Delta t^- = \Delta t_{AVB,x}^- \qquad 3.8$$

Thus, the change in credit throughout the $\Delta t$ period fulfills the requirement.

$$V^x(t) - V^x(s) = V^x(t) \qquad\qquad 3.9$$

$$= \Delta t^+ . I^x + \Delta t^+ . S^x \qquad\qquad 3.10$$

$$= \left(\Delta t - \left(\Delta t_{TT}^0 + \Delta t_{syn}^0\right)\right) . I^x - \Delta t_{AVB,x}^- . (I^x - S^x) \qquad 3.11$$

Since this is the case, we may utilize non-preemption mode to determine the relationship between the AVB Class x service times, the TT traffic times, and the syn band timings in any given time period $\Delta t$,

$$\Delta t_{AVB,x}^- = \frac{\left\{\left(\Delta t - \left(\Delta t_{TT}^0 + \Delta t_{syn}^0\right)\right) . I^x - V^x(t)\right\}}{C} \qquad 3.12$$

Furthermore, under the worst-case scenario, the output frames of TT traffic during the time interval $\Delta t$ may be calculated using

$$O_{TT}^{h*}(t) - O_{TT}^{h*}(t) = O_{TT}^{h*}(t) - O_{TT}^h(s) = C . \Delta t_{TT}^0 \qquad 3.13$$

Similarly, the time lost owing to syn bands during $\Delta t$ is a loss of time.

$$O_{syn}^{h*}(t) - O_{syn}^{h*}(t) = O_{syn}^{h*}(t) - O_{syn}^h(s) = C . \Delta t_{syn}^0 \qquad 3.14$$

So, $\Delta t^0 = \Delta t_{TT}^0 + \Delta t_{syn}^0$ in order to limit

$$\Delta t^0 \leq \left\{\left(O_{TT}^h(t) + O_{syn}^h(t)\right) - \left(O_{TT}^h(s) + O_{syn}^h(s)\right)\right\}/C \qquad 3.15$$

$$\leq \frac{\alpha_{syn+TT}^{h,CBS+TAS}(\Delta t)}{C} \qquad 3.16$$

Then, based on (3.13) and (3.15), the AVB Class x output frames over the interval t are constrained by

$$O_x^{h*}(t) - O_x^{h*}(s) = C . \Delta t_{AVB,x}^- \qquad 3.17$$

$$\geq I^x \left[\Delta t - \frac{\alpha_{syn+TT}^{h,CBS+TAS}(\Delta t)}{C} - \frac{V_{max}^x}{I^x}\right] \qquad 3.18$$

$$\beta_{AVB,x}^{h,np,CBS+TAS}(t) = I^x \left[\Delta t - \frac{\alpha_{syn+TT}^{h,CBS+TAS}(\Delta t)}{C} - \frac{V_{max}^x}{I^x}\right]^+ \qquad 3.19$$

### 3.7.2 Service Curve for TT and AVB traffic with preemption in CBS+TAS

When using the preemption mode, if an AVB frame is preempted, an overhead is imposed to the next AVB frame that is still active. TT traffic windows preempt AVB Class x frames in the worst-case scenario, as seen in Fig. 23. We assume for the moment that the length of the overhead along with syn-window is $L_{syn,j}$. Overheads may be considered as a distinct factor contributing to the delay of AVB data transmission. Because the overhead will only exist immediately after each TT traffic window, the overhead constraint curve may be calculated using the following Lemma 3.3.

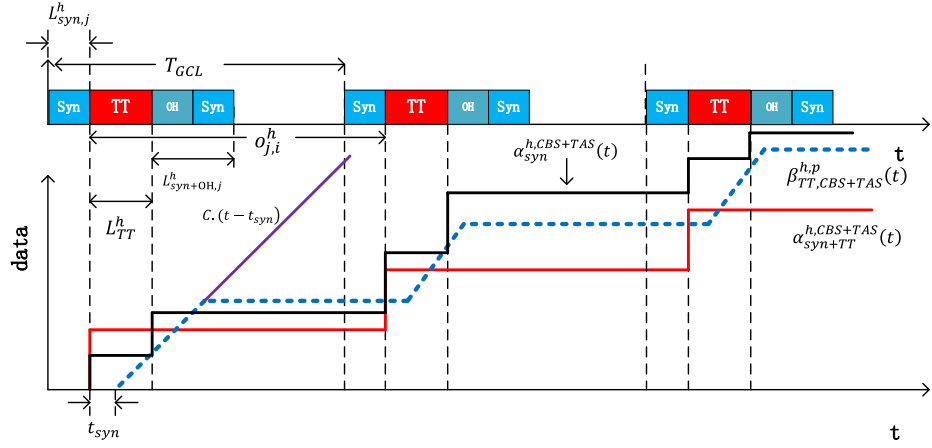

**Figure 23 Example of arrival curve and service curve corresponding windows and overhead bands**

**Lemma 3.3** Under preemption conditions, the aggregate arrival curve for intersecting TT flows and syn-band along with overhead in an output port h is given by the equation

$$\alpha_{syn}^{h,CBS+TAS}(t) = \max_{0 \le i \le N^h-1} \left\{ \sum_{j=i}^{i+N^h-1} L_{syn+OH} . C . \left\lceil \frac{\left(t - O_{j,i}^h - L_{TT,j} - L_{syn,j}\right)}{T_{GCL}} \right\rceil \right\} \qquad 3.20$$

Here, $L_{TT,j}$ is the TT traffic frame length in j position.

For consideration service curve of TT traffic of preemption mode is same as Lemma 3.2

**Theorem 3.2** *According to non-preemption mode, service curve of AVB class x where $x \in \{A, B\}$ in an output port of ethernet switch port h,*

$$\beta_{AVB,x}^{h,p,CBS+TAS}(t) = I^x \left[ \Delta t - \frac{\alpha_{TT}^{h,CBS+TAS}(\Delta t)}{C} - \frac{\alpha_{syn}^{h,CBAS+TAS}(\Delta t)}{I^x} - \frac{V_{max}^x}{I^x} \right]^+ \qquad 3.21$$

*Where p indicates the preemption integration mode.*

***Proof:*** s is the beginning of the most recent phase of server overload. All backlogs for all flows have been cleared at this point in time such as $O_x^{h*}(s) = O_x^h(s)$, $O_{TT}^{h*}(s) = O_{TT}^h(s)$, $O_{syn}^{h*}(s) = O_{syn}^h(s)$ and $V^x(s) = 0$

With respect to any given time $t \ge s$, the interval between two consecutive times $\Delta t = t - s$

may be decomposed as follows:

$$\Delta t^0 = \Delta t_{TT}^0 \qquad\qquad 3.22$$

Preemption-related overheads may be divided into two categories: the durations of frame transmission in AVB Class x and the durations of overheads owing to preemption.

$$\Delta t^- = \Delta t_{syn}^- + \Delta t_{AVB,x}^- \qquad\qquad 3.23$$

$\Delta t$ fulfills the credit variation requirement

$$V^x(t) - V^x(s) = V^x(t) \qquad\qquad 3.24$$

$$= \Delta t^+ . I^x + \Delta t^+ . S^x \qquad\qquad 3.25$$

$$= \left(\Delta t - (\Delta t_{TT}^0)\right) . I^x - (\Delta t_{syn}^- + \Delta t_{AVB,x}^-) . C \qquad\qquad 3.26$$

As a result, using preemption mode, we can determine the connection between service times for AVB Class x, extra service time for AVB overheads, and the length of TT windows in any interval $\Delta t$.

$$\Delta t_{AVB,x}^- = \frac{\left\{\left(\Delta t - (\Delta t_{TT}^0)\right) . I^x - \Delta t_{syn}^- . C - V^x(t)\right\}}{C} \qquad\qquad 3.27$$

The output frames of TT traffic during t are higher in the worst-case scenario.

$$O_{TT}^{h*}(t) - O_{TT}^{h*}(t) = O_{TT}^{h*}(t) - O_{TT}^{h}(s) = C . \Delta t_{TT}^0 \qquad\qquad 3.28$$

During the time period $\Delta t$, AVB syn band may be covered by an additional service offered by

$$O_{syn}^{h*}(t) - O_{syn}^{h*}(t) = O_{syn}^{h*}(t) - O_{syn}^{h}(s) = C . \Delta t_{syn}^0 \qquad\qquad 3.29$$

So, $\Delta t_{TT}^0$ in order to limit

$$\Delta t_{TT}^0 \leq \frac{\left\{O_{TT}^{h*}(t) - O_{TT}^{h}(s)\right\}}{C} \leq \frac{\alpha_{TT}^{h,CBS+TAS}(\Delta t)}{C} \qquad\qquad 3.30$$

in the same way that $\Delta t_{syn}^-$ fulfills,

$$\Delta t_{syn}^- \leq \frac{\alpha_{syn}^{h,CBS+TAS}(\Delta t)}{C} \qquad\qquad 3.31$$

According to (3.27), (3.29), Class x output frames over the interval t are limited by, respectively.

$$O_x^{h*}(t) - O_x^{h*}(s) = C . \Delta t_{AVB,x}^- \qquad\qquad 3.32$$

$$\geq I^x \left[\Delta t - \frac{\alpha_{TT}^{h,CBS+TAS}(\Delta t)}{C} - \frac{\alpha_{syn}^{h,CBS+TAS}(\Delta t)}{I^x} - \frac{V_{max}^x}{I^x}\right] \qquad\qquad 3.33$$

$$\beta_{AVB,x}^{h,p,CBS+TAS}(t) = I^x \left[\Delta t - \frac{\alpha_{TT}^{h,CBS+TAS}(\Delta t)}{C} - \frac{\alpha_{syn}^{h,CBS+TAS}(\Delta t)}{I^x} - \frac{V_{max}^x}{I^x}\right]^+ \qquad\qquad 3.34$$

### 3.7.3    Service Curve for CDT and AVB traffic with non-preemption in CBS+SP

The following theorem establishes the service curves provided by a CBS at a TSN node for AVB flows in the presence of CDT flows with an LB arrival curve. The authors of [8] calculate service curves for AVB flows in accordance with the IEEE AVB standard [10], i.e., without Control data traffic (CDT). Notably, [12] proposes service curves for AVB flows in TSN; however, credit reset is not included in their argument. We construct service curves that are distinct from those in [12] and utilize them to get tight delay bounds.

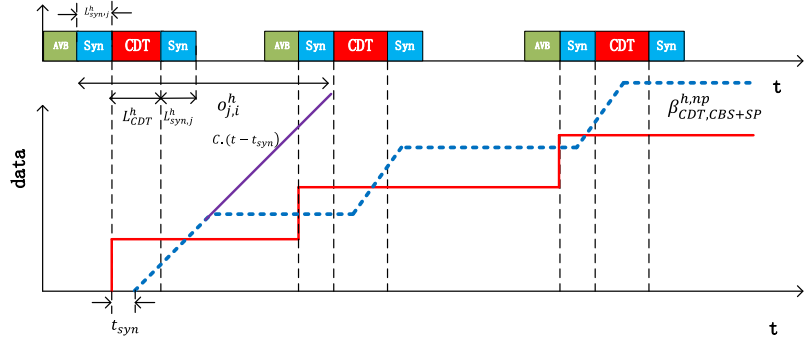

**Figure 24 Example of arrival curve and service curve corresponding windows and Guard bands**

**Lemma 3.4** Under non-preemption conditions, the aggregate arrival curve for intersecting CDT flows and syn window in an output port h is given by the equation

$$\alpha_{syn+CDT}^{h,CBS+SP}(t) = \max_{0 \le i \le N^h-1}\left\{\sum_{j=i}^{i+N^h-1}\left(2L_{CDT,j}^h\right).C.\left\lceil\left(t-o_{j,i}^h\right)\right\rceil\right\} \qquad 3.35$$

This is identical to the curves in equation 3.35. This is obtained by multiplying the worst-case number of bits generated by the twice of $L_{syn,j}^h$ syn-band by the maximum number of bits broadcast in the $L_{CDT,j}^h$ TT traffic window, both of which cause AVB traffic to be delayed until after the service for AVB traffic has been restored to normal. Aside from that, the distance between new "Syn+CDT" windows may vary as a result of the possibility of a varied length of syn-band.

**Lemma 3.5** Under non-preemption conditions, the Service curve for intersecting CDT flows and Syn. bands in an output port h is given by the equation

$$\beta_{CDT,CBS+SP}^{h,np}(t) = C.\max\left(\left\lfloor t-t_{syn}\right\rfloor L_{CDT}^h, t-\left\lceil t-t_{syn}\right\rceil\left(T-L_{CDT}^h\right)\right)^+ \qquad 3.36$$

In equation 3.34 refers that CDT frames are flowing according to opening CDT frame, on the other hands the seconds term of the equation 3.36 indicates when the CDT frame will be finished, the value of the will be frozen. An example of the remaining service curve for AVB flows is given by the blue dotted line in Fig. 24, by considering the aggregate arrival curve of CDT flows from Fig. 24.

**Theorem 3.3** *According to non-preemption mode and CBS+SP scheduling, service curve of AVB class x where $x \in \{A, B\}$ in an output port of ethernet switch port h,*

$$\beta_{AVB,x}^{h,np,CBS+SP}(t) = I^x \left[ \Delta t - \frac{\alpha_{syn+CDT}^{h,CBS+SP}(\Delta t)}{C} - \frac{V_{max}^x}{I^x} \right]^+ \qquad 3.37$$

*Where np indicates the non-preemption integration mode.*

**Proof:** s is the beginning of the most recent phase of server overload. All backlogs for all flows have been cleared at this point in time such as $O_x^{h*}(s) = O_x^h(s)$ , $O_{CDT}^{h*}(s) = O_{CDT}^h(s), O_{syn}^{h*}(s) = O_{syn}^h(s)$ and $V^x(s) = 0$

With respect to any given time $t \geq s$, the interval between two consecutive times $\Delta t = t - s$ may be decomposed as follows:

$$\Delta t = \Delta t^+ + \Delta t^- + \Delta t^0 \qquad 3.38$$

Because of syn bands and TT traffic windows in non-preemption mode, time zero is created by

$$\Delta t^0 = \Delta t_{TT}^0 + \Delta t_{syn}^0 \qquad 3.39$$

where t denotes the length of time between AVB Class x frames. As a result, we have

$$\Delta t^- = \Delta t_{AVB,x}^- \qquad 3.40$$

Thus, the change in credit throughout the $\Delta t$ period fulfills the requirement.

$$V^x(t) - V^x(s) = V^x(t) \qquad 3.41$$

$$= \Delta t^+ . I^x + \Delta t^+ . S^x \qquad 3.42$$

$$= \left( \Delta t - \left( \Delta t_{CDT}^0 + \Delta t_{syn}^0 \right) \right) . I^x - \Delta t_{AVB,x}^- . (I^x - S^x) \qquad 3.43$$

Since this is the case, we may utilize non-preemption mode to determine the relationship between the AVB Class x service times, the CDT traffic times, and the syn band timings in any given time period $\Delta t$,

$$\Delta t_{AVB,x}^- = \frac{\left\{ \left( \Delta t - \left( \Delta t_{CDT}^0 + \Delta t_{syn}^0 \right) \right) . I^x - V^x(t) \right\}}{C} \qquad 3.44$$

Furthermore, under the worst-case scenario, the output frames of CDT traffic during the time interval $\Delta t$ may be calculated using

$$O_{CDT}^{h*}(t) - O_{CDT}^{h*}(t) = O_{CDT}^{h*}(t) - O_{CDT}^h(s) = C.\Delta t_{CDT}^0 \qquad 3.45$$

Similarly, the time lost owing to syn bands during $\Delta t$ is a loss of time.

$$O_{syn}^{h*}(t) - O_{syn}^{h*}(t) = O_{syn}^{h*}(t) - O_{syn}^h(s) = C.\Delta t_{syn}^0 \qquad 3.46$$

So, $\Delta t^0 = \Delta t_{CDT}^0 + \Delta t_{syn}^0$ in order to limit

$$\Delta t^0 \leq \left\{ \left( O_{CDT}^h(t) + O_{syn}^h(t) \right) - \left( O_{CDT}^h(s) + O_{syn}^h(s) \right) \right\} / C \qquad 3.47$$

$$\leq \frac{\alpha_{syn+CDT}^{h,CBS+SP}(\Delta t)}{C} \qquad 3.48$$

Then, based on (3.44) and (3.48), the AVB Class x output frames over the interval t are constrained by

$$O_x^{h*}(t) - O_x^{h*}(s) = C.\Delta t_{AVB,x}^- \qquad 3.49$$

$$\geq I^x \left[ \Delta t - \frac{\alpha_{syn+CDT}^{h,CBS+SP}(\Delta t)}{C} - \frac{V_{max}^x}{I^x} \right] \qquad 3.50$$

$$\beta_{AVB,x}^{h,np,CBS+SP}(t) = I^x \left[ \Delta t - \frac{\alpha_{syn+CDT}^{h,CBS+SP}(\Delta t)}{C} - \frac{V_{max}^x}{I^x} \right]^+ \qquad 3.51$$

### 3.7.4    Service Curve for CDT and AVB traffic with preemption in CBS+SP

When using the preemption mode, if an AVB frame is preempted, an overhead is imposed to the next AVB frame that is still active. CDT traffic frame preempts AVB Class x frames in the worst-case scenario, as seen in Fig. 25. We assume for the moment that the length of the overhead along with syn-window is $L_{syn,j}$. Overheads may be considered as a distinct factor contributing to the delay of AVB data transmission. Because the overhead will only exist immediately after each CDT traffic window, the overhead constraint curve may be calculated using the following Lemma 3.6.

**Lemma 3.6** Under preemption conditions, the aggregate arrival curve for intersecting CDT flows and syn window along with overhead band in an output port h is given by the equation

$$\alpha_{syn}^{h,CBS+SP}(t) = \max_{0 \leq i \leq N^h - 1} \left\{ \sum_{j=i}^{i+N^h-1} L_{syn+OH}.C.\left[ \left( t - O_{j,i}^h - L_{CDT,j} - L_{syn,j} \right) \right] \right\} \qquad 3.52$$

Here, $L_{TT,j}$ is the CDT traffic frame length in j position.

For consideration service curve of CDT traffic of preemption mode is same as Lemma 3.2

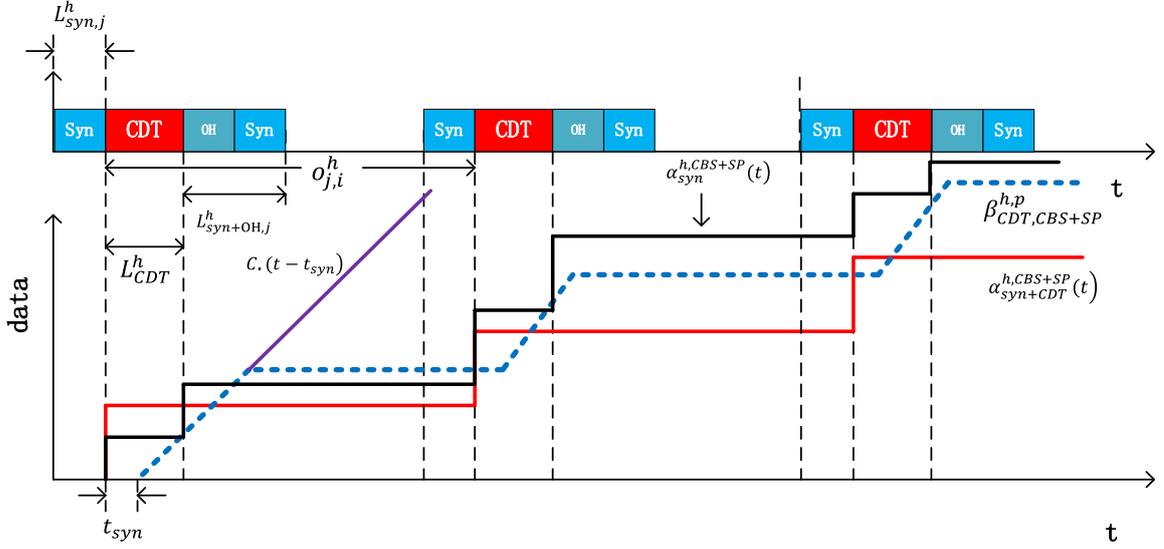

**Figure 25 Example of arrival curve and service curve corresponding windows and overhead bands**

**Theorem 3.4** *According to preemption mode, service curve of AVB class x where $x \in \{A, B\}$ in an output port of ethernet switch port h,*

$$\beta_{AVB,x}^{h,p,CBS+SP}(t) = I^x \left[ \Delta t - \frac{\alpha_{CDT}^{h,CBS+SP}(\Delta t)}{C} - \frac{\alpha_{syn}^{h,CBS+SP}(\Delta t)}{I^x} - \frac{V_{max}^x}{I^x} \right]^+ \qquad 3.53$$

*Where p indicates the preemption integration mode.*

***Proof:*** s is the beginning of the most recent phase of server overload. All backlogs for all flows have been cleared at this point in time such as $O_x^{h*}(s) = O_x^h(s)$ , $O_{CDT}^{h*}(s) = O_{CDT}^h(s), O_{syn}^{h*}(s) = O_{syn}^h(s)$ and $V^x(s) = 0$

With respect to any given time $t \geq s$, the interval between two consecutive times $\Delta t = t - s$ may be decomposed as follows:

$$\Delta t^0 = \Delta t_{TT}^0 \qquad 3.54$$

Preemption-related overheads may be divided into two categories: the durations of frame transmission in AVB Class x and the durations of overheads owing to preemption.

$$\Delta t^- = \Delta t_{syn}^- + \Delta t_{AVB,x}^- \qquad 3.55$$

$\Delta t$ fulfills the credit variation requirement

$$V^x(t) - V^x(s) = V^x(t) \qquad 3.56$$

$$= \Delta t^+ . I^x + \Delta t^+ . S^x \tag{3.57}$$

$$= \left( \Delta t - (\Delta t_{TT}^0) \right) . I^x - (\Delta t_{syn}^- + \ \Delta t_{AVB,x}) . C \tag{3.58}$$

As a result, using preemption mode, we can determine the connection between service times for AVB Class x, extra service time for AVB overheads, and the length of CDT windows in any interval $\Delta t$.

$$\Delta t_{AVB,x}^- = \frac{\left\{ \left( \Delta t - (\Delta t_{CDT}^0) \right) . I^x - \Delta t_{syn}^- . C - V^x(t) \right\}}{C} \tag{3.59}$$

The output frames of TT traffic during t are higher in the worst-case scenario.

$$O_{CDT}^{h*}(t) - O_{CDT}^{h*}(t) = \ O_{CDT}^{h*}(t) - O_{CDT}^h(s) = C . \Delta t_{CDT}^0 \tag{3.60}$$

During the time period $\Delta t$, AVB syn band may be covered by an additional service offered by

$$O_{syn}^{h*}(t) - O_{syn}^{h*}(t) = \ O_{syn}^{h*}(t) - O_{syn}^h(s) = C . \Delta t_{syn}^0 \tag{3.61}$$

So, $\Delta t_{CDT}^0$ in order to limit

$$\Delta t_{CDT}^0 \leq \frac{\left\{ O_{CDT}^{h*}(t) - O_{CDT}^h(s) \right\}}{C} \leq \ \frac{\alpha_{CDT}^{h,CBS+SP}(\Delta t)}{C} \tag{3.62}$$

in the same way that $\Delta t_{syn}^-$ fulfills,

$$\Delta t_{syn}^- \leq \frac{\alpha_{syn}^{h,CBS+SP}(\Delta t)}{C} \tag{3.63}$$

According to (3.59), (3.62) and (3.63), Class x output frames over the interval t are limited by, respectively.

$$O_x^{h*}(t) - O_x^{h*}(s) = \ C . \Delta t_{AVB,x}^- \tag{3.64}$$

$$\geq I^x \left[ \Delta t - \frac{\alpha_{CDT}^{h,CBS+SP}(\Delta t)}{C} - \frac{\alpha_{syn}^{h,CBS+SP}(\Delta t)}{I^x} - \frac{V_{max}^x}{I^x} \right] \tag{3.65}$$

$$\beta_{AVB,x}^{h,p,CBS+SP}(t) = \ I^x \left[ \Delta t - \frac{\alpha_{CDT}^{h,CBS+SP}(\Delta t)}{C} - \frac{\alpha_{syn}^{h,CBS+SP}(\Delta t)}{I^x} - \frac{V_{max}^x}{I^x} \right]^+ \tag{3.66}$$

**Table 13 Several Service Curve for CBS+TAS & CBS+SP**

| Service Curve | Several Service Curve of AVB Traffic |
|---|---|
| Non-preemption mode CBS+TAS | $\beta_{AVB,x}^{h,np,CBS+TAS}(t) = I^x \left[ \Delta t - \frac{\alpha_{syn+TT}^{h,CBS+TAS}(\Delta t)}{C} - \frac{V_{max}^x}{I^x} \right]^+$ |
| Preemption mode CBS+TAS | $\beta_{AVB,x}^{h,p,CBS+TAS}(t) = I^x \left[ \Delta t - \frac{\alpha_{TT}^{h,CBS+TAS}(\Delta t)}{C} - \frac{\alpha_{syn}^{h,CBS+TAS}(\Delta t)}{I^x} - \frac{V_{max}^x}{I^x} \right]^+$ |
| Non-preemption mode CBS+SP | $\beta_{AVB,x}^{h,np,CBS+SP}(t) = I^x \left[ \Delta t - \dfrac{\alpha_{syn+CDT}^{h,CBS+SP}(\Delta t)}{C} - \dfrac{V_{max}^x}{I^x} \right]^+$ |
| Preemption mode CBS+SP | $\beta_{AVB,x}^{h,p,CBS+SP}(t) = I^x \left[ \Delta t - \dfrac{\alpha_{CDT}^{h,CBS+SP}(\Delta t)}{C} - \dfrac{\alpha_{syn}^{h,CBS+SP}(\Delta t)}{I^x} - \dfrac{V_{max}^x}{I^x} \right]^+$ |

In following fig. 26 is shown that sum up of the service curve of two scheduler according to table 13. In general comparison, TT traffic can be more delay in CBS+TAS scheduler with compare to CBS+SP because CBS+TAS scheduler depending GCL gate but CBS+SP is an event triggered scheduler, so TT can transfer when it arrival in queue. On the other hand, lower priority or middle priority traffic can experience more delay if TT traffic transfer frequently. In details comparison has been shown in section 6.

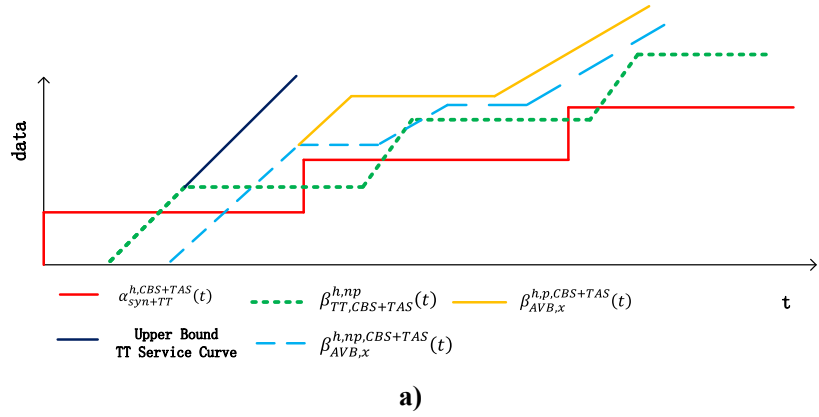

**a)**

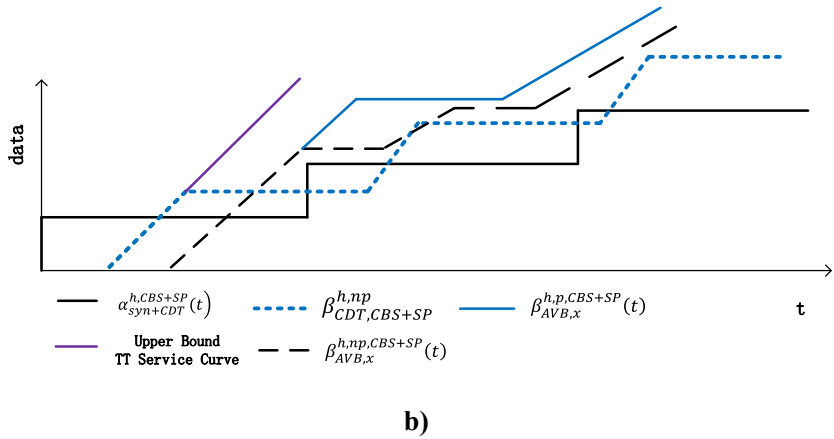

**b)**

**Figure 26 Example of several Service Curve for two schedulers**

# 4.    Fault-Resilient Topology and Traffic Configuration

In recent years, there has been a rise in the need for distributed embedded systems for safety-critical applications such as automotive and avionics [74][78]. In this kind of system, a large number of sensors and actuators are controlled by a large number of Electronic Control Units at the same time (ECUs). This type of technologies cannot be realized without the development of networks that provide dependable and real-time communication. Time Sensitive Networking (TSN) is a collection of IEEE standards that specify the specifications for Ethernet-based switching networks that are designed to meet the dependability and timing requirements of safety-critical applications [79]. TSN-based networks meet the system's real-time needs by enabling a type of traffic known as Time-Triggered (TT) traffic, also known as planned traffic.

Transparent switching networks on the other hand, use the concept of seamless redundancy in order to satisfy the dependability requirement. Every critical message, in particular, is transmitted via multiple channels that are not linked, guaranteeing that communication continues even if one of the links fails. Many copies of a network are often utilized to offer seamless redundancy across multiple sites. Because of current constraints on system space and power, this kind of redundancy may not always be possible despite the fact that it may provide the necessary level of dependability. As stated in [80] academics and system designers have lately started investigating the feasibility of including the need for fault-resilience into the network planning stage as early as possible. That is, designing a fault-tolerant network architecture that allows for several paths for critical signals to be delivered at the same time is essential. In line with expectations, preliminary results suggest that such integration offers the required level of dependability in a more efficient manner than full network redundancy. Because it reduces the total number of components in the networks, this synthesis lowers the overall cost of the network as a result.

The dependence between the network topology and the traffic supplied by the network is a source of intrinsic complexity associated with fault-resilient synthesis. If the size of the messages and the necessary redundancy level are both large, the number of distinct routes that the topology may offer between nodes $v_k$ and $v_j$ is limited. Accordingly, as described in [80] combined topology and routing synthesis are required to provide viable routing. Only non-scheduled traffic is considered by the suggested TSN network synthesis method in (i.e., not TT).

When TT traffic is taken into consideration, extra restrictions are associated with scheduling that must be divided into groups. In the case of TT traffic, these restrictions impose additional symbiotic relationships between the network architecture and traffic. The issue, on the other hand, is that scheduling feasibility cannot be ensured unless the schedule is determined in conjunction with the topology synthesis process. As previously stated, addressing each of the topology or schedule synthesis problems is an NP-hard task [35], [81], and as a result, the combined synthesis will suffer from severe scaling constraints as a result. Because of this, methods that are based on ILP and SMT, such as the ones suggested in [82], respectively, are not capable of dealing with actual large-scale networks.

## 4.1 Features of Fault Resilience

Time-Triggered (TT) communication is supported by Ethernet, allowing for low latency and deterministic timing behavior. TSN adapts the concept of uninterrupted redundancy to ensure unobstructed flexibility. Our goal is to integrate the network topology that supports the uninterrupted transmission of TT messages. As a result, for a general topology, routing, and timeline, we offer a greedy heuristic approach. The suggested technique can design a fault-tolerant topology that assures proper TT traffic routing and scheduling. Specifically, the architecture is designed in such a way that all messages are routed through an isolated path with a potential schedule and network costs are reduced. To do this, the structure fitting issue is reformulated as a recurrent path selection problem. Starting with a loaded unstructured graph representing a fully connected primary network, the cost of using each link is set as the weight of the arcs in the graph. Next, we adapt the yen algorithm to iteratively search for the least expensive paths of TT messages. In the following passages, we have discussed some famous algorithms which is adopted by several researchers for ensuring fault-resilience of TT messages.

When it comes to safety-critical applications, communication reliability is essential. TSN fault-resilience is a subject that has received a lot of attention in recent research. The authors of [13] offer a good summary of the various fault-resilience ideas for IEEE 802.1 TSN networks, which is particularly useful for network designers and administrators.

It is feasible to enhance transmission reliability by including temporal redundancy into the system. Recent research [83] shown how to utilize the temporal redundancy technique to develop a TSN network routing algorithm that is aware of the importance of dependability. It is necessary to satisfy the MTTDE's mean time to detect error criterion by identifying the message's route and redundancy level using an ILP-based formulation, which is implemented in this approach. Because all replicas are sent via the same route, it is difficult to accept a permanent link failure scenario, regardless of how robust the temporal redundancy is against temporary transmission faults.

It is proposed in [84] that heuristics as well as constraint programming-based methods for fault-tolerant topology planning and traffic routing may be used for TSN network topology planning and traffic routing. A particular type of rate-constraint traffic, referred to as Urgency-Based Scheduler (UBS), is addressed by the suggested methods, and scheduling tables are not required. However, traffic is routed via a limited number of discontinuous routes in order to keep network costs and area to a minimum.

These methods, on the other hand, are not appropriate for traffic over TT networks. Furthermore, the approach of combined topology and routing synthesis for time-triggered networks does not ensure that scheduling is possible.

Laursen and colleagues [85] propose a schedule-ability-aware routing for time-sensitive traffic (TT traffic) on TSN networks, which generates routes that improve the likelihood of finding a schedule that is both feasible and efficient. Using a set of ILP constraints, the authors construct the routing issue of a multi-hop TT network that takes into consideration an extra parameter, namely, the maximum planned traffic load. Conflicts between TT messages are less likely to occur when this option is set to true. As an additional point of interest, many studies have been published that deal with the synthesis of combined routing and scheduling. In [80], it is proposed to use ILP formulations to simultaneously address the routing and scheduling issues of TT traffic, which have the advantage of providing more optimization capability.

In contrast, these ILP-based methods are restricted in their applicability due to their scalability limitations, which means that they are only relevant to issues of modest scale.

## 4.2 System Model

Specifically, we examine an Ethernet-based multi-hop switching architecture that complies with the IEEE TSN standard in this thesis. Consider a group of ECUs E that communicate with one another through a group of bridges B, exchanging a group of TT messages M. A schedule should be established for the messages in M to be delivered across the network such that the output ports of the bridges along the message's route are allocated to that message before it arrives. As input to the issue, the set of ECUs E, as well as the set of TT messages M, have already been specified. The number of bridges in the network, as well as the physical link between these bridges, are, on the other hand, decision variables that are derived by the suggested method. All physical connections in the network operate at full duplex and at the same speed.

As a weighted undirected graph $G (V, W)$, the network topology is represented in fig. 27 as a collection of vertex-edge-arc pairs, with V being the set of vertices and W denoting the set of weighted arches. Bridges in the library are denoted by the letter $A$, and these bridges may be utilized to build the topology. In $A$, each element's monetary cost and the number of ports it has are specified by the letter $A$. Table 14, we can see an example of a bridge library. This tuple defines the message $m \in M$ for each of the messages m Where m.src and m.dest are the message source and destination, respectively, and m.size, m.period are the message length. frame size and time between frames. In the end, $m.t_1$ represents the Tolerance Level (TL) of the message, which determines the minimum number of discontinuous routes via which the message should be routed. In PTP block, there is a synchronization clock and another is the scheduler where synchronization clock ensures Yen's algorithm to update adjacent hops and finding free and shortest path.

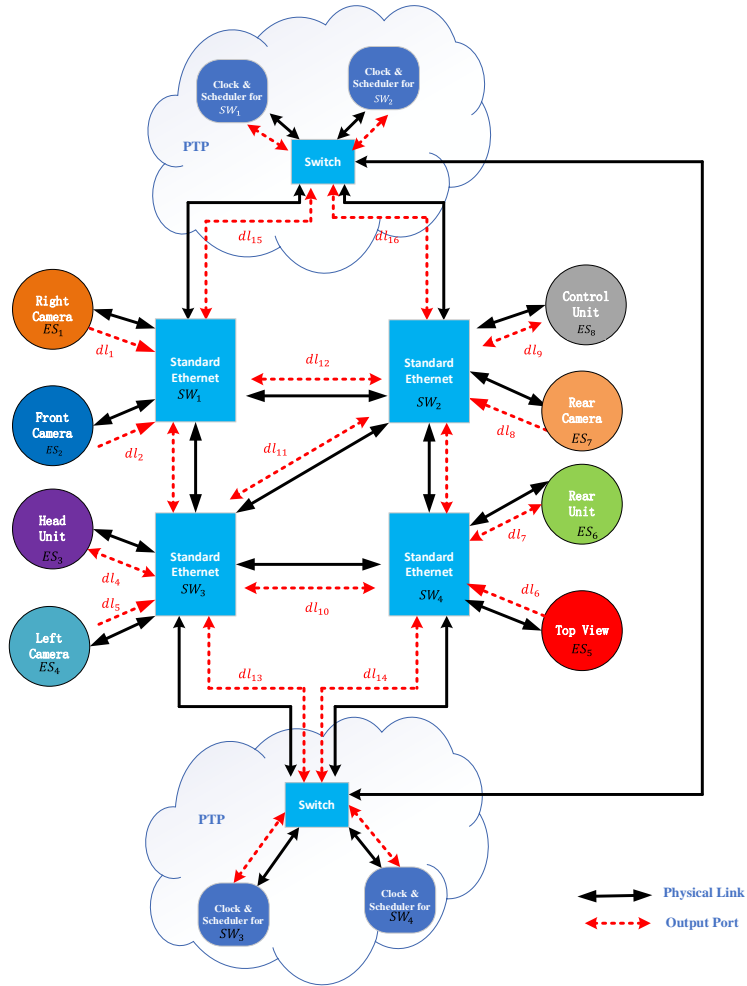

**Figure 27 Network Topology for Fault-Resilient**

**Table 14 Example of source to destination message information**

| Index | (m.src, m.dest) | m.size(KB) | m.period(ms) | $m.t_1$ |
|-------|-----------------|------------|--------------|---------|
| $M_1$ | $(ECU_3, ECU_2)$ | 1.5 | 3 | 1 |
| $M_2$ | $(ECU_2, ECU_2)$ | 0.5 | 3 | 2 |
| $M_3$ | $(ECU_1, ECU_2)$ | 0.5 | 2 | 2 |
| $M_4$ | $(ECU_4, ECU_2)$ | 1 | 2 | 1 |

Our presumption is that the message time limit is defined by the criticality of the program, which is set by the application's developers. The same message is delivered many times via a number of distinct yet linked channels, $m^i$ where $i \in m.rl$. When it starts at the origin and finish at the destination, you have created a route, which is represented by an ordered sequence of linked vertices. The traffic routing, denoted by the letter R, is made up of all of the messages that flow via the network's different routes. Message $m^i$ is sent on each of the connections along $r_m^i \in R_m$ according to a global schedule, S, which defines the transmission instance of the message.

## 4.3 Topology Synthesis

We utilize the graph model to frame the synthesis of network topology as a labeling problem, which is then solved using the graph model. It should keep in mind that the value $B = \{b_1, b_2, b_3, b_4, \ldots, b_k\}$ indicates the bridge in the network architecture $L = \{l_1, l_2, l_3, l_4, \ldots, b_N\}$ indicates the also bridge in the network architecture. $k$ implies the upper bound number for network bridge, $N$ fully-connected network has a number of connections equal to $K(K-1)/2$, where K is the maximum number of bridges selected by the user and $N$ is the total number of connections in the network. Some basic network symbol along with description in table 15

**Table 15 Network Symbol and Descriptions**

| Symbols | Descriptions |
|---------|--------------|
| $B = \{b_1, b_2, b_3, b_4, \ldots, b_k\}$ | Bridge in the Network topology |
| $L = \{l_1, l_2, l_3, l_4, \ldots, b_N\}$ | Connection of the bridge |
| $K$ | Upper bound of number of bridges |
| $N = K(K-1)/2$ | Fully-Connected Number of bridges |
| $D \in B \cup L$ | Decision Element |
| $T = \{t_{d_1}, t_{d_2}, t_{d_3}, \ldots, t_{d_{(K+N)}}\}$ | Labeling Vector |
| $t_{d_i}$ | Assignment of label |
| $d_i \in D$ | Element of assignment |
| $t_{d_i} \in T$ | Element of T |
| $A$ | Library of the bridge |
| $t_{d_i} = \emptyset$ | Discarded Element |
| $C(T, R, S)$ | Cost Function |

The network topology, as well as the traffic configuration, are optimized in accordance with the cost function C(T), which is designed to include both the monetary cost of the topology, the latency of the messages, and the schedule-ability of the traffic, as stated in equation 4.1

$$C(T, R, S) = C_{cost}(T) + \alpha . C_{hops}(R) + C_{overlap}(S) \tag{4.1}$$

The importance of $C_{cost}(T)$ in respect to $C_{hops}(R)$ is regulated by the coefficient, which is non-negative in nature $\alpha$. The monetary cost $C_{cost}(T)$ is used to express the total amount of topological equation 4.2.

$$C_{cost}(T) = \sum_{d_i \in D} A\{t_{d_i}\} \tag{4.2}$$

Where $A\{t_{d_i}\}$ is the cost of element $d_i$ according to the library A, while the discarded element does not count, i.e., A $\{\emptyset\} = 0$. $C_{hops}(R)$ is a protocol that attempts to reduce message latency by encouraging traffic to be routed via shorter routes. $C_{hops}(R)$, in particular, is equal to the sum of the entire length of all routes in R. If we assume that the connections are equal in terms of speed, then the number of hops reflects the amount of time it takes for the message to arrive. Last term $C_{overlap}(S)$ guarantees that M is schedulable, i.e., $C_{overlap}(S) = 0$ if a feasible overlap free schedule exists, and $C_{overlap}(S) = \infty$ if no such schedule exists.

## 4.4 Routing Algorithm

The routing technique we have created estimates the cost of the route in order to discover nodes that are troublesome. As long as all of a network's PTP clocks are synced and connected to one another, a PTP clock can determine which route has a high cost and which route has a low cost. In the network, this is the main path selection. Since SW and ES are considered as intermediate nodes for the purpose of computing route cost, specific attributes must be gathered from these nodes. Furthermore, the cost of the route is completely reliant on the SW and ES variables. To compute parameters such as $C_{hops}(R)$, $C_{overlap}(S)$, and $C_{cost}(T)$, a collection signal should be sent to a nearby node. However, because $C_{cost}(T)$ is regarded a fixed cost, it is not required to compute $C_{cost}(T)$.

In order to send this information to its own PTP clock and scheduler, each SW must first get the value of the aforementioned parameter. PTP clocks, on the other hand, share this information with one another, as a result of which every PTP clock is aware of the route selection and cost calculation. Using the quickest route determined by path calculation, the PTP clock transmits data in accordance with the schedule. In table 16, PTP clock is selecting the best routing path in the following algorithm.

**Table 16 Path Selection Algorithm**

| | Path Selection Algorithm |
|---|---|
| 1 | *// Initial time State* |
| 2 | *Every switch & End Device send connection signal to Adjacent device* |
| 3 | *do* |
| 4 | *if    adjacent device receives connection signal* |
| 5 | *Every node and device send $C_{hops}(R)$ & $C_{overlap}(S)$* |
| 6 | *if    Overlapped free scheduling* |
| 7 | *$C_{overlap}(S)$= 0* |
| 8 | *end* |
| 9 | *end* |
| 10 | *else* |
| 11 | *$C_{overlap}(S) = \infty$* |
| 12 | *end* |
| 13 | *elseif   Do not receive any connection signal in initial time state* |
| 14 | *Node is considered as faulty* |
| 15 | *end* |
| 16 | *until    Collection $C_{hops}(R)$ & $C_{overlap}(S)$* |
| 17 | *do* |
| 18 | *if    ES and SW collect   $C_{hops}(R)$ & $C_{overlap}(S)$* |
| 19 | *Send to the PTP clock* |
| 20 | *$C(T,R,S) = C_{cost}(T) + \alpha . C_{hops}(R) + C_{overlap}(S)$* |
| 21 | *$C_{cost}(T) = \sum_{d_i \in D} A\{t_{d_i}\}$* |
| 22 | *end* |
| 23 | *until    Initial time finish* |
| 24 | *while    $C(T,R,S)$  calculate* |
| 25 | *if    $C_{node_1} > C_{node_2}$* |
| 26 | *Send data through  $C_{node_1}$* |

| 27 | | *end* |
| 28 | | *else* |
| 29 | | | *Send data through* $C_{node_1}$ |
| 30 | | *end* |
| 31 | *end* | |

## 4.5 Experiment Result

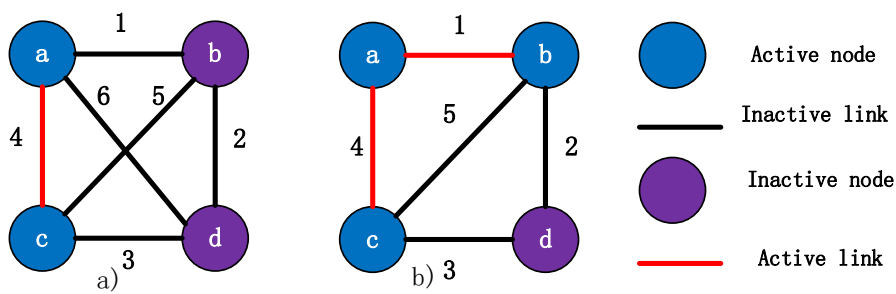

**Figure 28 Example of active and inactive node**

Fig. 28 depicts an example of a route selection method for a data transfer between an active and idle node. Even if node is physically connected to three ports, consider the cost of the nodes 4,6,1 that are connected to ports <a,c>,<a,d>,<a,b> before updating the remaining nodes. Following the logic of the diagram, we can see that the cost of the b node is quite low, but the cost of connecting to d is extremely high, making it almost difficult for A to link with B, while connecting with D is nearly impossible. The above procedure should be followed by b in order to connect two more nodes. If we follow the steps, it is possible for node b to connect with node d, even if it has already connected with node a, if the procedure is followed. This implies that, as of this update, it will only be able to join node d and will not automatically connect to nodes b and c as previously did.

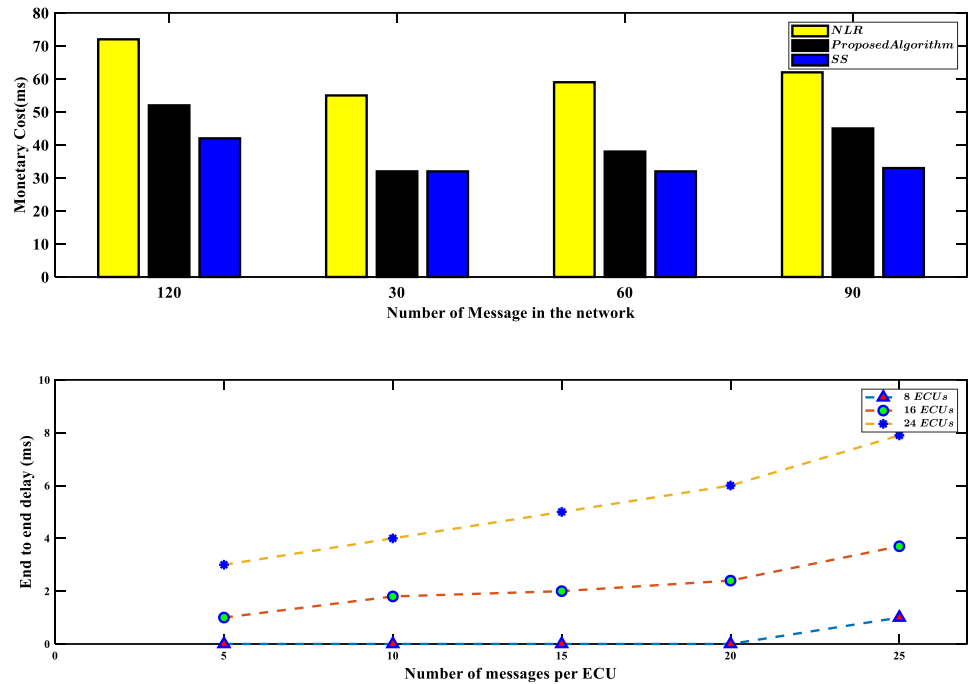

**Figure 29 Monetary cost and end to end delay**

To determine how efficient the proposed strategy is, we first compare it to two alternative approaches. The first option, according to [86], is based on Network-Level Redundancy (NLR) (NLR). Several copies of the network architecture are employed to realize the idea of many pathways, which allows for multi-path transmission to be accomplished. It is built on the basis of the path-selection algorithm, which was discussed in Section 4.4. In truth, the work mentioned in [87] has had a significant influence on this method. Six ECUs are randomly assigned to an 80-synthetic test scenario with 30 to 120 messages in order to evaluate the two approaches. It is required that messages have a tolerance level (TL) of 2 and a necessary maximum payload of 1,518 bytes, with a period of 1, 2, 3, and 10 milliseconds. All of the experiments make use of 100 Mbit/s connections. First of fig. 29, shows the average cost as well as the percentage of feasible scheduling options. The greater cost linked with the NLR approach is unquestionably a downside of the technique. As a result, since routing and scheduling are not always handled jointly in some situations, a realistic timetable was not always offered. Although the path-selection strategy does not address the potential of message scheduling, it has the lowest topological cost of all the techniques tested. Unavoidably, this results in an unrealistic timeframe for a major section of the population. In this circumstance, it may be essential to resynthesize the topology by constructing new bridges and links as needed. Alternatively, the proposed is more cost-effective than the NLR approach while still allowing for realistic scheduling to occur. We employed 300 synthetic test cases to examine the scalability of the proposed technique, with each example covering the following scenarios. 8, 16, and 24 ECUs are used to construct the problems, with the number of messages sent by each ECU ranging from 5 to 25 messages per ECU. In addition to the maximum 1,500 Byte payload (which is the Ethernet Maximum Transmission Unit (MTU), all messages include periods of 1, 2, 3, and 10 milliseconds. While the needed tolerance level is determined to be 2 or 3, the acceptable tolerance level is determined to be 2. A random number generator is used to determine the origin and destination of the communications. Second of fig. 29 depicts an average runtime for each size of instance for which data is available. In addition, the findings demonstrate that the proposed algorithm has excellent scalability, as shown by its capacity to handle large-scale problems, such as 24 ECUs with 600 messages, in a relatively short amount of time.

# 5. Data and Frame Indexing Transmission

## 5.1 Related Work

When using ordinary Ethernet technology, it is possible to distribute hard real-time applications across many networks. Due to the use of various scheduling algorithms and multiple queuing, the IEEE 802.1 Working Task Group developed the Transmission System Network (TSN). TSN is intended to decrease end-to-end delays and jitter while also preventing packet loss. Data flow methods are used in the TSN network to guarantee that capacity and latency are kept within acceptable limits [57]. The Time-Aware Shaper (TAS) and the Credit-Based Shaper (CBS) are two of the most popular and commonly utilized ways in the TSN networks, and they are also the most frequently used methods in the TSN networks (CBS). TAS is heavily involved in the control data traffic (CDT) in order to ensure that it is not interfered with by any other traffic. A 'periodic time frame' is provided by the GCL in order to govern this TAS process. Even though CBS is a credit-based approach that can transport both audio and video data depending on its class and priority, it is not the greatest solution for streaming video and audio. The importance of audio data above other types of data is well recognized, and it is designated as class A data. Consequently, video is designated as an activity with a lower priority level, known as class B. The key objectives of this CBS approach are to prevent the depletion of best-effort data and to improve deterministic data transmission for higher classes. With the proliferation of modern businesses, there has been a significant growth in the use of data on an unprecedented scale. The use of end devices is also continuing to grow at a rapid pace. The TSN network gets overburdened as a result of the increased and rising volume of traffic flows, as well as the presence of critical real-time applications on the network. The growing amount of data and the increasing number of traffic have made it harder to ensure real-time application. To prepare for the worst-case scenario, the TSN network is working to enhance latency and jitter. Whenever it comes to delivering on real-time communication commitments, scheduling and shaping procedures are often challenged with challenging circumstances. Example: The BE class in the CBS shaper is unable to provide a regular timeline because to the nature of the work. However, by using data compression technologies, the considerable delay and jitter associated with TSN may be decreased. The use of data compression is suggested in this research.

## 5.2 Basic system model

Each cluster in a switch network (SW) network contributes to the overall network's performance. The end systems (ESes) in each cluster are linked by connections and switches to form a network (SWs). A buffer for the output port is included in each ES, which connects to a single input port on the SW. Each SW has no input buffers on any of its input ports, but does have an output buffer on each of its output channels. ES or another switch input port is linked to one of the output ports of SW. There is full duplex communication on the connections, which allows for communication in both directions, and the networks may be multi-hop in nature. An example cluster consisting of four ESes ($ES_1$ to $ES_8$) and three SWs ($SW_1$ to $SW_3$). Table 17 demonstrate network symbol and descriptions

**Table 17 Network Symbol and Descriptions**

| Symbol | Descriptions |
|---|---|
| $G\ (\boldsymbol{E},\ \boldsymbol{V})$ | Undirected Graph |
| $\boldsymbol{V} = \boldsymbol{ES} \cup \boldsymbol{SW}$ | $\boldsymbol{V} = \{ES_1, ES_2, ES_3, ES_4, ES_5, ES_6, ES_7, ES_8\}$ $\cup \{SW_1, SW_2\}$ |
| $EW$ | End Systems |
| $SW$ | Standard Ethernet Switches |
| $dl_i = [v_a, v_b] \in \boldsymbol{L}$ | Dataflow Link, $\boldsymbol{L} = Set\ of\ data\ flow$ |
| $v_a \rightarrow v_b$ | Direction edge $v_a\ to\ v_b$ |
| $v_a \& v_b \in \boldsymbol{V}$ | ESes & SWes |
| $dl_i.C$ | Physical Link Rate |
| $h$ | Output Port |
| $\delta^S = d_{max}^S - d_{min}^S$ | Maximum jitter |
| $d_{max}^V$ | Message process real-time Units |
| $t_0$ | Nonfaulty at real-time |
| $d_{max}^E$ | All messages dispatched |
| $d_{max} = d_{max}^E + d_{min}^V$ | Maximum Delay |
| $P_{TT_n}, n \in 1,2,3,\dots,n$ | Priority of TT traffic |

## 5.3 Application Model

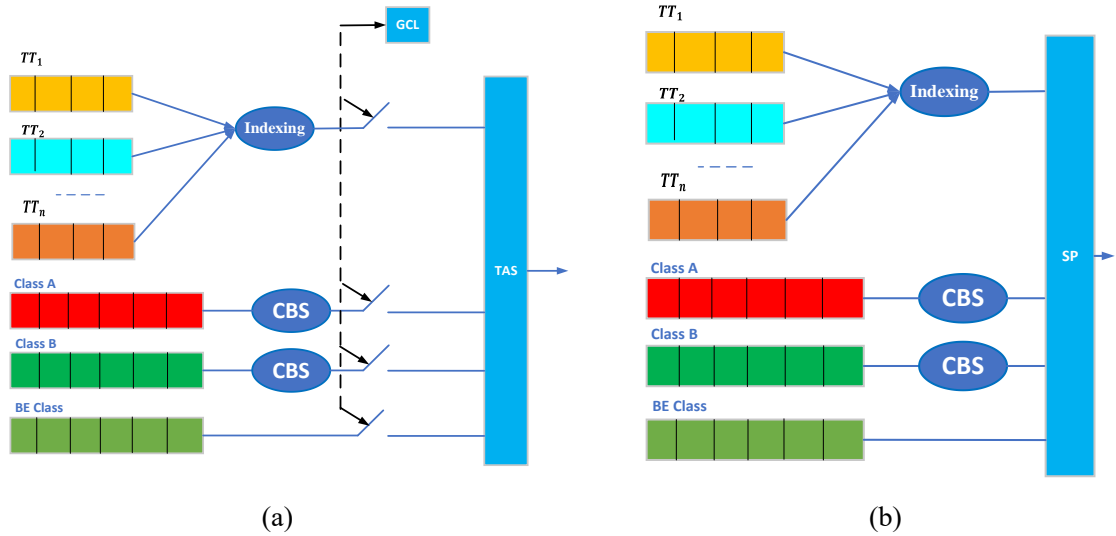

**Figure 30 Output Scheduling of the Network**

The main goals of the application model are the transfer of data from one ES to one or more ESes (destination) via a physical link between the two ESs.TT, AVB class A, AVB class B (some networks also offer AVB class C) and BE traffic types are all supported by TSN networks, as are a number of other kinds of traffic. Example of a TSN model for CBS is shown in Fig. 30. Design engineers, we have believed, are responsible for determining the number of individuals who will utilize a model. The following notation has been developed for our system in order to make it simpler to understand.

## 5.4 Data and Frame Indexing

Consider the following scenario: $TT_1, TT_2, TT_3, \ldots, TT_n$ are the higher priority traffic, while all other traffics are equivalent in importance(Same priority), such as $P_{TT_1} = P_{TT_2} = P_{TT_3} = \cdots = P_{TT_n}$; The quantity of TT is rising day by day in contemporary industrial communication, and the worst situation is commonly seen in real-time communication. To understand of practical type of same priority TT traffic, we can say temperature, weather, humidity, chemical parameters, and other types of sensors are primarily indicated by TT, for example.

Consider the following scenario: four TT traffic streams are accessible. For indexing, (seen fig.30), we should assume some condition, like, all of the hops have installed statistical data of the industry. Now question is what is the statistical data? Statistical data refer that all of the TT frame data of any industry keep arrange in m*n matrix except Control Data Traffic (CDT) data in each hop. In table 18, there is a $Index_{i,j}$ which indicate that each row of statistical data represents by $Index_{i,j}$. Also assume that statistical data update in offline but in run-time it will be remain unchanged. Large number statistical TT data ensure that arrival TT frame pattern and previous store TT frame pattern in each hop will be same. If some TT frame pattern will not match with store TT frame, then TT frame will be transfer traditional method which is already discuss in section 3. In additional, all shrink TT will be transfer all of the end devices and in the end-device will decode the shrink frame to desire frame.

**Table 18 Shrink of TT traffic**

| $Index_{i,j}$ | $TT_1$ | $TT_2$ | $TT_3$ | $TT_4$ |
|---|---|---|---|---|
| $Index_{1,1}$ | $TT_{1,1}$ | $TT_{2,1}$ | $TT_{3,1}$ | $TT_{4,1}$ |
| $Index_{1,2}$ | $TT_{1,2}$ | $TT_{2,2}$ | $TT_{3,2}$ | $TT_{4,2}$ |
| $Index_{1,3}$ | $TT_{1,3}$ | $TT_{2,3}$ | $TT_{3,3}$ | $TT_{4,3}$ |
| $Index_{1,4}$ | $TT_{1,4}$ | $TT_{2,4}$ | $TT_{3,4}$ | $TT_{4,4}$ |

In table 18, $TT_{k,l}$ represents the numerical value of each traffic which is collected industrial database. $TT_{k,l}$ have to import in the payload of the ethernet frame. $Index_{i,j}$ represents a unique numerical value assignment which implies the combination different TT messages. Size of the message is always $Index_{i,j} < \sum_{k=1} TT_{k,l}$. The table will be update after each $t\ \mu s$.

i = Number of Table, j = Number of Index in each table

Here, in table 18, combination of TT traffics are compacted by indexing value. Each combination represents in unique index.

The ratio of $Index_{i,j}$ and $TT_{k,l}$ represent shrink of the actual message is $r_{index}$ for example $r_{index} = \frac{Index_{i,j}}{\sum_{k=1} TT_{k,l}}$. As $Index_{i,j} < \sum_{k=1} TT_{k,l}$ so $r_{index} < 1$. Consider the length of the TT message is 'a'.

If the actual message shrink successively, so the geometric progression will be $a, r, r^2, r^3, \ldots \infty$, a is unity that means a=1.So the sum of the shrink of the series. $\sum_{z=0}^{\infty} r_{index}^z = \frac{1}{1-r_{index}^z}$.In consequence, the duration of shrink message will be $\frac{1}{1-r_{index}^z} - 1 = \frac{r_{index}^z}{1-r_{index}^z}$.

In shrink duration ($\frac{r_{index}^z}{1-r_{index}^z}$) the equivalent service rate is amplified to $\frac{1}{r_{index}^z}$

## 5.5 Shrink Service Curve for TT and AVB traffic with CBS+TAS

### 5.5.1    Non-preemption mode

In fig. 31, we can observe that shrink TT frame can reduce of the TT window size beside the frame size of the TT traffic will vary $a \leq TT_{frame} \leq b$ where a is minimum size of the shrink TT frame and b is the maximum size of the TT frame as a consequence the frame size of the TT traffic will not exceed the TT windows size though how many bytes will transfer in each window depending on window size.

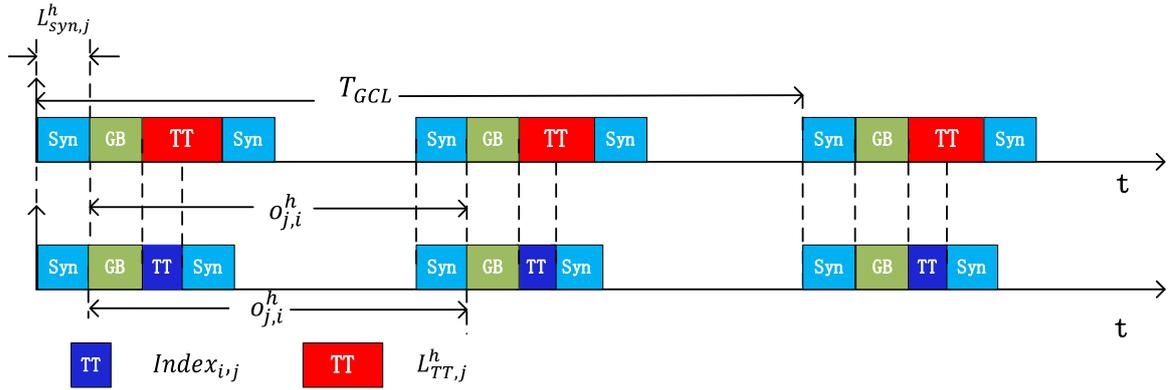

**Figure 31 Shrink TT traffic windows with Guard Band**

**Lemma 5.1** Under non-preemption conditions, the aggregate arrival curve for intersecting TT flows and syn-bands in an output port h is given by the equation

$$\alpha_{syn+TT,index}^{h,CBS+TAS}(t) = max_{0 \leq i \leq N^h-1} \left\{ \sum_{j=l}^{i+N^h-1} (r_{index}^z \sum L_{TT_{k-l}}^h + L_{syn+GB,j}^h ).C.\left\lceil \left\lceil \frac{(t-o_{j,i}^h+L_{syn,j}^h-L_{syn+GB,j}^h)}{T_{GCL}} \right\rceil \right\rceil \right\}$$  5.1

This is identical to the curves in equation 5.1. This is obtained by multiplying the worst-case number of bits generated by the twice of $L_{syn,j}^h$ syn-band by the maximum number of bits broadcast in the $L_{TT,j}^h$ TT traffic window, both of which cause AVB traffic to be delayed until after the service for AVB traffic has been restored to normal. Aside from that, the distance between new "Syn+TT" windows may vary as a result of the possibility of a varied length of syn-band.

**Lemma 5.2** Under non-preemption conditions, the Service curve for intersecting TT flows and Syn. bands in an output port h is given by the equation

$$\beta_{TT,CBS+TAS}^{h,np,index}(t) = C.\max\left(\left\lceil\frac{t-t_{syn}}{T_{GCL}}\right\rceil r_{index}^z \sum L_{TT_{k-l}}^h, t - \left\lceil\frac{t-t_{syn}}{T_{GCL}}\right\rceil \left(T_{GCL} - r_{index}^z \sum L_{TT_{k-l}}^h\right)\right)^+ \qquad 5.2$$

Here, $t_{syn} = synchronization\ time$

In equation 5.2 refers that TT frames are flowing according to opening TT window, on the other hands the seconds term of the equation 5.2 indicates when the TT windows will be finished, the value of the will be frozen. In fig. 32, an example of the TT window and AVB window along with guard band and syn window for non-preemption mode.

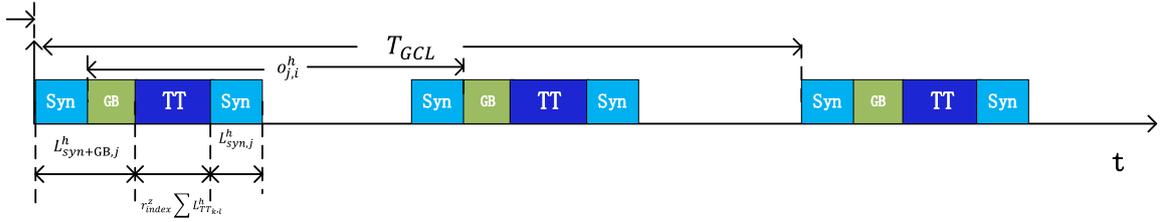

**Figure 32 Shrink TT traffic output in h port with Guard Band**

***Theorem 5.1*** *According to non-preemption mode, service curve of AVB class x where* $x \in \{A, B\}$ *in an output port of ethernet switch port h,*

$$\beta_{AVB,x,index}^{h,np,CBS+TAS}(t) = I^x\left[\Delta t - \frac{\alpha_{syn+TT,index}^{h,CBS+TAS}(\Delta t)}{C} - \frac{V_{max}^x}{I^x}\right]^+ \qquad 5.3$$

*Where np indicates the non-preemption integration mode.*

***Proof:*** s is the beginning of the most recent phase of server overload. All backlogs for all flows have been cleared at this point in time such as $O_x^{h*}(s) = O_x^h(s),\ O_{TT}^{h*}(s) = O_{TT}^h(s), O_{syn}^{h*}(s) = O_{syn}^h(s)$ and $V^x(s) = 0$

With respect to any given time $t \geq s$, the interval between two consecutive times $\Delta t = t - s$ may be decomposed as follows:

$$\Delta t = \Delta t^+ + \Delta t^- + \Delta t^0 \qquad 5.4$$

Because of syn bands and TT traffic windows in non-preemption mode, time zero is created by

$$\Delta t^0 = \Delta t_{TT}^0 + \Delta t_{syn}^0 \qquad 5.5$$

where t denotes the length of time between AVB Class x frames. As a result, we have

$$\Delta t^- = \Delta t_{AVB,x}^- \qquad 5.6$$

Thus, the change in credit throughout the $\Delta t$ period fulfills the requirement.

$$V^x(t) - V^x(s) = V^x(t) \qquad 5.7$$

$$= \Delta t^+ . I^x + \Delta t^+ . S^x \qquad 5.8$$

$$= \left(\Delta t - \left(\Delta t_{TT}^0 + \Delta t_{syn}^0\right)\right) . I^x - \Delta t_{AVB,x}^- . (I^x - S^x) \qquad 5.9$$

Since this is the case, we may utilize non-preemption mode to determine the relationship between the AVB Class x service times, the TT traffic times, and the syn band timings in any given time period $\Delta t$,

$$\Delta t_{AVB,x}^- = \frac{\left\{\left(\Delta t - \left(\Delta t_{TT}^0 + \Delta t_{syn}^0\right)\right) . I^x - V^x(t)\right\}}{C} \qquad 5.10$$

Furthermore, under the worst-case scenario, the output frames of TT traffic during the time interval $\Delta t$ may be calculated using

$$O_{TT}^{h*}(t) - O_{TT}^{h*}(t) = O_{TT}^{h*}(t) - O_{TT}^h(s) = C . \Delta t_{TT}^0 \qquad 5.11$$

Similarly, the time lost owing to syn bands during $\Delta$t is a loss of time.

$$O_{syn}^{h*}(t) - O_{syn}^{h*}(t) = O_{syn}^{h*}(t) - O_{syn}^h(s) = C . \Delta t_{syn}^0 \qquad 5.12$$

So, $\Delta t^0 = \Delta t_{TT}^0 + \Delta t_{syn}^0$ in order to limit

$$\Delta t^0 \leq \left\{\left(O_{TT}^h(t) + O_{syn}^h(t)\right) - \left(O_{TT}^h(s) + O_{syn}^h(s)\right)\right\}/C \qquad 5.13$$

$$\leq \frac{\alpha_{syn+TT}^{h,CBS+TAS}(\Delta t)}{C} \qquad 5.14$$

Then, based on (5.10) and (5.14), the AVB Class x output frames over the interval t are constrained by

$$O_x^{h*}(t) - O_x^{h*}(s) = C . \Delta t_{AVB,x}^- \qquad 5.15$$

$$\geq I^x \left[\Delta t - \frac{\alpha_{syn+TT}^{h,CBS+TAS}(\Delta t)}{C} - \frac{V_{max}^x}{I^x}\right] \qquad 5.16$$

$$\beta_{AVB,x}^{h,np}(t) = I^x \left[\Delta t - \frac{\alpha_{syn+TT}^{h,CBS+TAS}(\Delta t)}{C} - \frac{V_{max}^x}{I^x}\right]^+ \qquad 5.17$$

**Corollary 5.1:** Amplified service rate of $AVB_x$, $\frac{1}{r_{index}}$ times $\beta_{AVB,x}^{h,np,amp}(t) =$

$\frac{I^x}{r_{index}}\left[\sup_{0 \le u \le t}\left\{u - \frac{\alpha_{syn+TT}^{h,CBS+TAS}(u)}{C} - \frac{V_{max}^x}{I^x}\right\}\right]^+$ Here, *amp* indicates that amplified service curve

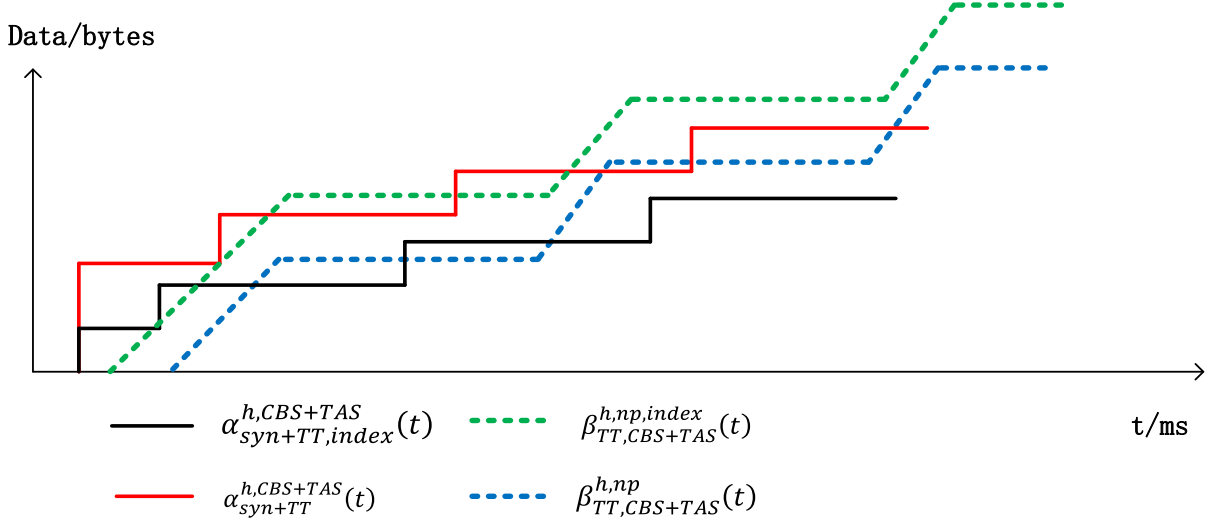

**Figure 33 Several TT arrival and Service Curve Non-preemption mode**

## 5.5.2 Preemption mode

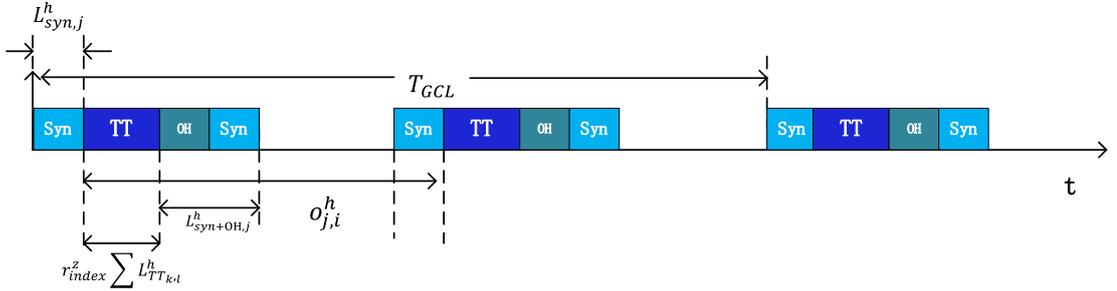

**Figure 34 Shrink TT traffic output in h port**

When using the preemption mode, if an AVB frame is preempted, an overhead is imposed to the next AVB frame that is still active. TT traffic windows preempt AVB Class x frames in the worst-case scenario, and window design is shown in fig. 34. We assume for the moment that the length of the overhead along with syn-band is $L_{syn,j}$. Overheads may be considered as a distinct factor contributing to the delay of AVB data transmission. Because the overhead will only exist immediately after each TT traffic window, the overhead constraint curve may be calculated using the following Lemma 5.3.

**Lemma 5.3** Under preemption conditions, the aggregate arrival curve for intersecting TT flows and syn-bands along with overheads in an output port h is given by the equation

$$\alpha_{syn}^{h,CBS+TAS}(t) = \max_{0 \le i \le N^h - 1} \left\{ \sum_{j=i}^{i+N^h-1} L_{syn+OH}.C.\left\lceil \frac{\left(t - O_{j,i}^h - r_{index}^z \sum L_{TT_{k,l}}^h - L_{syn,j}\right)}{T_{GCL}} \right\rceil \right\}$$

Here, $L_{TT,j}$ is the TT traffic frame length in j position.

For consideration service curve of TT traffic of preemption mode is same as Lemma 5.2

**Theorem 5.2** *According to non-preemption mode, service curve of AVB class x where* $x \in \{A, B\}$ *in an output port of ethernet switch port h,*

$$\beta_{AVB,x,index}^{h,p,CBS+TAS}(t) = I^x \left[ \Delta t - \frac{\alpha_{TT}^{h,CBS+TAS}(\Delta t)}{C} - \frac{\alpha_{syn,index}^{h,CBAS+TAS}(\Delta t)}{I^x} - \frac{V_{max}^x}{I^x} \right]^+ \qquad 5.18$$

*Where p indicates the preemption integration mode.*

**Proof:** s is the beginning of the most recent phase of server overload. All backlogs for all flows have been cleared at this point in time such as $O_x^{h*}(s) = O_x^h(s)$, $O_{TT}^{h*}(s) = O_{TT}^h(s)$, $O_{syn}^{h*}(s) = O_{syn}^h(s)$ and $V^x(s) = 0$

With respect to any given time$t \ge s$, the interval between two consecutive times $\Delta t = t - s$ may be decomposed as follows:

$$\Delta t^0 = \Delta t_{TT}^0 \qquad 5.19$$

Preemption-related overheads may be divided into two categories: the durations of frame transmission in AVB Class x and the durations of overheads owing to preemption.

$$\Delta t^- = \Delta t_{syn}^- + \Delta t_{AVB,x}^- \qquad 5.20$$

$\Delta t$ fulfills the credit variation requirement

$$V^x(t) - V^x(s) = V^x(t) \qquad 5.21$$

$$= \Delta t^+ . I^x + \Delta t^+ . S^x \qquad 5.22$$

$$= \left(\Delta t - (\Delta t_{TT}^0)\right). I^x - (\Delta t_{syn}^- + \Delta t_{AVB,x}^-).C \qquad 5.23$$

As a result, using preemption mode, we can determine the connection between service times for AVB Class x, extra service time for AVB overheads, and the length of TT windows in any interval $\Delta t$.

$$\Delta t_{AVB,x}^- = \frac{\left\{\left(\Delta t - (\Delta t_{TT}^0)\right). I^x - \Delta t_{syn}^-. C - V^x(t)\right\}}{C} \qquad 5.24$$

The output frames of TT traffic during t are higher in the worst-case scenario.

$$O_{TT}^{h*}(t) - O_{TT}^{h*}(t) = O_{TT}^{h*}(t) - O_{TT}^{h}(s) = C.\Delta t_{TT}^0 \qquad 5.25$$

During the time period $\Delta t$, AVB syn band may be covered by an additional service offered by

$$O_{syn}^{h*}(t) - O_{syn}^{h*}(t) = O_{syn}^{h*}(t) - O_{syn}^{h}(s) = C.\Delta t_{syn}^0 \qquad 5.26$$

So, $\Delta t_{TT}^0$ in order to limit

$$\Delta t_{TT}^0 \leq \frac{\left\{O_{TT}^{h*}(t) - O_{TT}^{h}(s)\right\}}{C} \leq \frac{\alpha_{TT}^{h,CBS+TAS}(\Delta t)}{C} \qquad 5.27$$

in the same way that $\Delta t_{syn}^-$ fulfills,

$$\Delta t_{syn}^- \leq \frac{\alpha_{syn}^{h,CBS+TAS}(\Delta t)}{C} \qquad 5.28$$

According to (5.24), (5.27) and (5.28), Class x output frames over the interval t are limited by, respectively.

$$O_x^{h*}(t) - O_x^{h*}(s) = C.\Delta t_{AVB,x}^- \qquad 5.29$$

$$\geq I^x \left[\Delta t - \frac{\alpha_{TT}^{h,CBS+TAS}(\Delta t)}{C} - \frac{\alpha_{syn}^{h,CBS+TAS}(\Delta t)}{I^x} - \frac{V_{max}^x}{I^x}\right] \qquad 5.30$$

$$\beta_{AVB,x}^{h,p}(t) = I^x \left[\Delta t - \frac{\alpha_{TT}^{h,CBS+TAS}(\Delta t)}{C} - \frac{\alpha_{syn}^{h,CBS+TAS}(\Delta t)}{I^x} - \frac{V_{max}^x}{I^x}\right]^+ \qquad 5.31$$

**Corollary 5.2:** Amplified service rate of $AVB_x$, $\frac{1}{r_{index}}$ times $\beta_{AVB,x}^{h,p,amp}(t) = \frac{I^x}{r_{index}}\left[\sup_{0 \leq u \leq t}\left\{u - \frac{\alpha_{TT}^{h,CBS+TAS}(u)}{C} - \frac{\alpha_{syn}^{h,CBS+TAS}(u)}{I^x} - \frac{V_{max}^x}{I^x}\right\}\right]^+$ Here, *amp* indicates that amplified service curve. In fig. 35, we demonstrate that all of the arrival curve and service curve for TT and AVB traffic.

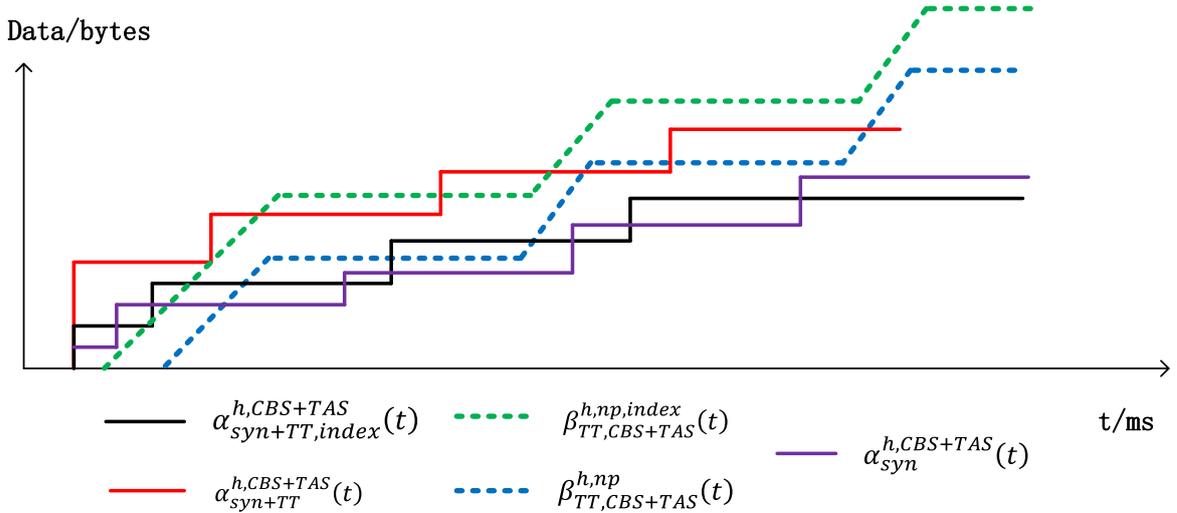

**Figure 35 Several TT arrival and Service Curve preemption mode**

## 5.6 Shrink Service Curve for CDT and AVB traffic with CBS+SP

### 5.6.1    Non-preemption

The following theorem establishes the service curves provided by a CBS at a TSN node for AVB flows in the presence of CDT flows with an LB arrival curve. The authors of [8] calculate service curves for AVB flows in accordance with the IEEE AVB standard [10], i.e., without CDT. Notably, [12] proposes service curves for AVB flows in TSN; however, credit reset is not included in their argument. We construct service curves that are distinct from those in [12] and utilize them to get tight delay bounds.

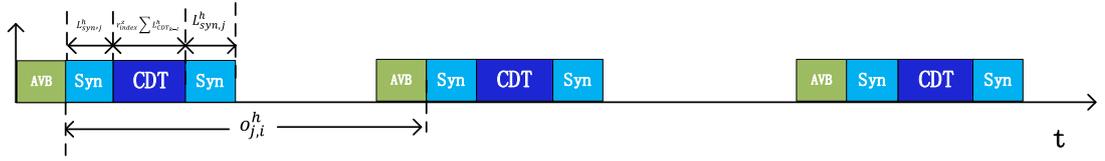

**Figure 36 Shrink TT traffic output in h port Guard Band**

**Lemma 5.4** Under non-preemption conditions, the aggregate arrival curve for intersecting CDT flows and syn bands in an output port h is given by the equation

$$\alpha_{syn+CDT}^{h,CBS+SP}(t) = max_{0 \le i \le N^h-1} \left\{ \sum_{j=i}^{(i+N^h-1)} \left( r_{index}^z \sum L_{CDT_{k,l}}^h + L_{syn,j}^h \right) . C . \lceil (t-o_{j,i}^h) \rceil \right\} \qquad 5.32$$

This is identical to the curves in equation 5.32. This is obtained by multiplying the worst-case number of bits generated by the twice of $L_{syn,j}^h$ syn-band by the maximum number of bits broadcast in the $L_{CDT,j}^h$ CDT traffic window, both of which cause AVB traffic to be delayed until after the service for AVB traffic has been restored to normal. Aside from that, the distance between new "Syn+CDT",frame may vary as a result of the possibility of a varied length of syn-band.

**Lemma 5.5** Under non-preemption conditions, the Service curve for intersecting CDT flows and Syn. bands in an output port h is given by the equation

$$\beta_{CDT,CBS+SP}^{h,np,index}(t) = C . \max \left( \lfloor t - t_{syn} \rfloor r_{index}^z \sum L_{CDT_{k,l}}^h, t - \lceil t - t_{syn} \rceil \left( T - r_{index}^z \sum L_{CDT_{k,l}}^h \right) \right)^+ \qquad 5.33$$

In equation 5.33 refers that TT frames are flowing according to opening CDT window, on the other hands the seconds term of the equation 5.33 indicates when the CDT windows will be finished, the value of the will be frozen. In fig.36, an example of the CDT window and AVB window along with guard band and syn window for non-preemption mode.

**Theorem 5.3** *According to non-preemption mode and CBS+SP scheduling, service curve of AVB class x where* $x \in \{A, B\}$ *in an output port of ethernet switch port h,*

$$\beta_{AVB,x,index}^{h,np,CBS+SP}(t) = I^x \left[ \Delta t - \frac{\alpha_{syn+CDT}^{h,CBS+SP}(\Delta t)}{C} - \frac{V_{max}^x}{I^x} \right]^+ \qquad 5.34$$

*Where np indicates the non-preemption integration mode.*

**Proof:** s is the beginning of the most recent phase of server overload. All backlogs for all flows have been cleared at this point in time such as $O_x^{h*}(s) = O_x^h(s)$ , $O_{CDT}^{h*}(s) = O_{CDT}^h(s)$, $O_{syn}^{h*}(s) = O_{syn}^h(s)$ and $V^x(s) = 0$

With respect to any given time$t \geq s$, the interval between two consecutive times $\Delta t = t - s$ may be decomposed as follows:

$$\Delta t = \Delta t^+ + \Delta t^- + \Delta t^0 \qquad 5.35$$

Because of syn bands and CDT traffic windows in non-preemption mode, time zero is created by

$$\Delta t^0 = \Delta t_{CDT}^0 + \Delta t_{syn}^0 \qquad 5.36$$

where t denotes the length of time between AVB Class x frames. As a result, we have

$$\Delta t^- = \Delta t_{AVB,x}^- \qquad 5.37$$

Thus, the change in credit throughout the $\Delta t$ period fulfills the requirement.

$$V^x(t) - V^x(s) = V^x(t) \qquad 5.38$$

$$= \Delta t^+ . I^x + \Delta t^+ . S^x \qquad 5.39$$

$$= \left( \Delta t - \left( \Delta t_{CDT}^0 + \Delta t_{syn}^0 \right) \right) . I^x - \Delta t_{AVB,x}^- . (I^x - S^x) \qquad 5.40$$

Since this is the case, we may utilize non-preemption mode to determine the relationship between the AVB Class x service times, the CDT traffic times, and the syn band timings in any given time period $\Delta t$,

$$\Delta t_{AVB,x}^- = \frac{\left\{ \left( \Delta t - \left( \Delta t_{CDT}^0 + \Delta t_{syn}^0 \right) \right) . I^x - V^x(t) \right\}}{C} \qquad 5.41$$

Furthermore, under the worst-case scenario, the output frames of CDT traffic during the time interval $\Delta t$ may be calculated using

$$O_{CDT}^{h*}(t) - O_{CDT}^{h*}(t) = O_{CDT}^{h*}(t) - O_{CDT}^h(s) = C. \Delta t_{CDT}^0 \qquad 5.42$$

Similarly, the time lost owing to syn bands during $\Delta t$ is a loss of time.

$$O_{syn}^{h*}(t) - O_{syn}^{h*}(t) = O_{syn}^{h*}(t) - O_{syn}^h(s) = C.\Delta t_{syn}^0 \qquad 5.43$$

So, $\Delta t^0 = \Delta t_{CDT}^0 + \Delta t_{syn}^0$ in order to limit

$$\Delta t^0 \leq \left\{ \left( O_{CDT}^h(t) + O_{syn}^h(t) \right) - \left( O_{CDT}^h(s) + O_{syn}^h(s) \right) \right\}/C \qquad 5.44$$

$$\leq \frac{\alpha_{syn+CDT}^{h,CBS+SP}(\Delta t)}{C} \qquad 5.45$$

Then, based on (5.41) and (5.44), the AVB Class x output frames over the interval t are constrained by

$$O_x^{h*}(t) - O_x^{h*}(s) = C.\Delta t_{AVB,x}^- \qquad 5.46$$

$$\geq I^x \left[ \Delta t - \frac{\alpha_{syn+CDT}^{h,CBS+SP}(\Delta t)}{C} - \frac{V_{max}^x}{I^x} \right] \qquad 5.47$$

$$\beta_{AVB,x}^{h,np,CBS+SP}(t) = I^x \left[ \Delta t - \frac{\alpha_{syn+CDT}^{h,CBS+SP}(\Delta t)}{C} - \frac{V_{max}^x}{I^x} \right]^+ \qquad 5.48$$

**Corollary 5.3:** Amplified service rate of $AVB_x, \frac{1}{r_{index}}$ times $\beta_{AVB,x,index}^{h,np,amp,CBS+SP}(t) =$

$\frac{I^x}{r_{index}} \left[ \sup\limits_{0 \leq u \leq t} \left\{ u - \frac{\alpha_{syn+CDT}^{h,CBS+SP}(u)}{C} - \frac{V_{max}^x}{I^x} \right\} \right]^+$ Here, *amp* indicates that amplified service curve. In

fig.37, we demonstrate that all of the arrival curve and service curve for CDT and AVB traffic.

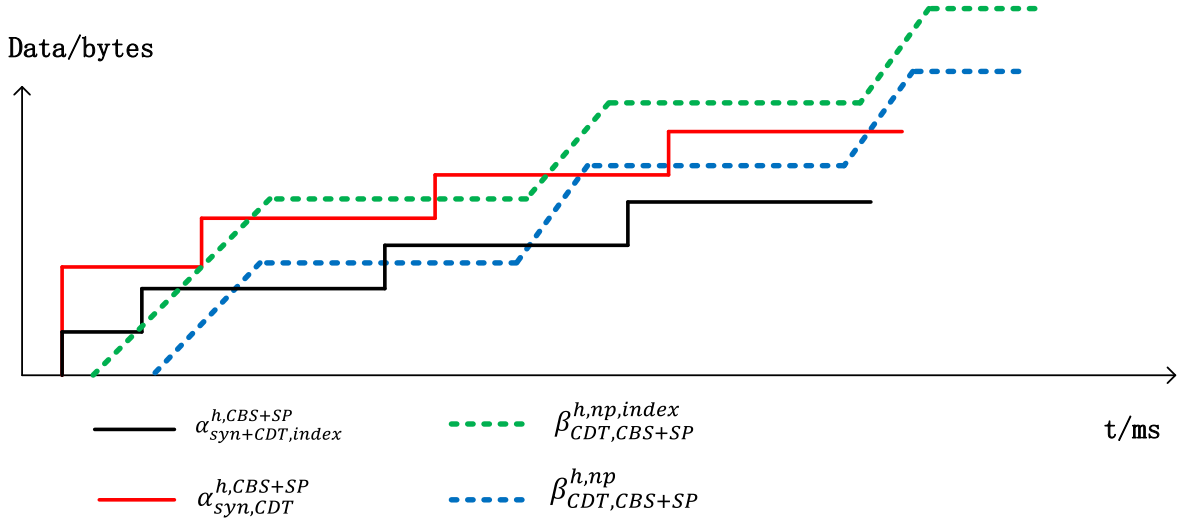

**Figure 37 Several CDT arrival and Service Curve non-preemption mode**

### 5.6.2    Preemption mode

When using the preemption mode, if an AVB frame is preempted, an overhead is imposed to the next AVB frame that is still active. CDT traffic windows preempt AVB Class M frames in the worst-case scenario, and fig. 38 has shown in shrink CDT traffic along with overhead and syn window. We'll assume for the moment that the length of the overhead along with syn-band is $L_{syn,j}$. Overheads may be considered as a distinct factor contributing to the delay of AVB data transmission. Because the overhead will only exist immediately after each CDT traffic window, the overhead constraint curve may be calculated using the following Lemma 5.4.

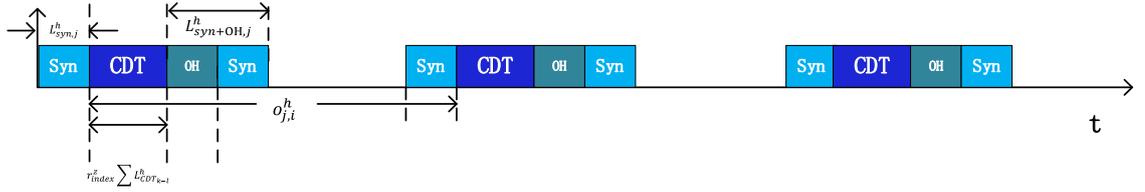

**Figure 38 Shrink CDT traffic output in h port with overhead**

**Lemma 5.6** Under preemption conditions, the aggregate arrival curve for intersecting CDT flows and syn-bands along with overheads in an output port h is given by the equation

$$\alpha_{syn}^{h,CBS+SP}(t) = \max_{0 \le i \le N^h - 1} \left\{ \sum_{j=i}^{i+N^h-1} L_{syn} . C . \left\lceil \frac{(t - O_{j,i}^h - L_{CDT,j} - L_{syn,j})}{T} \right\rceil \right\} \qquad 5.49$$

Here, $L_{TT,j}$ is the TT traffic frame length in j position.

For consideration service curve of CDT traffic of preemption mode is same as Lemma 5.5

**Theorem 5.4** *According to non-preemption mode, service curve of AVB class x where $x \in \{A, B\}$ in an output port of ethernet switch port h,*

$$\beta_{AVB,x,index}^{h,p,CBS+SP}(t) = I^x \left[ \Delta t - \frac{\alpha_{CDT}^{h,CBS+SP}(\Delta t)}{C} - \frac{\alpha_{syn}^{h,CBS+SP}(\Delta t)}{I^x} - \frac{V_{max}^x}{I^x} \right]^+ \qquad 5.50$$

*Where p indicates the preemption integration mode.*

**Proof:** s is the beginning of the most recent phase of server overload. All backlogs for all flows have been cleared at this point in time such as $O_x^{h*}(s) = O_x^h(s)$ , $O_{CDT}^{h*}(s) = O_{CDT}^h(s), O_{syn}^{h*}(s) = O_{syn}^h(s)$ and $V^x(s) = 0$

With respect to any given time $t \ge s$, the interval between two consecutive times $\Delta t = t - s$ may be decomposed as follows:

$$\Delta t^0 = \Delta t_{CDT}^0 \qquad 5.51$$

Preemption-related overheads may be divided into two categories: the durations of frame transmission in AVB Class x and the durations of overheads owing to preemption.

$$\Delta t^- = \Delta t_{syn}^- + \Delta t_{AVB,x}^- \qquad 5.52$$

$\Delta t$ fulfills the credit variation requirement

$$V^x(t) - V^x(s) = V^x(t) \qquad 5.53$$

$$= \Delta t^+ . I^x + \Delta t^+ . S^x \qquad 5.54$$

$$= \left(\Delta t - (\Delta t_{CDT}^0)\right) . I^x - (\Delta t_{syn}^- + \Delta t_{AVB,x}^-) . C \qquad 5.55$$

As a result, using preemption mode, we can determine the connection between service times for AVB Class x, extra service time for AVB overheads, and the length of CDT windows in any interval $\Delta t$.

$$\Delta t_{AVB,x}^- = \frac{\left\{\left(\Delta t - (\Delta t_{CDT}^0)\right) . I^x - \Delta t_{syn}^- . C - V^x(t)\right\}}{C} \qquad 5.56$$

The output frames of CDT traffic during t are higher in the worst-case scenario.

$$O_{CDT}^{h*}(t) - O_{CDT}^{h*}(t) = O_{CDT}^{h*}(t) - O_{CDT}^h(s) = C . \Delta t_{CDT}^0 \qquad 5.57$$

During the time period $\Delta t$, AVB syn band may be covered by an additional service offered by

$$O_{syn}^{h*}(t) - O_{syn}^{h*}(t) = O_{syn}^{h*}(t) - O_{syn}^h(s) = C . \Delta t_{syn}^0 \qquad 5.58$$

So, $\Delta t_{CDT}^0$ in order to limit

$$\Delta t_{CDT}^0 \leq \frac{\left\{O_{CDT}^{h*}(t) - O_{CDT}^h(s)\right\}}{C} \leq \frac{\alpha_{CDT}^{h,CBS+SP}(\Delta t)}{C} \qquad 5.59$$

in the same way that $\Delta t_{syn}^-$ fulfills,

$$\Delta t_{syn}^- \leq \frac{\alpha_{syn}^{h,CBS+SP}(\Delta t)}{C} \qquad 5.60$$

According to (5.56), (5.59) and (5.60), Class x output frames over the interval t are limited by, respectively.

$$O_x^{h*}(t) - O_x^{h*}(s) = C . \Delta t_{AVB,x}^- \qquad 5.61$$

$$\geq I^x \left[\Delta t - \frac{\alpha_{CDT}^{h,CBS+SP}(\Delta t)}{C} - \frac{\alpha_{syn}^{h,CBS+SP}(\Delta t)}{I^x} - \frac{V_{max}^x}{I^x}\right] \qquad 5.62$$

$$\beta_{AVB,x}^{h,p,CBS+SP}(t) = I^x \left[\Delta t - \frac{\alpha_{CDT}^{h,CBS+SP}(\Delta t)}{C} - \frac{\alpha_{syn}^{h,CBS+SP}(\Delta t)}{I^x} - \frac{V_{max}^x}{I^x}\right]^+ \qquad 5.63$$

**Corollary 5.4:** Amplified service rate of $AVB_x$, $\frac{1}{r_{index}}$ times $\beta_{AVB,x,index}^{h,p,amp,CBS+SP}(t) = \frac{I^x}{r_{index}}\left[\sup_{0 \le u \le t}\left\{u - \frac{\alpha_{CDT}^{h,CBS+SP}(\Delta t)}{C} - \frac{\alpha_{syn}^{h,CBS+SP}(\Delta t)}{I^x} - \frac{V_{max}^x}{I^x}\right\}\right]^+$ Here, *amp* indicates that amplified service curve. In fig. 39, we demonstrate that all of the arrival curve and service curve for CDT and AVB traffic.

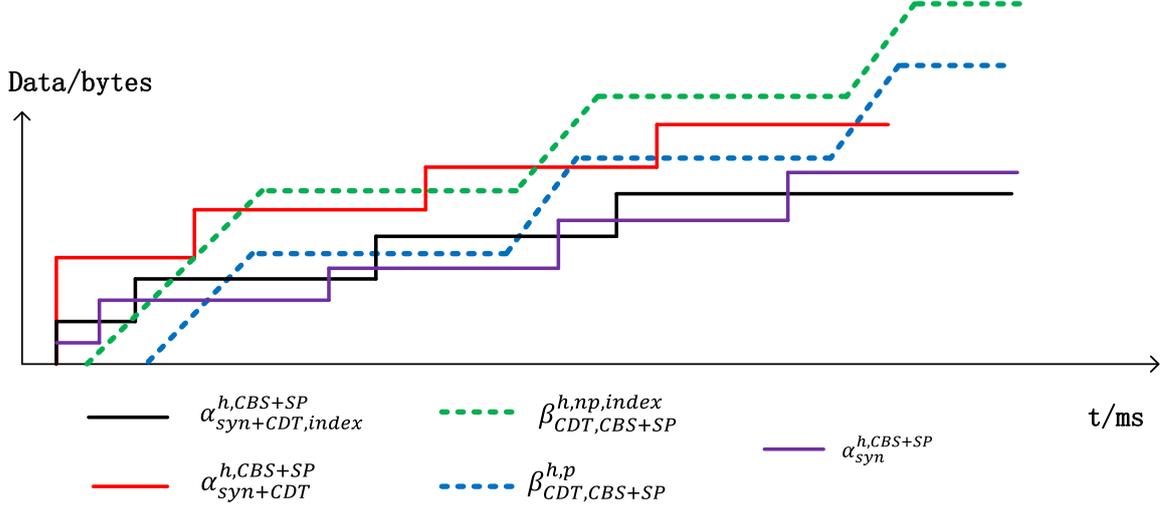

**Figure 39 Several CDT arrival and Service Curve preemption mode**

In table 19 and fig. 40, we have shown all of the service curve for different scheduling mechanism

**Table 19 Several Service Curve of AVB Traffic**

| Service Curve | Several Service Curve of AVB Traffic |
|---|---|
| Non-preemption mode CBS+TAS | $\beta_{AVB,x,index}^{h,np,CBS+TAS}(t) = I^x\left[\Delta t - \frac{\alpha_{syn+TT,index}^{h,CBS+TAS}(\Delta t)}{C} - \frac{V_{max}^x}{I^x}\right]^+$ |
| Non-preemption mode CBS+TAS, amplified | $\beta_{AVB,x,CBS+TAS}^{h,np,amp}(t) = \frac{I^x}{r_{index}}\left[\sup_{0 \le u \le t}\left\{u - \frac{\alpha_{syn+TT}^{h,CBS+TAS}(u)}{C} - \frac{V_{max}^x}{I^x}\right\}\right]^+$ |
| Preemption mode CBS+TAS | $\beta_{AVB,x,index}^{h,p,CBS+TAS}(t) = I^x\left[\Delta t - \frac{\alpha_{TT}^{h,CBS+TAS}(\Delta t)}{C} - \frac{\alpha_{syn,index}^{h,CBAS+TAS}(\Delta t)}{I^x} - \frac{V_{max}^x}{I^x}\right]^+$ |
| Preemption mode CBS+TAS, amplified | $\beta_{AVB,x,CBS+TAS}^{h,p,amp}(t) = \frac{I^x}{r_{index}}\left[\sup_{0 \le u \le t}\left\{u - \frac{\alpha_{TT}^{h,CBS+TAS}(u)}{C} - \frac{\alpha_{syn}^{h,CBS+TAS}(u)}{I^x} - \frac{V_{max}^x}{I^x}\right\}\right]^+$ |
| Non-preemption mode CBS+SP | $\beta_{AVB,x,index}^{h,np,CBS+SP}(t) = I^x\left[\Delta t - \frac{\alpha_{syn+CDT}^{h,CBS+SP}(\Delta t)}{C} - \frac{V_{max}^x}{I^x}\right]^+$ |
| Non-preemption mode CBS+SP, amplified | $\beta_{AVB,x,index}^{h,np,amp,CBS+SP}(t) = \frac{I^x}{r_{index}}\left[\sup_{0 \le u \le t}\left\{u - \frac{\alpha_{syn+CDT}^{h,CBS+SP}(u)}{C} - \frac{V_{max}^x}{I^x}\right\}\right]^+$ |
| Preemption mode CBS+SP | $\beta_{AVB,x,index}^{h,p,CBS+SP}(t) = I^x\left[\Delta t - \frac{\alpha_{CDT}^{h,CBS+SP}(\Delta t)}{C} - \frac{\alpha_{syn}^{h,CBS+SP}(\Delta t)}{I^x} - \frac{V_{max}^x}{I^x}\right]^+$ |

| Preemption mode CBS+SP, amplified | $\beta_{AVB,x,index}^{h,p,amp,CBS+SP}(t) = \frac{I^x}{r_{index}} \left[ \sup_{0 \leq u \leq t} \left\{ u - \frac{\alpha_{CDT}^{h,CBS+SP}(\Delta t)}{C} - \frac{\alpha_{syn}^{h,CBS+SP}(\Delta t)}{I^x} - \frac{V_{max}^x}{I^x} \right\} \right]^+$ |
|---|---|

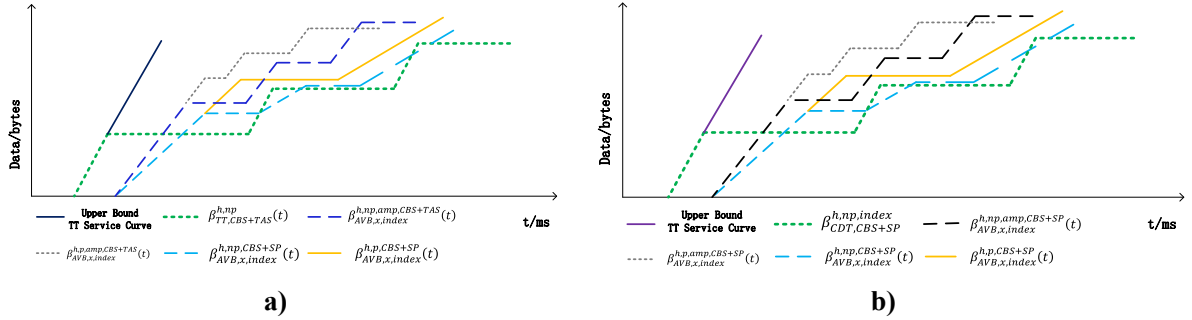

a)                                                          b)

**Figure 40 Shrink higher Priority service curve for different scheduling**

## 5.7 Upper bound delay for data and frame indexing Transmission

According to network calculus theory, the upper limit latency of a Class x flow in the output port h is determined by the greatest horizontal deviation between the arrival curve of AVB and the service curve of AVB in the output port h which is shown in fig. 41.

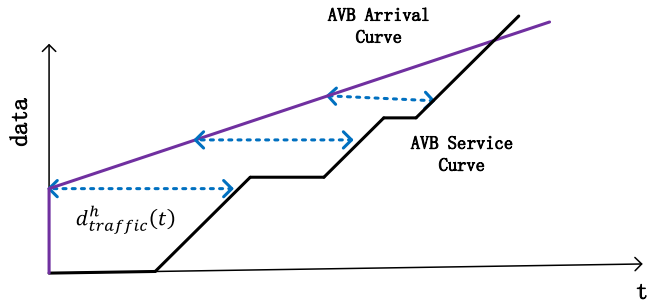

**Figure 41 Horizontal Delay Calculation**

$$d_{traffic}^h(t) = h\left(\alpha_{traffic}^h(t), \beta_{traffic}^h(t)\right), traffic \in \{TT, A, B\} \qquad 5.64$$

Fig. 42 illustrates the different delays experienced by packets as they travel from one end device to another end device for the purpose of estimating end-to-end latency, which is gathered from this paper [77]. If each node has a FIFO queue and TSN scheduling is available, and each device is linked to the network through a physical Ethernet wire, the following scenario is possible: There are three considerations in this equation: first, the FIFO queuing time, second, the scheduling time ($d_{traffic}^h(t)$, and thirdly, the physical cable speed time ($\frac{L_{traffic}}{C}$). The overall end-to-end latency for a single intermediate node is represented by Equation 5.64. There is more than one node in this case, and the intermediate delay will be summed proportionally to the number of nodes in the network.

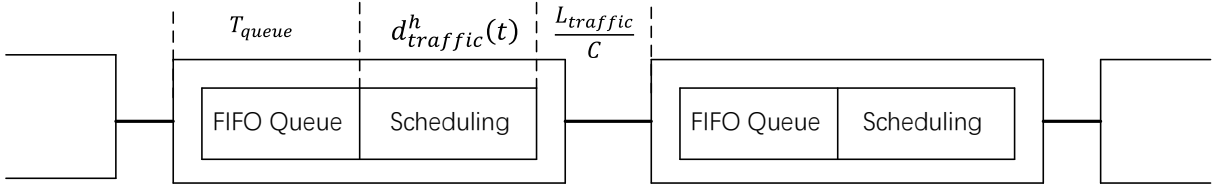

**Figure 42 End to end delay demonstration**

$$D_{total_{traffic}}^{e2e}(t) = T_{queue} + d_{traffic}^h(t) + \frac{L_{traffic}}{C} \qquad 5.65$$

In equation 5.65 may be vary according to changing time $d_{traffic}^h(t)$ because $T_{queue}$ and $T_{queue}$ remain same.

# 6. Experiment and Simulation Results

For performance evaluation, we have accumulated 3 basic concept such as network calculus (NC) which is theoretical analysis, secondly, we have used network simulation, thirdly, we have done experiment to verify simulation process. TSN standards for IEEE 802.1Qav and IEEE 802.1Qbv have been used for evaluation by using standard ethernet switch.

## 6.1 TSN over Standard Switch NC Approach

For the purpose of evaluating our approach, we have used four test cases based on the Orion Crew Exploration Vehicle (CEV) [28] that have been modified to use TSN, which are referred to as TC2 in this paper. CEV's architecture consists of 31 ESes, 15 SWs, and 39 routes, all of which are linked by dataflow connections capable of transmitting at 1 Gbps. Implementation of our suggested analysis is done in C++ using the RTC toolbox [29], which is executed on a machine with an Intel Core i5-3520M processor operating at 2.90 GHz and 8 GB of RAM.

**Table 20 Flow of TT Traffic include with Frame size, Periods**

| Flow | Size (B) | Periods(ms) | Flow | Size (B) | Periods(ms) |
|------|----------|-------------|------|----------|-------------|
| $TT_1$ | 145 | 62.5 | $TT_9$ | 186 | 37.5 |
| $TT_2$ | 508 | 125 | $TT_{10}$ | 1420 | 37.5 |
| $TT_3$ | 1268 | 37.5 | $TT_{11}$ | 197 | 25 |
| $TT_4$ | 888 | 37.5 | $TT_{12}$ | 246 | 12.5 |
| $TT_5$ | 170 | 25 | $TT_{13}$ | 103 | 12.5 |
| $TT_6$ | 1527 | 37.5 | $TT_{14}$ | 1053 | 37.5 |
| $TT_7$ | 578 | 62.5 | $TT_{15}$ | 913 | 75 |
| $TT_8$ | 908 | 75 | | | |

**Table 21 Flow of AVB Traffic include with Frame size, Periods**

| Flow | Size (B) | Periods(ms) | type | Flow | Size(B) | Period(ms) | type |
|------|----------|-------------|------|------|---------|------------|------|
| $RC_1$ | 163 | 125 | A | $RC_{18}$ | 147 | 125 | B |
| $RC_2$ | 157 | 125 | A | $RC_{19}$ | 116 | 125 | A |
| $RC_3$ | 170 | 125 | B | $RC_{20}$ | 118 | 125 | A |
| $RC_4$ | 125 | 125 | A | $RC_{21}$ | 138 | 125 | A |
| $RC_5$ | 114 | 125 | B | $RC_{22}$ | 118 | 125 | A |
| $RC_6$ | 119 | 125 | B | $RC_{23}$ | 148 | 125 | B |
| $RC_7$ | 156 | 125 | A | $RC_{24}$ | 118 | 125 | A |
| $RC_8$ | 112 | 125 | A | $RC_{25}$ | 149 | 125 | A |
| $RC_9$ | 110 | 125 | B | $RC_{26}$ | 116 | 125 | B |
| $RC_{10}$ | 155 | 125 | A | $RC_{27}$ | 110 | 125 | A |

| $RC_{11}$ | 126 | 125 | A | $RC_{28}$ | 124 | 125 | A |
| $RC_{12}$ | 114 | 125 | B | $RC_{29}$ | 114 | 125 | A |
| $RC_{13}$ | 140 | 125 | A | $RC_{30}$ | 119 | 125 | A |
| $RC_{14}$ | 114 | 125 | A | $RC_{31}$ | 168 | 125 | A |
| $RC_{15}$ | 114 | 125 | B | $RC_{32}$ | 157 | 125 | A |
| $RC_{16}$ | 187 | 125 | A | $RC_{33}$ | 116 | 125 | B |
| $RC_{17}$ | 114 | 125 | A | $RC_{34}$ | 117 | 125 | A |

In TC2, we are interested in determining how our approach handles the two integration modes, namely, non-preemption and preemption, in which it is implemented.TC2 is a broader scenario, with 15 TT flows, 20 AVB flows of Class A, and 14 AVB flows of Class B flowing on it at any one moment (including multicast flows). The specifics of the TT and AVB flows are reported in table 20 and 21, and the GCLs were created using the algorithm which is explained in Appendix B. The idle slopes of Class A and B are respectively 60 percent and 15 percent of the total bandwidth, and we assume that the parameters of CBS satisfy $I^x - S^x = C$ in this experiment, and the idle slopes of Class A & B are 60 percent and 15 percent of the whole bandwidth respectively.

### 6.1.1 Worst-Case Delay

To evaluate of different TSN scheduling namely CBS+TAS and CBS+SP along with 2 integration mode, we have demonstrated in fig. 43. The evaluation will be done in standard network switch with synchronization clock. In fig. 43, we can observe that non-preemption takes more time to flow AVB traffic with compare to preemption mode. This statement is equally true for both scheduling mechanisms. Because, during non-preemption mode, syn+GB bands are remaining idle and that time neither AVB traffic can transfer nor TT traffic can flow to the desire destination. On the other hands, CBS+SP scheduling delay is more unpredictable with compare to CBS+TAS scheduling delay because CBS+SP is event triggered that means when higher priority traffic is waiting then lower traffic will be in queue until the higher priority traffics are transferred as a consequences worst case delay is quite unpredictable. On the other hands, CBS+TAS scheduling delay is quite predictable because it is time-triggered scheduling mechanism and several traffics are controlled by timely opening gate where the gates are controlled by GCL. In fig.43, in above statement is mirror reflected in the standard network switch in the context of TSN.

### 6.1.2    Traffic Flow Calculation

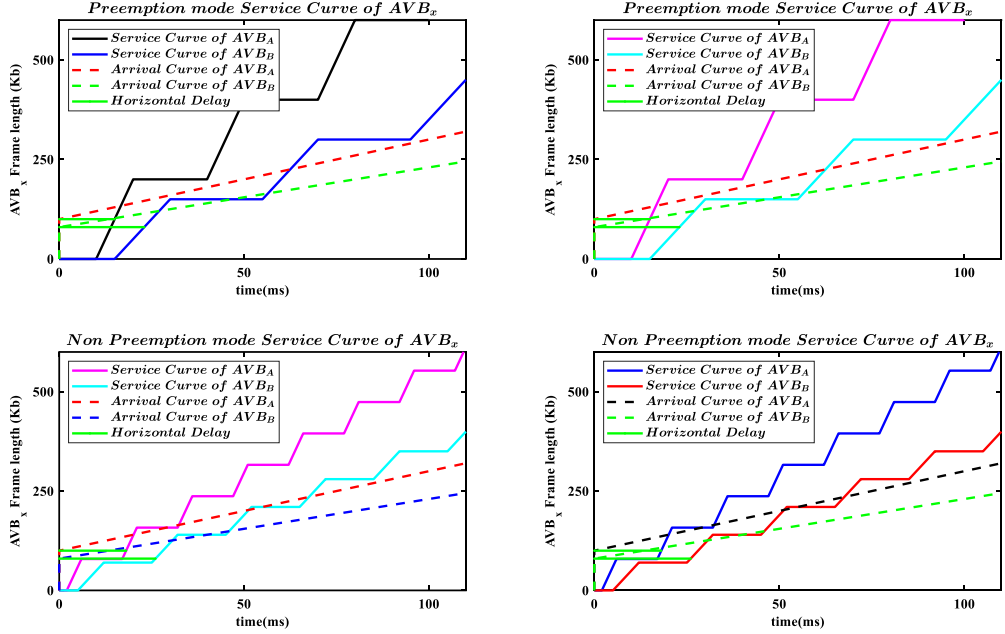

**Figure 43 AVB traffic accumulation for several mode and TAS+CBS & CBS+SP scheduling**

To compare our technique based on network calculus with the current methodology in [48], we utilize a test case TC2 taken directly from [23], which has 15 TT flows and 34 AVB flows of Class A and class B. According to leaky bucket arrival which is shown in equation 3.65, the burst of about avb traffic is 90kb.Besides, TT windows is $L_{TT} = 120\mu s$ or $L_{CDT} = 120\mu s$, AVB window is $L_{avb} = 15\mu s$, syn windows along with guard band is $L_{syn+GB} = 5\mu s$, physical link speed is 100 Mbit/s, offset is $o^h = 60\mu s$, hyper-periods is $T_{GCL} = 250\mu s$ though hyper-periods is only applicable for TAS+CBS scheduling, specific sys window is $L_{sys} = 2\mu s$. In table 20 and 21, we have provided framing information for TT and AVB traffic where we can see that maximum frame of TT traffic is 1420 bytes and our TT window is $L_{TT} = 120\mu s$ or $L_{CDT} = 120\mu s$ and physical speed is 100 Mbits/s so that $L_{TT}$ is sufficient for transfer the maximum frame of TT, similarly, maximum size of avb traffic is 187 bytes which is given in table 21 and   avb windows is   $L_{avb} = 15\mu s$  so, this windows is also sufficient for transferring maximum avb frame at a time.

Worst case delay of the TSN switch are comparatively low with the considering standard switches but the delay difference of the switch is in acceptable range where the fig. 44 clearly shown. If the standard switch uses for TSN network, network should be well synchronous as well as network configuration should be static and this statement is always for small network and large network. In fig. 44, we can observe that minimum delay of the TT traffic is $2\,ms$ and maximum delay is about $13\,ms$. Similarly, minimum delay of the AVB traffic is $10\,ms$ in preemption mode, $12\,ms$ in non-preemption where both of the modes take more time with compare to TSN switch. In similar trend, maximum delay is $26\,ms$ preemption, and $32\,ms$ in non-preemption. To sum up the whole flow by using standard switch, the delay is reliable for TSN network.

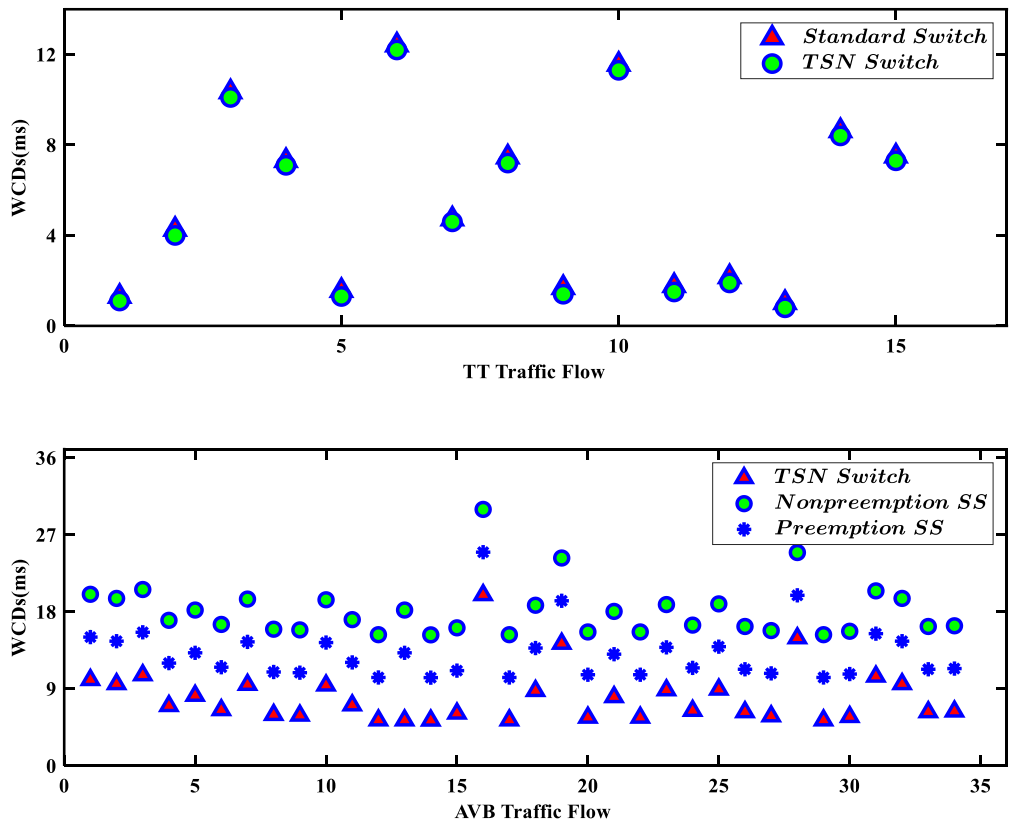

**Figure 44 TT and AVB traffic flow for different mode and scheduling**

### 6.1.3 Idle and Send Slope Relationship

Idle-slope setting can significantly affect the delay analysis results, increase $I^A$ and reduce the class a traffic delay; When $I^A$ is reduced, the class B flow delay becomes larger. Generally speaking, the larger the class A credit volume increase rate idle-slope B, the greater the logical bandwidth occupied by such traffic and the shorter the credit volume accumulation time. Therefore, there are more transmission opportunities for frames on the port. From the perspective of a single node, if the window time slot allocated to class A / B is fixed, increasing idle-slope of B can reduce the delay of this kind of traffic.

Next, the above traffic configuration is used as the input to calculate the delay of single node of different idle-slope. For TT traffic, set the frame length $L^{TT} = 120\mu s$, periods of TT is 4ms, $T_{GCL} = 250\mu s$ .For other types of traffic, set class a frame length $L^A = 1000 bytes$, period $L^A$ is 10ms, a total of 10; Class B frame length $L^B = 1500 bytes$, period $L^B$ is 20ms, 10 in total. Class a load is 8%, class B load is 6%, and the ratio of the two is 4:3; Assuming that 10 class A / b traffic are from different links, the maximum bursts are 80Kb and 120KB respectively, and the ratio of the two is 2:3.

As shown in fig. 45, the relationship between single node delay and idle-slope is given. Here, $I^A + I^B = 0.75C$ is given. Therefore, if idle-slope increases, send-slope will decrease.

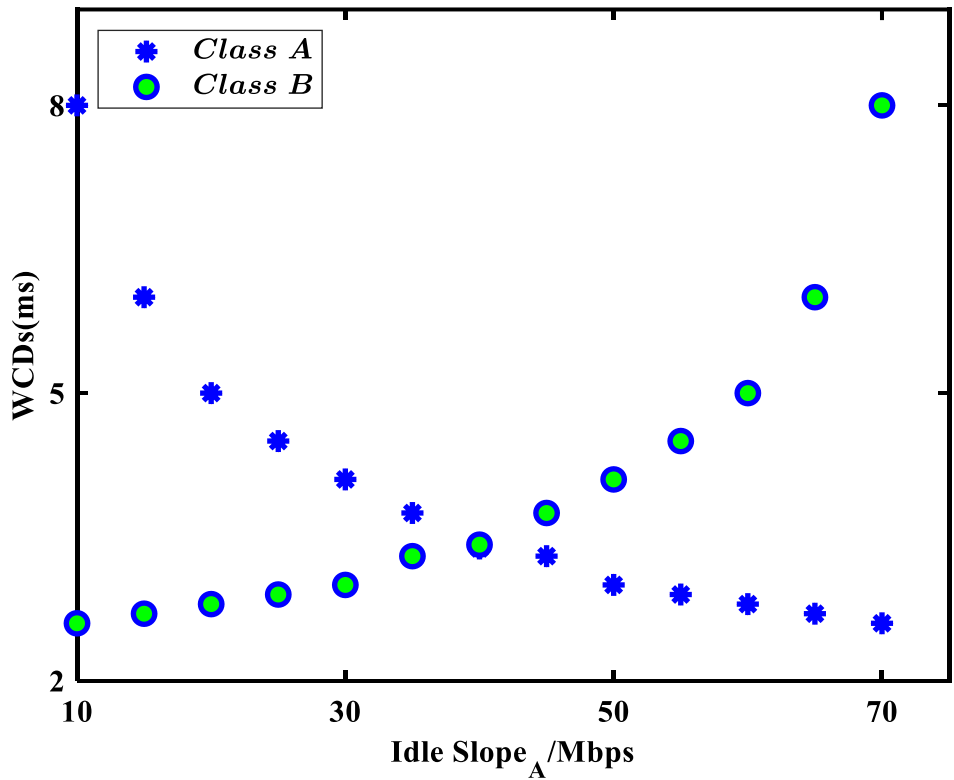

**Figure 45 Idle and send slope relationship**

### 6.1.4 End to End Delay in NC Approach

Along with TSN, the most common method of determining the end-to-end delay bound is to sum the delay limits of each switch in the course of a flow. In fig. 46, we have shown topological connection where 8 end devices are available and 4 standard network switches are connected each other. Moreover, 2 PTP clock are connected in edge of the network which is responsible for controlling scheduling mechanism according to algorithm. Between 2 PTP clock are connected each other. The ethernet connections are using duplex mode. In table 22, we have written the details topological path from each end to another end device. To performance evaluation, we have used in above table 20 and 21 frame length, flow number as well as traffic classification (TT, A, B).

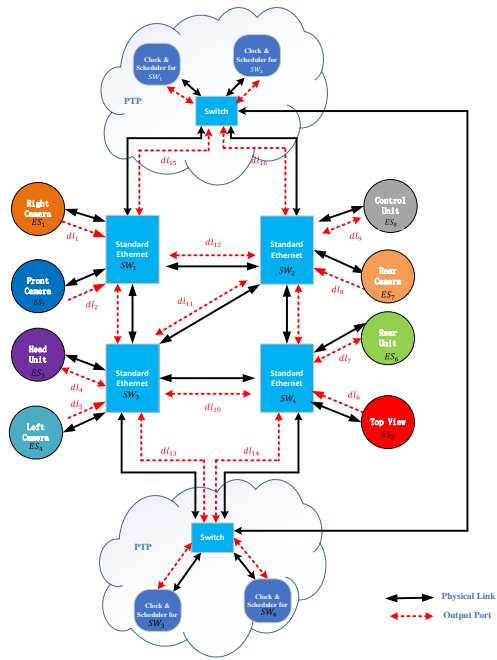

**Figure 46 Experiment and Simulation network topology**

**Table 22 Network Configuration for TT and AVB traffic**

| Data Flow Path | Network Configuration | Type | Data Flow Path | Network Configuration | Type |
|---|---|---|---|---|---|
| 1 | $[ES_1, SW_1]$ | A&B | 7 | $[ES_8, SW_2]$ | TT |
| | $[SW_1, SW_2]$ | | | $[SW_2, SW_1]$ | |
| | $[SW_2, SW_4]$ | | | $[SW_1, SW_3]$ | |
| | $[SW_4, ES_5]$ | | | $[SW_3, ES_3]$ | |
| 2 | $[ES_4, SW_3]$ | A&B | 8 | $[ES_8, SW_2]$ | TT |
| | $[SW_3, SW_4]$ | | | $[SW_2, SW_1]$ | |
| | $[SW_4, ES_5]$ | | | $[SW_1, SW_3]$ | |
| 3 | $[ES_5, SW_4]$ | A&B | | $[SW_3, ES_4]$ | |
| | $[SW_4, SW_3]$ | | | $[ES_8, SW_2]$ | |
| | $[SW_3, ES_3]$ | | 9 | $[SW_2, SW_4]$ | TT |
| 4 | $[ES_3, SW_3]$ | A&B | | $[SW_4, ES_5]$ | |
| | $[SW_3, SW_4]$ | | 10 | $[ES_8, SW_2]$ | TT |
| | $[SW_4, ES_6]$ | | | $[SW_2, SW_4]$ | |
| 5 | $[ES_8, SW_2]$ | TT | | $[SW_4, ES_5]$ | |
| | $[SW_2, SW_1]$ | | 11 | $[ES_8, SW_4]$ | TT |
| | $[SW_1, ES_1]$ | | | $[SW_4, SW_2]$ | |
| 6 | $[ES_8, SW_2]$ | TT | | $[SW_2, ES_6]$ | |
| | $[SW_2, SW_1]$ | | 12 | $[ES_8, SW_2]$ | TT |
| | $[SW_1, ES_2]$ | | | $[SW_2, ES_2]$ | |

In . 47, we can observe that non-preemption delay is comparatively higher in all cases because in non-preemption mode has used extra guard band along with syn windows as a consequences TT flow in both cases non-preemption delay is higher and the maximum delay of the of the TT flow around 11ms in both cases such as TAS+CBS and CBS+SP. In the context of AVB traffic, non-preemption mode has also experienced more delay but the end to end delay the of AVB traffic using standard switch (SS) meet the requirement of the real-time communication and according to our algorithm and frame length as well as scheduling mechanism, the maximum delay of AVB traffic is about 13ms .To performance evaluation, we have used TSN switch end to end delay which is mentioned in this paper[88]. In fig. 47, We can see the TSN switch delay in all of the cases are low because in TSN switch does not need to synchronizing frame to send and receiving data. Moreover, in TSN switch can handle algorithm without using PTP clock. But at a glance, difference of the TSN switch output and SS output are quite negotiable.

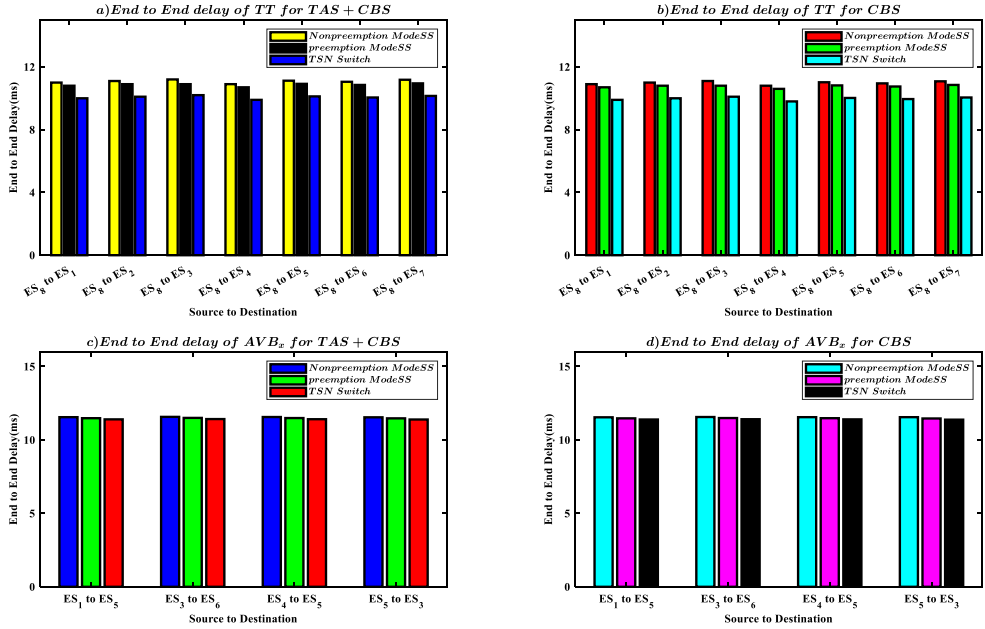

**Figure 47 End to end delay for TT and AVB traffic**

## 6.2 Omnet++ Output

Further performance evaluation of Time sensitive network over standard switch, we have used another environment which can virtual calculate end to end delay. There are several network simulators used calculating end to end delay. Among of the network simulator, Omnet++ is one of the reliable open source platforms where we can easily simulate our desire network. However, according to our basic requirement of the network simulation, we have designed standard switch, end devices. In fig. 48a, we have illustrated our network topology, and in fig. 48b implies that standard host (end device) has been designed, in fig. 48 c & d imply that ethernet switch is designed according to our requirement.

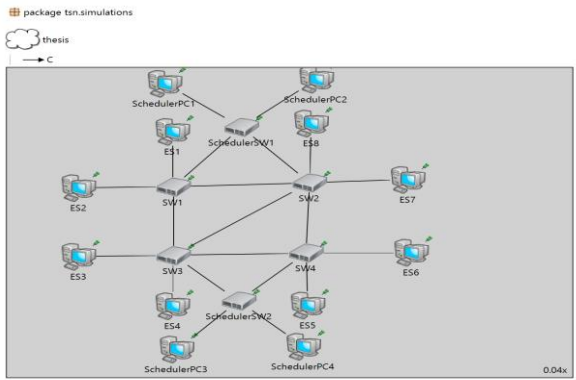

a)

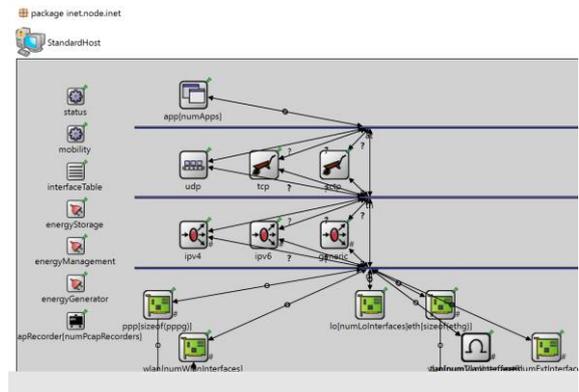

b)

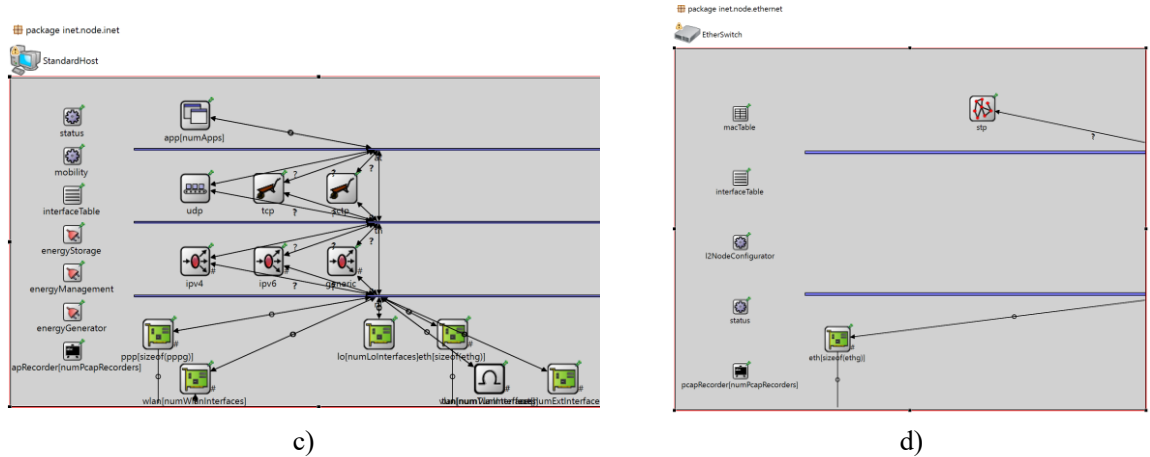

c)                                                                    d)

**Figure 48 Network Evaluation using Omnet++ simulation**

Non-preemption delay is comparatively higher in all cases, as shown in fig 49, due to the fact that non-preemption mode makes use of an additional grand band in addition to the syn-band. As a result, TT flow is higher in both cases, with the maximum non-preemption delay in both cases being around $11.5m$s in both cases, i.e. TAS+CBS and CBS+SP. In the context of AVB traffic, non-preemption mode has also experienced increased delay, but the end-to-end delay of AVB traffic using a standard switch (SS) meets the requirements of real-time communication, and according to our algorithm, frame length, and scheduling mechanism, the maximum delay of AVB traffic is approximately 14 $m$s. We employed TSN switch end to end delay, which is stated in this document [88], for performance assessment and comparison. Since the TSN switch does not need a synchronizing frame to transmit and receive data, the TSN switch latency is minimal in all of the scenarios, as can be shown in fig. 46. TSN switches, on the other hand, are capable of handling algorithms without the need of the PTP clock. Nonetheless, at first appearance, there is little difference between the TSN switch output and the SS output.

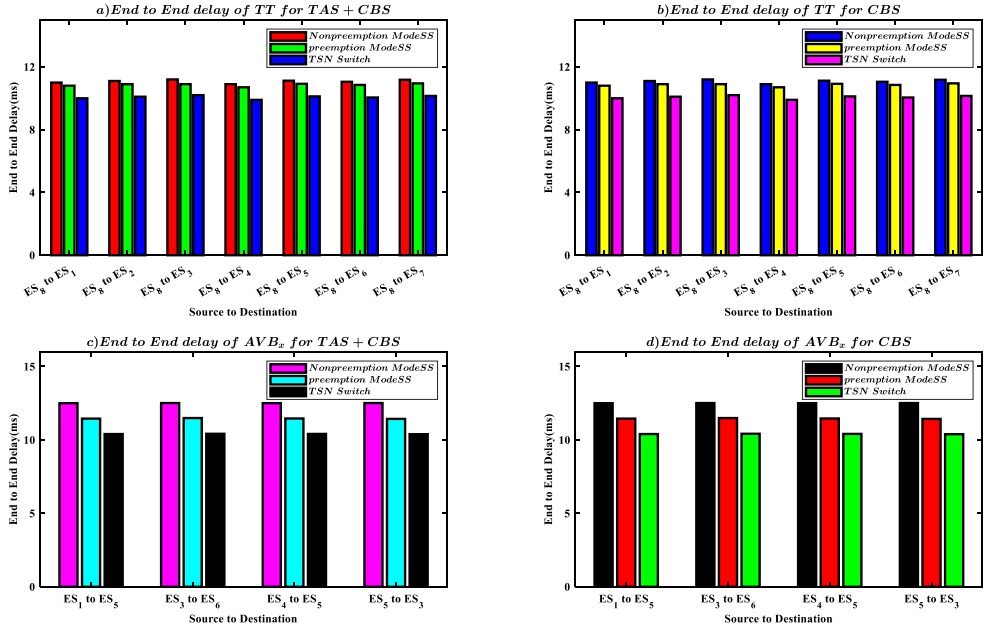

**Figure 49 End to end delay for TT and AVB traffic using Omnet++**

## 6.3 Experiments Results

### 6.3.1    Experiment Setup

To performance evaluation of standard switch in the context of TSN, experimental analysis is the final step. We have used 8 end devices and 4 standard switches. In fig. 50, end devices consist of 8GB RAM, intel core i5 processor with 2.90GHz bus speed. For sending and receiving data visualization, we have used MATLAB 2018b.For time measurement, we have also used two profishark tap which is connected outer port of the switches. Timing difference between two tap implies end to end delay. In fig. 50, Wireshrank software is used for gathering all of the time and protocols which is using the data transmission/IP protocol has been used for data communication.

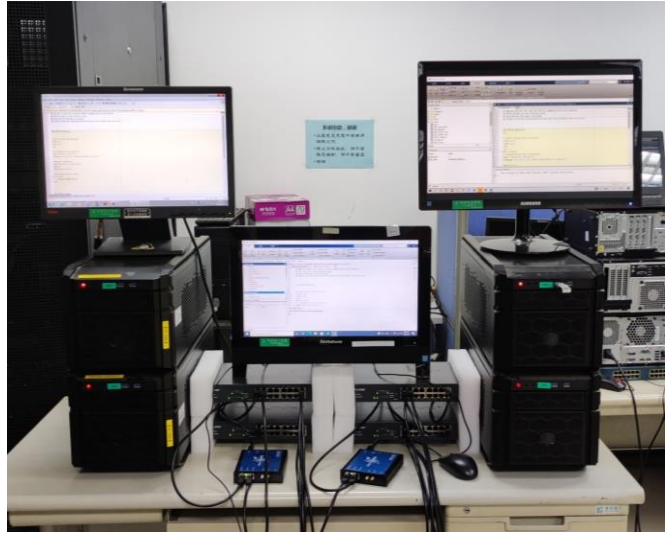

**a) Network Setup**

| No. | Time | Source | Destination | Protocol | Length | Info |
|---|---|---|---|---|---|---|
| 1 | 0.000000000 | 192.168.1.11 | 192.168.1.12 | TCP | | 70 49859 → 4013 [SYN] Seq=0 Win=64240 Len=0 MSS=1460 WS=256 SACK_PERM=1 |
| 2 | 0.000587015 | 192.168.1.12 | 192.168.1.11 | TCP | | 70 4013 → 49859 [SYN, ACK] Seq=0 Ack=1 Win=65535 Len=0 MSS=1460 WS=256 SACK_PERM=1 |
| 3 | 0.001028265 | 192.168.1.11 | 192.168.1.12 | TCP | | 64 49859 → 4013 [ACK] Seq=1 Ack=1 Win=262656 Len=0 |
| 4 | 0.001092070 | 192.168.1.11 | 192.168.1.12 | TCP | | 64 [TCP Window Update] 49859 → 4013 [ACK] Seq=1 Ack=1 Win=2097152 Len=0 |
| 5 | 0.006148010 | 192.168.1.11 | 192.168.1.12 | TCP | | 64 49859 → 4013 [PSH, ACK] Seq=1 Ack=1 Win=2097152 Len=2 |
| 6 | 0.016926975 | 192.168.1.11 | 192.168.1.11 | TCP | | 78 49859 → 4013 [FIN, PSH, ACK] Seq=3 Ack=1 Win=2097152 Len=20 |
| 7 | 0.017078590 | 192.168.1.12 | 192.168.1.11 | TCP | | 64 4013 → 49859 [ACK] Seq=1 Ack=24 Win=2102272 Len=0 |

**b) WireShrank Output**

**Figure 50 Experimental Setup**

### 6.3.2 End to end delay comparison in all environment

This part contains examples of end-to-end latency in a variety of environments, including network calculus technique, omnet++ simulation, and experimental approach, amongst others. Fig. 48 shows that the delay in three environments is relatively comparable, with some differences within an acceptable range, despite the fact that the experiment delay is larger than the delay in NC and Omnet++ environments. Despite the fact that some technological latency contributes to the increased experimental delay, this technical latency has no impact on real-time communication, and data may be sent before the deadline.

TT flow is shown in fig. 51 as having a larger experimental delay in both situations, with the greatest delay of the TT flow being about 13.5ms in both cases (TAS+CBS and CBS+SP), as shown. In the context of AVB traffic, network calculus has also experienced more delay, but the end-to-end delay of AVB traffic using a standard switch (SS) meets the requirements of real-time communication, and according to our algorithm, frame length, and scheduling mechanism, the maximum delay of AVB traffic is approximately 14.8ms.

According to fig. 51, the TT flow simulation and NC delay are almost same in both circumstances, with the highest delay of the TT flow averaging approximately 11.5ms and 10ms respectively in both cases, such as TAS+CBS and CBS+SP, and the minimum delay of the TT flow around 8ms. However, the end-to-end latency of AVB traffic utilizing standard switch (SS) meets the requirements of real-time communication, and according to our algorithm and frame length as well as scheduling mechanism, the maximum delay of AVB traffic is around 12ms and 13 ms respectively.

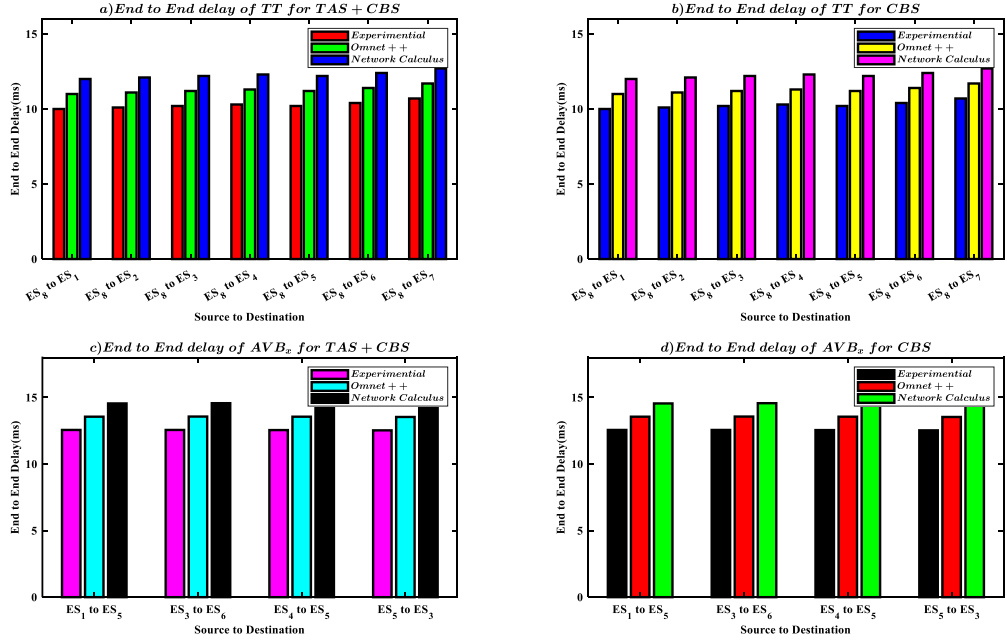

**Figure 51 End to end delay comparison for all environment**

## 6.4 Shrink Service Curve

### 6.4.1 Network Calculus Approach

As shown in fig. 52, we were able to compare and contrast two alternative TSN scheduling schemes: CBS+TAS and CBS+SP, as well as two different integration modes. The examination will be carried out using a standard network switch in conjunction with a synchronization clock. Non-preemption mode has a longer time to flow AVB traffic than preemption mode, as can be seen in the comparison between the two modes. This claim holds true for both scheduling systems when they are used in conjunction with one another. Because, when non-preemption mode is enabled, the syn+GB window+bands remain idle, and at such times, neither AVB traffic nor TT traffic can transfer to the desired destination, non-preemption mode is advantageous. When comparing CBS+SP scheduling delay to CBS+TAS scheduling delay, the CBS+SP scheduling delay is much more variable. As a consequence of the fact that CBS+SP is event triggered, while higher priority traffic is waiting, lower priority traffic will be put in queue until the higher priority traffics are transferred, resulting in a worst-case delay that is very unpredictable in the worst-case scenario. GCL scheduling delays, on the other hand, are rather predictable since CBS+TAS is a time-triggered scheduling mechanism and because many traffics are governed by timely opening gates, which are controlled by GCL. Fig. 49 depicts a conventional network with the mirror depicted in the preceding sentence.

The benefit of the shrunk TT frame is seen in fig. 52. We can see that the service curves for all of the modes, as well as the scheduler, have been greatly improved in this case. Furthermore, the dotted lines imply that if we adopt a shrunk service curve, we will save time in the long run.

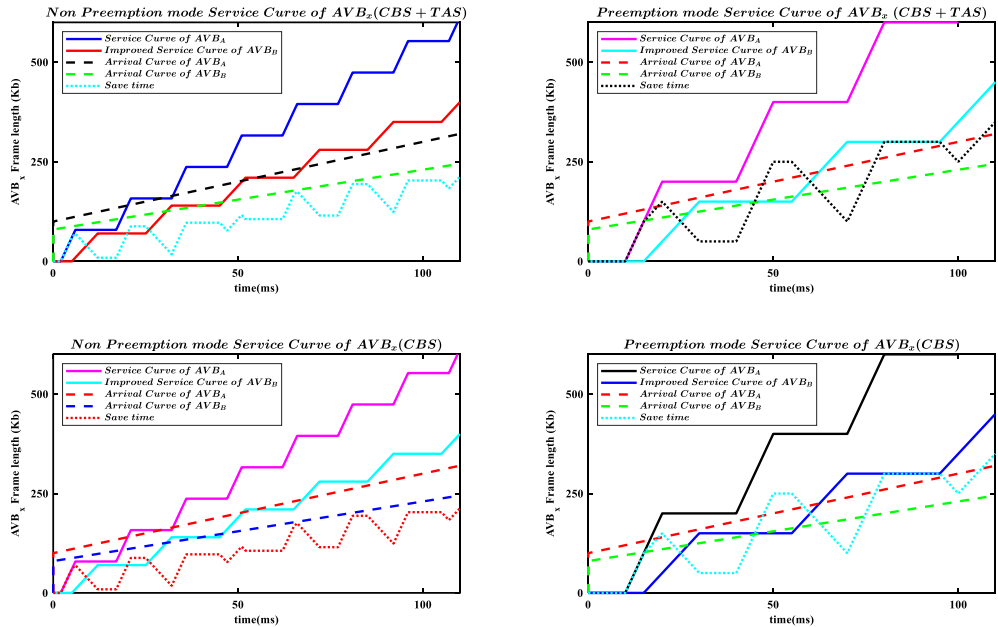

**Figure 52 AVB traffic accumulation for several mode and scheduling**

To compare our technique based on network calculus with the current methodology, we utilize a test case TC2 taken directly from [23], which has 15 TT flows and 34 AVB flows of Class A and class B. According to leaky bucket arrival which is shown in equation 5.65, the burst of about avb traffic is 90kb. Besides, TT windows is $L_{TT} = 50\mu s$, AVB window is $L_{avb} = 15\mu s$, syn windows along with guard band is $L_{syn+GB} = 5\mu s$, physical link speed is 100 Mbit/s, offset is $o^h = 60\mu s$, hyper-periods is $T_{GCL} = 250\mu s$ though hyper-periods is only applicable for TAS+CBS scheduling, specific sys window is $L_{sys} = 2\mu s$. In table 20 and 21, we have provided framing information for TT and AVB traffic where we can see that maximum frame of TT traffic is 1420 bytes and our TT window is $L_{TT} = 50\mu s$ and physical speed is 100 Mbits/s so that $L_{TT}$ is sufficient for transfer the maximum frame of TT, similarly, maximum size of avb traffic is 187 bytes which is given in table 21 and avb windows is $L_{avb} = 15\mu s$ so, this windows is also sufficient for transferring maximum avb frame at a time.

Worst case delay of the TSN switch are comparatively low with the considering standard switches but the delay difference of the switch is in acceptable range where the fig. 53 clearly shown. If the standard switch uses for TSN network, network should be well synchronous as well as network configuration should be static and this statement is always for small network and large network. In fig. 53, we can observe that minimum delay of the TT traffic is $2ms$ and maximum delay is about $7\ ms$.

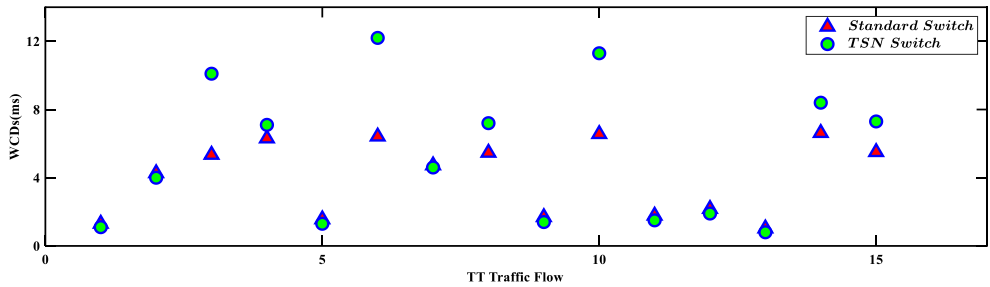

**Figure 53 Shrink TT traffic flow**

## 6.4.2    End to end delay

**Table 23 Flow of TT Traffic include with Shrink Frame size, Periods**

| Flow | Size (B) | Periods(ms) | Flow | Size (B) | Periods(ms) |
|------|----------|-------------|------|----------|-------------|
| $TT_1$ | 70 | 62.5 | $TT_9$ | 72 | 37.5 |
| $TT_2$ | 71 | 125 | $TT_{10}$ | 80 | 37.5 |
| $TT_3$ | 72 | 37.5 | $TT_{11}$ | 70 | 25 |
| $TT_4$ | 77 | 37.5 | $TT_{12}$ | 75 | 12.5 |
| $TT_5$ | 73 | 25 | $TT_{13}$ | 76 | 12.5 |
| $TT_6$ | 80 | 37.5 | $TT_{14}$ | 77 | 37.5 |
| $TT_7$ | 75 | 62.5 | $TT_{15}$ | 76 | 75 |
| $TT_8$ | 74 | 75 | | | |

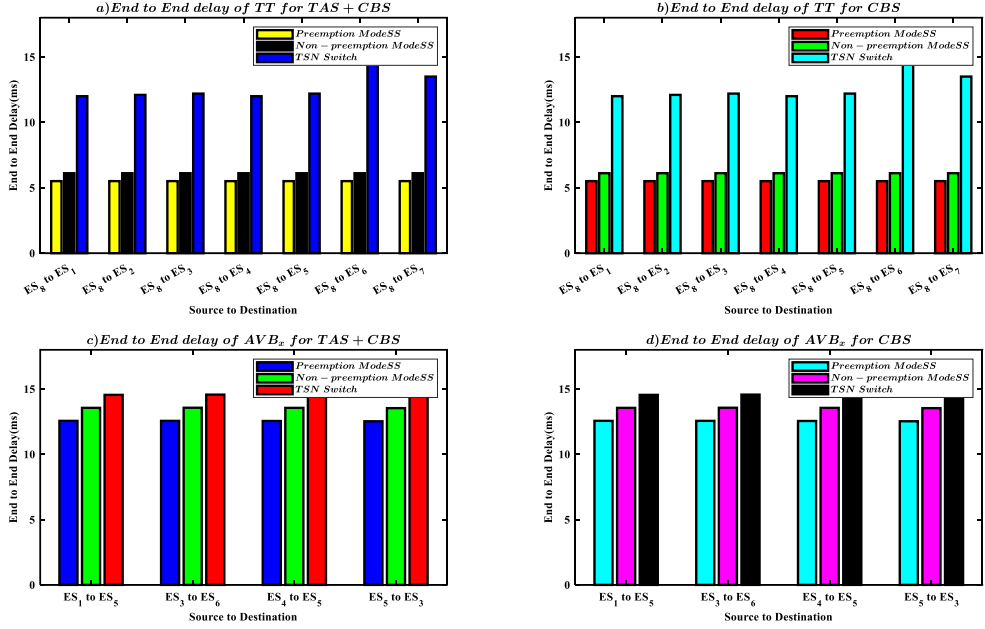

**Figure 54 End to end Comparison for Shrink TT and AVB**

Fig. 54, for measuring end to end, we have used table 22 for topology and table 23 frame information. TT windows is $L_{TT} = 50\mu s$, AVB window is $L_{avb} = 15\mu s$, syn windows along with guard band is $L_{syn+GB} = 5\mu s$, physical link speed is 100 Mbit/s, offset is $o^h = 60\mu s$, hyper-periods is $T_{GCL} = 250\mu s$ though hyper-periods is only applicable for TAS+CBS scheduling, specific sys window is $L_{sys} = 2\mu s$. In table 23, we have provided framing information for TT and table 21 AVB traffic where we can see that maximum frame of TT traffic is 1420 bytes and our TT window is $L_{TT} = 50\mu s$ and physical speed is 100 Mbits/s so that $L_{TT}$ is sufficient for transfer the maximum frame of TT, similarly, maximum size of avb traffic is 187 bytes which is given in table 22 and avb windows is $L_{avb} = 15\mu s$ so, this windows is also sufficient for transferring maximum avb frame at a time. In fig. 54, we can observe that end to end delay of TT traffic is significantly minimize if we have used shrink TT frame. Moreover, end to end delay all destination are remain same though negligible variation can experience. Because, indexing frame sizes are very concise and all of the index frame sizes are more or less small. Non-preemption delay is comparatively higher in all cases, as shown in fig. 54, due to the fact that non-preemption mode makes use of an additional grand band in addition to the syn-band. As a result, TT flow is higher in both cases, with the maximum delay

of the TT flow being around 6 ms in both cases, such as TAS+CBS and CBS+SP, respectively. In the context of AVB traffic, non-preemption mode has also experienced increased delay, but the end-to-end delay of AVB traffic using a standard switch (SS) meets the requirements of real-time communication, and according to our algorithm, frame length, and scheduling mechanism, the maximum delay of AVB traffic is approximately 13ms. The end-to-end latency of a TSN switch, which is discussed in this document [88], was utilized to evaluate performance. We can see in fig. 54 that the TSN switch latency is negligible in all of the scenarios because the TSN switch does not need to synchronize frames to transmit and receive data in the TSN switch environment. TSN switches, on the other hand, are capable of handling algorithms without the need of the PTP clock. Nonetheless, at first appearance, there is little difference between the TSN switch output and the SS output. Eventually, shrink TT traffics are significantly improved end-to-end not only the TT traffic but also avb traffics.

# 7. Conclusion

We have demonstrated uses of standard ethernet switch in the context of the time sensitive analyze. Though time sensitive network adopts several scheduling mechanisms, we have illustrated only two type scheduling mechanisms are explained such as IEEE 802.1Qbu and IEEE 802. 1Qbv.Besides, this scheduling mechanisms also have some traffic mode for instance preemption mode and non-preemption mode, we also considered both modes to generalize each output of each traffic. We have analyzed worst-case delay for each traffic by using network calculus and further validation we have simulated as well as experiment to get accuracy. Besides, we have considered fault resilience topology for minimizing end-to-end delay, though several fault tolerance algorithms are available in this era, but all of the algorithms are not suited for TSN network, especially, deterministic network over standard switch. On the other hand, we have proposed TT traffic frame shrink for removing traffic latency of the network, improving end-to-end delay not only TT traffic but also AVB traffic and ensure negligible jitter introduce in this case. Calculation of three main objectives, we have used basically network calculus, but to more clarification, we have simulated the network by using Omnet++ and to get precise time, we have been experimented in mention network.

The outcome of the proposed algorithms for using standard ethernet in the field of TSN, we have got that standard ethernet can be used for TSN network if the traffic soft real time communication or mid-hard real time communication. But one hardest thing should be ensuring that frame and time synchronization should be accurate and PTP clock should strictly send fixed frame to the source and destination. If the synchronization is well enough, the output of the network such as worst-case delay, jitter, end-to-end delay, latency is the range of the real time communication and it can be preferable for industrial communication.

# Appendix A

We provide a worst-case credit-based shaping scenario for each of the methods discussed in this paper. Say, for example, that there are two SR classes: A and B. In this situation, A is given precedence over B. Class A traffic has the greatest priority for data transfer. If there is enough credit available, Class A frames may be delivered back-to-back without interruption. The worst-case scenario includes sending the maximum number of frames even if the credit is negative. After the maximum number of frames have been sent, the bare minimum credit is received.

**Table 24 Mathematical Symbols and Descriptions**

| Symbols | Descriptions |
|---|---|
| $V_{max}^x, x \in A, B$ | Maximum Credit of Class A, B |
| $V_{min}^x, x \in A, B$ | Minimum Credit of Class A, B |
| $I^x, x \in A, B$ | Idle-Slope of Class A, B |
| $S^x, x \in A, B$ | Send-Slope of Class A, B |
| $V_{same}^x, x \in A, B$ | Intersection Credit of Class A, B |
| $\Delta t_n^{x+}, n = 1,2,3, \dots.$ | Increase Credit Time of Class A, B |
| $\Delta t_n^{x-}, n = 1,2,3, \dots.$ | Decrease Credit Time of Class A, B |

We must first explain the circumstances of the AVB traffic before proceeding to the mathematical assessment. The linear CBS method is used in this case.

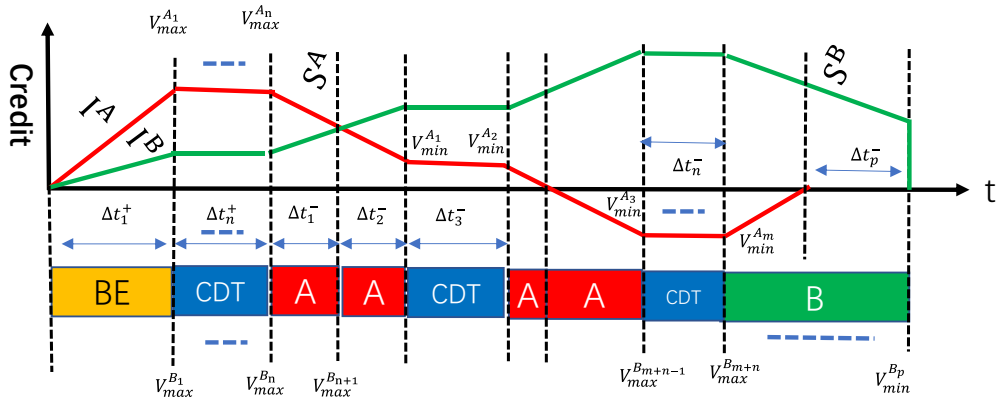

**Figure 55 Credit-Based Shaping Example**

$$C = I^x - S^x \qquad (A1)$$

$I_x$ is considered the idle-slope and $S_x$ is considered the send-slope where $C$ is the connection speed or transmission rate (Mbits/s). An example is shown in fig. 55.

If we now examine the fact that $V_A(t)$ is the amount of credit that class A owes the class A is $\Delta t^{A+} = \frac{\overline{L_A^{max}}}{C}$; $\{\overline{L_A^{max}} = max(L^B, L^{BE})\}$.

This is what we get if we use the credit for idle slope definition: For Class A

Increasing Credit of A = time intervals × Idle-Slope

$$V_{max}^A = \Delta t^{A+} \times I^A \qquad (A2)$$

Obtain the bare minimum credit for AVB traffic of class A; if $\Delta t^{A-} = \frac{L_A^{max}}{C}$ $\{L_A^{max}$ is the maximum frame size of class A$\}$ So, we can say that

$$V_{min}^A = \frac{L_A^{max}}{C} \times S^A \qquad (A3)$$

for AVB traffic, receive the absolute minimum credit of class B

$\Delta t^{B+} = \frac{\overline{L_B^{max}}}{C}$ $\{\overline{L_B^{max}} = max(L^{BE})\}$ So, Class B's idle-slope value is what we may call it.

$$V_{max}^B = I^B \times \left\{ \frac{\overline{L_B^{max}}}{C} + \frac{L_A^{max}}{C} - \frac{\overline{L_A^{max}}}{C} \times \frac{I^A}{S^A} \right\} \qquad (A4)$$

AVB traffic of class B receives the minimal credit if and only if the following conditions are met: $\Delta t^{B-} = \frac{L_B^{max}}{C}$ and $L_B^{max}$ is the maximum frame size of class B. We may call the value of class B's send-slope. we can say that

$$V_{min}^B = \frac{L_B^{max}}{C} \times S^B \qquad (A5)$$

**Theorem A1:** *If CDT classes often travel through the TSN switch, the maximum idle credits will be $\Delta t^{A+} = \frac{V_{max}^{An}}{I^A}$, maximum sending credit time will be $\Delta t^{A-} = \frac{V_{max}^{An} - V_{min}^{Am}}{S^A}$*

**Proof of Theorem A1:**

$$\Delta t_1^+ = \left(\frac{V_{max}^{A_1}}{I^A} - \frac{0}{I^A}\right); \ \Delta t_2^+ = \left(\frac{V_{max}^{A_2}}{I^A} - \frac{V_{max}^{A_1}}{I^A}\right); \tag{A6}$$

$$\Delta t_3^+ = \left(\frac{V_{max}^{A_3}}{I^A} - \frac{V_{max}^{A_2}}{I^A}\right); \ \dots \dots; \Delta t_n^+ = \left(\frac{V_{max}^{A_n}}{I^A} - \frac{V_{max}^{A_{n-1}}}{I^A}\right)$$

From fig.55. we get time,

So, combing all of the equations from A6, we will get,

$$\Delta t^{A+} = \Delta t_1^+ + \Delta t_2^+ + \Delta t_3^+ + \cdots + \Delta t_n^+ \tag{A7}$$

$$\Rightarrow \Delta t^{A+} = \frac{1}{I^A}\left[(V_{max}^{A_1} - 0) + (V_{max}^{A_2} - V_{max}^{A_1}) + (V_{max}^{A_3} - V_{max}^{A_2}) + \cdots + V_{max}^{A_n} - V_{max}^{A_{n-1}}\right] \tag{A8}$$

$$\Rightarrow \Delta t^{A+} = \frac{V_{max}^{A_n}}{I^x} \tag{A9}$$

Similarly,

$$\Delta t_1^- = \left(\frac{V_{min}^{A_n}}{S^A} - \frac{V_{min}^{A_1}}{S^A}\right); \ \Delta t_2^+ = \left(\frac{V_{min}^{A_1}}{S^A} - \frac{V_{min}^{A_2}}{S^A}\right); \tag{A10}$$

$$\Delta t_3^+ = \left(\frac{V_{min}^{A_2}}{S^A} - \frac{V_{min}^{A_3}}{S^A}\right); \dots \dots \dots; \Delta t_n^+ = \left(\frac{V_{min}^{A_{m-1}}}{S^A} - \frac{V_{min}^{A_m}}{S^A}\right)$$

$$\Rightarrow \Delta t^{A-} = \Delta t_1^- + \Delta t_2^- + \Delta t_3^- + \cdots + \Delta t_n^- \tag{A11}$$

$$\Rightarrow \Delta t^{A-} = \frac{1}{S^A}\left[(V_{max}^{A_n} - V_{min}^{A_1}) + (V_{min}^{A} - V_{min}^{A_1}) + (V_{min}^{A_2} - V_{min}^{A_3}) + \cdots + V_{min}^{A_{m-1}} \right. \tag{A12}$$

$$\left. - V_{min}^{A_m}\right]$$

$$\Rightarrow \Delta t^{A-} = \frac{V_{max}^{A_n} - V_{min}^{A_m}}{S^A} \tag{A13}$$

For total CDT time will be,

$$\Delta t^{CDT} = \Delta t_1^0 + \Delta t_2^0 + \Delta t_3^0 + \cdots + \Delta t_n^0 \tag{A14}$$

$$\Rightarrow \Delta t^{CDT} = \sum_{i=1}^{u} \Delta t_i^u \tag{A15}$$

**Theorem A2:** *If CDT class are passing through the TSN switch very frequently, then overall maximum Idle credit's will be* $\Delta t^{B+} = \frac{V_{max}^{B_{m+n}}}{I^B}$ *and maximum sending credit time will be* $\Delta t^{B-} = \frac{V_{max}^{B_{m+n}} - V_{min}^{B_p}}{S^B}$

**Proof of Theorem A2:**

From fig.55. we get time,

So, combing all of the equations from 16, we will get,

$$\Delta t_1^+ = \left(\frac{V_{max}^{B_1}}{I^B} - \frac{0}{I^B}\right); \ \Delta t_2^+ = \left(\frac{V_{max}^{B_2}}{I^B} - \frac{V_{max}^{B_1}}{I^B}\right);$$

(A16)

$$\Delta t_3^+ = \left(\frac{V_{max}^{B_3}}{I^B} - \frac{V_{max}^{B_2}}{I^B}\right); \ldots \ldots; \Delta t_{m+n}^+ = \left(\frac{V_{max}^{B_{m+n}}}{I^B} - \frac{V_{max}^{B_{m+n-1}}}{I^B}\right)$$

$$\Delta t^{B+} = \Delta t_1^+ + \Delta t_2^+ + \Delta t_3^+ + \cdots + \Delta t_m^+$$

(A17)

$$\Rightarrow \Delta t^{B+} = \frac{1}{I^B}\left[\left(V_{max}^{B_1} - 0\right) + \left(V_{max}^{B_2} - V_{max}^{B_1}\right) + \left(V_{max}^{B_3} - V_{max}^{B_2}\right) + \cdots + V_{max}^{B_{m+n}}\right.$$

(A18)

$$\left. - V_{max}^{B_{m+n-1}}\right]$$

$$\Rightarrow \Delta t^{B+} = \frac{V_{max}^{B_{m+n}}}{I^B}$$

(A19)

Similarly,

$$\Delta t_1^- = \left(\frac{V_{min}^{B_{m+n}}}{S^A} - \frac{V_{min}^{B_1}}{S^A}\right); \ \Delta t_2^+ = \left(\frac{V_{min}^{B_1}}{S^A} - \frac{V_{min}^{B_2}}{S^A}\right);$$

(A20)

$$\Delta t_3^+ = \left(\frac{V_{min}^{B_2}}{S^A} - \frac{V_{min}^{B_3}}{S^A}\right); \ldots \ldots \ldots; \Delta t_p^+ = \left(\frac{V_{min}^{B_{p-1}}}{S^A} - \frac{V_{min}^{B_p}}{S^A}\right)$$

$$\Rightarrow \Delta t^{B-} = \Delta t_1^- + \Delta t_2^- + \Delta t_3^- + \cdots + \Delta t_p^-$$

(A21)

$$\Rightarrow \Delta t^{B-} = \frac{1}{S^B}\left[\left(V_{max}^{B_{m+n}} - V_{min}^{B_1}\right) + \left(V_{min}^{B_1} - V_{min}^{B_2}\right) + \left(V_{min}^{B_3} - V_{min}^{B_4}\right) + \cdots + \left(V_{min}^{B_{p-1}}\right.\right.$$

(A22)

$$\left.\left. - V_{min}^{B_p}\right)\right]$$

$$\Rightarrow \Delta t^{B-} = \frac{V_{max}^{B_{m+n}} - V_{min}^{B_p}}{S^B}$$

(A23)

**Lemma A1:** *Timing difference between AVB class A Idle-Slope & AVB class B Idle-Slope will be*

$$\Delta t^{B+} - \Delta t^{A+} = \frac{I^A \cdot \left(V_{max}^{B_{m+n}} - V_{max}^{A_n}\right) - V_{max}^{A_n} \cdot \left(S^B - S^A\right)}{I^A(C - S^B)}$$

(A24)

*and similarly, time difference for send-slope will be*

$$\Delta t^{B-} - \Delta t^{A-} = \frac{S^A.\left(V_{max}^{B_{m+n}} - V_{max}^{B_p}\right) - \left(V_{max}^{A_n} - V_{min}^{A_m}\right).S^B}{S^A.S^B} \tag{A25}$$

**Proof of Lemma A1:**

Idle-slope time difference between AVB class A & B,

$$\Delta t^{B+} - \Delta t^{A+} = \frac{V_{max}^{B_{m+n}}}{I^B} - \frac{V_{max}^{A_n}}{I^A} \tag{A26}$$

$$\Longrightarrow \Delta t^{B+} - \Delta t^{A+} = \frac{I^A.V_{max}^{B_{m+n}} - I^B.V_{max}^{A_n}}{I^A.I^B} \tag{A27}$$

But, according band reservation of class A&B will be

$$C = I^A - S^A = I^B - S^B \tag{A28}$$

$$\Longrightarrow I^B = I^A - I^A - S^B = C + S^B \tag{A29}$$

Using equation A29 in equation A27, we get

$$\Delta t^{B+} - \Delta t^{A+} = \frac{I^A.V_{max}^{B_{m+n}} - (I^A - S^A + S^B).V_{max}^{A_n}}{I^A.(I^A - S^A + S^B)} \tag{A30}$$

$$\Longrightarrow \Delta t^{B+} - \Delta t^{A+} = \frac{I^A.\left(V_{max}^{B_{m+n}} - V_{max}^{A_n}\right) - V_{max}^{A_n}.(S^B - S^A)}{I^A(C - S^B)} \tag{A31}$$

Send-slope time difference between AVB class A & B,

$$\Delta t^{B-} - \Delta t^{A-} = \frac{V_{max}^{B_{m+n}} - V_{max}^{B_p}}{S^B} - \frac{V_{max}^{A_n} - V_{min}^{A_m}}{S^A} \tag{A32}$$

$$\Longrightarrow \Delta t^{B-} - \Delta t^{A-} = \frac{S^A.(V_{max}^{B_{m+n}} - V_{max}^{B_p}) - S^B.(V_{max}^{A_n} - V_{min}^{A_m})}{S^A.S^B} \tag{A33}$$

**Lemma A2:** *Timing difference between Idle-Slope & Send-Slope of AVB class A & B will be*

$$\Delta t^{A-} - \Delta t^{A+} = \frac{I^A.\left(V_{min}^{A_m}\right) + C.(V_{max}^{A_n})}{I^A.(C - I^A)} \tag{A34}$$

*and*

$$\Delta t^{B-} - \Delta t^{B+} = \frac{C.(V_{max}^{B_{m+n}}) - I^B.\left(V_{min}^{B_p}\right)}{I^B.(I^B - C)} \tag{A35}$$

**Proof of Lemma A2:**

$$\Delta t^{A-} - \Delta t^{A+} = \frac{V_{max}^{A_n} - V_{min}^{A_m}}{S^A} - \frac{V_{max}^{A_n}}{I^A} \tag{A36}$$

$$= \frac{\left(V_{max}^{A_n} - V_{min}^{A_m}\right).I^A - V_{max}^{A_n}.S^A}{I^A.S^A} \tag{A37}$$

$$= \frac{\left(V_{max}^{A_n} - V_{min}^{A_m}\right).I^A - V_{max}^{A_n}.(I^A - C)}{I^A.(I^A - C)} \tag{A38}$$

We know that $S^A = (I^A - C)$, we get that,

$$\Delta t^{A-} - \Delta t^{A+} = \frac{I^A.\left(V_{min}^{A_m}\right) + C.\left(V_{max}^{A_n}\right)}{I^A.(C - I^A)} \tag{A39}$$

Similarly,

$$\Delta t^{B-} - \Delta t^{B+} = \frac{V_{max}^{B_{m+n}} - V_{min}^{B_p}}{S^B} - \frac{V_{max}^{B_{m+n}}}{I^B} \tag{A40}$$

$$= \frac{\left(V_{max}^{B_{m+n}} - V_{min}^{B_p}\right).I^B - V_{max}^{B_{m+n}}.S^A}{I^A.S^A} \tag{A41}$$

$$= \frac{\left(V_{max}^{B_{m+n}} - V_{min}^{B_p}\right).I^B - V_{max}^{B_{m+n}}.(I^B - C)}{I^B.(I^B - C)} \tag{A42}$$

We know that $S^B = (I^B - C)$, we get that,

$$\Delta t^{B-} - \Delta t^{B+} = \frac{C.\left(V_{max}^{B_{m+n}}\right) - I^B.\left(V_{min}^{B_p}\right)}{I^B.(I^B - C)} \tag{A43}$$

**Theorem A3:** *When credit of class A are increasing, at that moment credit of class B are decreasing (if CDT is not transferring at that moment). On the other hand, credit of class A are decreasing, at that time, credit of class B are increasing. In a particular time, credit of the both classes will be same which is*

$$V_{same}^x = V_{max}^{A_m} + \frac{I^A(V_{max}^{B_m} - V_{max}^{A_m})}{I^A - I^B} \tag{A44}$$

*and the optimal time will be*

$$t_{same} = t_m + \frac{(V_{max}^{B_m} - V_{max}^{A_m})}{I^A - I^B} \tag{A45}$$

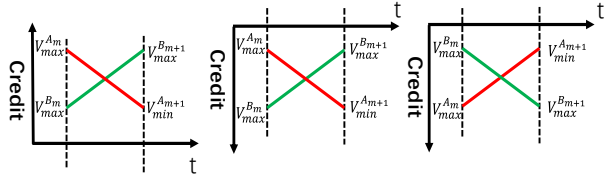

**Figure 56 Intersection Scenario of credit of class A & B**

In above fig.56, we get the four co-ordinates of the four position. So

$$(t_m, V_{max}^{A_m}), (t_{m+1}, V_{min}^{A_{m+1}}), (t_m, V_{max}^{B_m}), (t_{m+1}, V_{max}^{B_{m+1}})$$

According to straight line formula, we get,

$$\frac{V^x - V_{max}^{A_m}}{V_{max}^{A_m} - V_{min}^{A_{m+1}}} = \frac{t - t_m}{t_m - t_{m+1}} \qquad (A46)$$

$$\Rightarrow V^x = I^A.(t - t_m) + V_{max}^{A_m} \qquad (A47)$$

Similarly,

$$(V^x - V_{max}^{B_m}) = I^B(t - t_m) \qquad (A48)$$

Using equation A48, then we get,

$$I^A(t - t_m) + V_{max}^{A_m} - V_{max}^{B_m} = I^B(t - t_m) \qquad (A49)$$

$$\Rightarrow t = t_{same} = t_m + \frac{(V_{max}^{B_m} - V_{max}^{A_m})}{I^A - I^B} \qquad (A50)$$

Again, we use equation A48, then

$$V^x = V_{same}^x = I^A.\left(t_m + \frac{(V_{max}^{B_m} - V_{max}^{A_m})}{I^A - I^B} - t_m\right) + V_{max}^{A_m} \qquad (A51)$$

$$V_{same}^x = V_{max}^{A_m} + \frac{I^A(V_{max}^{B_m} - V_{max}^{A_m})}{I^A - I^B} \qquad (A52)$$

**Lemma A3:** *Credit of two class A & B never same in the midpoint of* $t_m$ & $t_{m+1}$

**Proof of Lemma A3:**

According to credit concept $I^A > I^B$ and $I^A - S^A = C; S^A = (I^A - C)$ & $I^B - S^B = C; S^B = (I^B - C)$. If credit of class A & B are same in midpoint of $t_m$ & $t_{m+1}$, it should be satisfied in $S^A.I^B = -1$ or $S^B.I^A = -1$. But, $I^A or I^B$ is always positive and greater than

one,on the other hand, $S^A or S^B is$ always negative and magnitude greater than one. So, it should be either $S^A . I^B < -1$ or $S^A . I^B < -1$

**Theorem A4:** *When credit of class A or B reach to zero, at that time, credit of the AVB class x is*

$$V_{mid}^x = \frac{V_{max}^{x_1} + k_1 \times V_{max}^{x_2}}{k_1 + 1}; x \in A, B \qquad (A53)$$

$$Where, k_1 = \frac{t_{eq} - t_1}{t_2 - t_{eq}} \qquad (A54)$$

**Proof of Theorem A2:**

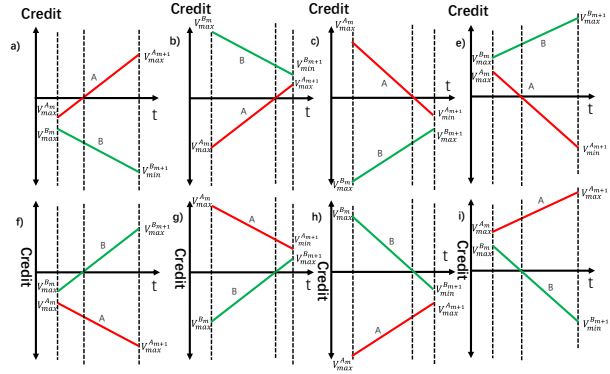

**Figure 57 Difference Scenario of zero credit of class A & B**

In fig 57,when credit of any class will be zero, then axis of x will divide into k:1, then we get according to straight line formula,

$$\frac{V_{max}^{A_1} + k \times V_{min}^{A_2}}{k + 1} = 0 \qquad (A55)$$

$$k = \frac{-V_{max}^{A_1}}{V_{min}^{A_2}} \qquad (A56)$$

When credit of class $x$ is reached in zero, then

$$t_{eq} = \frac{V_{min}^{A_2} \times t_1 - V_{max}^{A_1} \times t_2}{V_{max}^{A_1} - V_{min}^{A_2}} \qquad (A57)$$

For credit of Class B at the $t_{eq}$ time,

$$V_{mid}^B = \frac{V_{max}^{B_1} + k_1 \times V_{max}^{B_2}}{k_1 + 1} \qquad (A58)$$

$$t_{eq} = \frac{t_1 + k_1 \times t_2}{k_1 + 1} \tag{A59}$$

$$k_1 = \frac{t_{eq} - t_1}{t_2 - t_{eq}} \tag{A60}$$

# Appendix B

## B.1.1 Gate Control List (GCL)

The same gate may be used by different kinds of traffic, as well as traffic with the same priority. Gate overlap between separate kinds of traffic, as well as traffic of the same priority, is a major problem in this area. Accordingly, IEEE 802.1 ASrev has a considerable debate on the subject of global sync clocks.

Many researchers have employed the GCL in various configurations for scheduled traffic (ST), where the GCL has been used to offer more dependable data delivery [33]. Individual ST can flow consistently because to a unique GCLs configuration mechanism established by [89]. Because of this, individual ST may flow consistently in a predictable way when the jitter of the system approaches zero. ST traffic is differentiated from conventional traffic, which may flow in either the time domain or the space domain, in terms of its time-critical nature[90].

There are several disadvantages to using the above-mentioned strategy, as you will see. When it comes to space, for example, the GCL synthesis solution may take a long time to solve, especially as the network grows in size[91]. ST may be ready to accept some delay and decrease strictness as long as data is stored in the same queue, which will ultimately result in an increase in end-to-end latency. Scheduling determines when frames are delivered and received in a dispersed network environment, as previously mentioned. Thus, all of these actions may be carried out using a variety of ST queues.

This is the most critical flaw of previous systems, which rely on the synchronization of the transmitting nodes to work properly. The usage of off-the-shelf sensor nodes, for example, is common in many systems, although they are not always reliable. Thus, it's not always the case like that. Proposals for GCL systems based on classes have been made[19], since there is no demand for synchronized end systems, there is no need for per-flow scheduling, and there is no requirement for synchronized end systems However, despite the fact that it is a highly effective GCL system for the TSN network, it does have some limitations. For example, according to their scheduling models, they do not take into account offsets of the windows in several nodes, and they consider the opening window to be aligned with all switches, among other things.

To sum up, it is possible to conclude that some scheduling mechanisms are deterministic in small networks but semi-deterministic in large networks; and second, that any GCL synthesis can overlap either the same priority traffic gate or another priority class depending on the prioritization of the prioritization of that gate. Despite the fact that the [92] article addresses how to keep traffic moving when two or more streams of traffic are overlapping, the answer is always higher priority, unless lower priority classes are passing before incoming higher priority classes. A research by [92] explains how gate opening is synthesizable for the same priority classes, but it also discusses the problem of gate overlapping, which has been previously highlighted.

Our discussion has thus far been focused on the overlapping circumstances, which may be either the same class traffic or distinct class traffic, and we will demonstrate how overlapping for all classes in GCL can be avoided completely.

**Table 25 Mathematical Symbol and Descriptions**

| Symbol | Descriptions |
|---|---|
| $t_{TT}$ | TT Executed Time |
| $L_{TT}$ | TT Gate Open Length |
| $t_{AVB}$ | AVB Executed Time |
| $L_{AVB}$ | AVB Gate Open Length |
| $\mathcal{P}$ | Scheduled Set |
| | $\mathcal{P} = \{1, \dots, P\}$ of P Hyper-Periods |
| $a_{TT,k}$ | Gate of TT in kth Hyper-Periods |
| $a_{AVB,k}$ | Gate of AVB in kth Hyper-Periods |
| $\Pi$ | Set $\Pi = \{1, \dots, N\}$ |
| $\mathcal{E}$ | Set of Couples (TT, AVB) $\in \Pi^2$ |
| $a_{TT,p}$ | Gate of TT in pth Hyper-Periods |
| $a_{AVB,p}$ | Gate of AVB in pth Hyper-Periods |
| $\mathcal{E}_c$ | Set of gate couples |
| $\mathcal{E}_a$ | Set of gate couples |
| $\beta_{CBS\_TT}^h$ | Service Curve of TT |
| $I^x$ | Idle-Slope of class x |
| $\Delta t$ | time |
| $V^x(t)$ | Credit |
| $\alpha_{TT}^h$ | Arrival Curve of TT |
| $C$ | Transmission of rate |
| $\Delta t^{TT}$ | Time of TT |
| $S^x$ | Send Slope of class x |
| $O_{j,i}^h$ | Offset Length |
| $p_{GCL}^h$ | Hyper-Periods |
| $\beta^x(t)$ | Service of AVB traffic |
| $b$ | Burst rate of Arrival Curve |

## B.1.2 Problem Definitions and Modeling

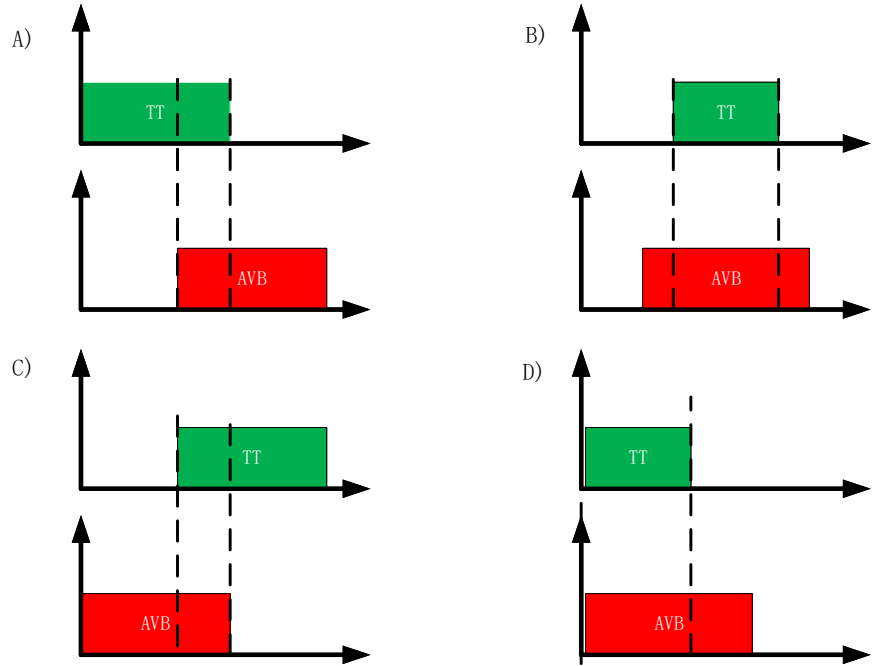

**Figure 58 Possible Scenario for Non-overlapping gate for different Traffic.**

***Theorem B1:*** *For two traffic gate TT and AVB, the intersections between intervals* $[t_{TT} + kT_{TT}, t_{TT} + kT_{TT} + L_{TT})$ *and* $[t_{AVB} + \ell T_{AVB}, t_{AVB} + \ell T_{AVB} + L_{AVB})$ *is empty,* $\forall (k, l) \in \mathbb{Z}^2$,*or in other words the two traffic gate will not be overlapped if and only if*

$$L_{TT} \leq (t_{AVB} - t_{TT}) \bmod g_{TT,AVB} \leq g_{TT,AVB} - L_{AVB} \tag{B.1}$$

where $g_{TT,AVB}$ denotes the greatest common divisor of $\text{T}_{TT}$ and $\text{T}_{AVB}$

**Proof:** In term of non-overlapped gate, illustrates the minimum distance from gate of TT, to gate of AVB. On the other hand, time duration of gate of TT, can allow being executed without interfacing with another gate like AVB.Fig.58 is demonstrate an example.

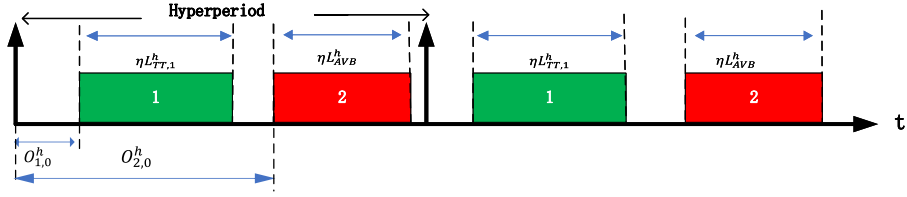

**Figure 59 Non-Overlapped Gate Opening time.**

**Corollary B1:** *Gate opening time order between i and j has no* effect. *Hence*

$$L_{TT} \le (t_{AVB} - t_{TT}) \bmod g_{TT,AVB} \le g_{TT,AVB} - L_{AVB} \; and$$

$$L_{TT} \le (t_{TT} - t_{AVB}) \bmod g_{TT,AVB} \le g_{TT,AVB} - L_{TT}$$

**Proof:** To prove the theorem, we should know that (-a) mod b = b-a mod b for a mod b>0

According to theorem 1, there is no chance to overlapping two gates in one hyper-period and some equivalent condition can be written,

$$\forall (TT, AVB) \in \Pi^2, \forall k \in \mathcal{P}$$

$$L_{TT} - (2 - a_{TT,k} - a_{AVB,k})Z \le (t_{AVB} - t_{TT}) \bmod g_{TT,AVB} \qquad \text{(B.2)}$$

$$\le g_{TT,AVB} - L_{AVB} + (2 - a_{TT,k} - a_{AVB,k})Z$$

Here, Z is considered as large constant which indicate that constraint is not active until $a_{TT,k} = a_{AVB,k} = 1$

## B.1.3 Optimization Criteria

As a result of this strategy, a workable schedule and the allocation of available resources to the gates are transformed into an optimization problem, allowing them to take advantage of additional execution time. In the case of the optimal solution (a,t), the coefficient is defined as the greatest amount by which all gates' execution times may be increased without making the schedule unworkable (i.e., without making the schedule unworkable, the coefficient is 1). Figure 60A depicts a timetable that has two divisions on a single gate.

The varied affects are shown in Fig. 60B. The smaller rectangles indicate the budget for starting time, while the larger filling rectangles show the maximum time restriction. Clearly, $\eta$ value of 1 suggests schedules with reduced periods and thus may be deemed ineffective. To put it another way, if the number is less than one, it means that no work can be done. To make sure that errors caused by undervalued split time budgets are avoided, and to provide system experts more flexibility in the event that Gate Time budgets are adjusted, the guarantees are enhanced.

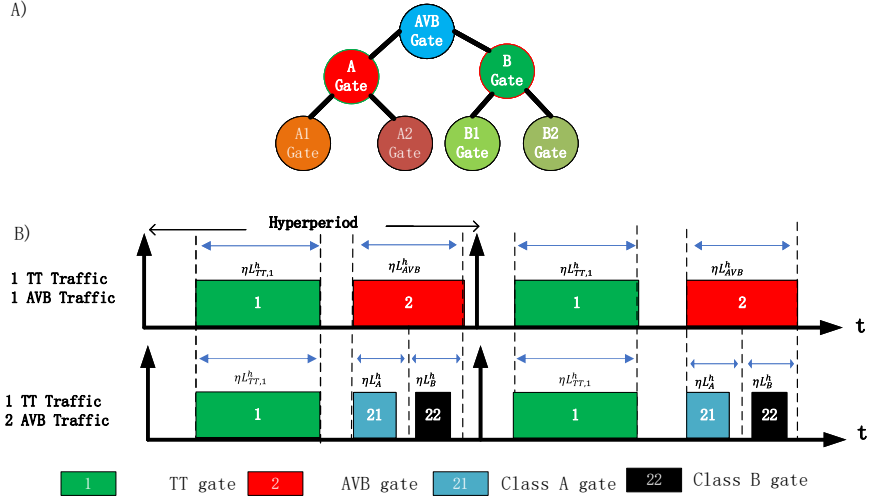

**Figure 60 A) Tree Representation of each gate B) Non-Overlapped Gate Opening time.**

### B.1.3.1 Exact Formulation

The aim of such a formulation is to optimize the system's minimum evolution potential to the greatest extent possible. GCL gates are the gates in a system (temporal execution-wise). It will be possible to come across gates that have the potential to develop with a potential equal to or greater than this minimum value, while others will have the chance to develop with a higher potential.

**Theorem B2**: *In one hyper-period operation, to make linear condition, we say that*

$$(t_{AVB} - t_{TT}) mod \ g_{TT,AVB} = (t_{AVB} - t_{TT}) - q_{TT,AVB} g_{TT,AVB} \qquad (B.3)$$

Here, $q_{TT,AVB} = \left\lfloor \frac{t_{AVB} - t_{TT}}{g_{TT,AVB}} \right\rfloor$ is representing that the quotient of modulo operation.

**Proof:** For exact linear condition, the quotient $q_{TT,AVB}$ have to support in following constraint for every gate couple and should be allocated in same hyper-periods.

$$0 < (t_{TT} - t_{AVB}) - q_{AVB,TT} \ g_{TT,AVB} < g_{TT,AVB} \qquad (B.4)$$

Certainly, $L_{TT} \ and \ L_{TT}$ is positive and we know from definition of floor function, we can get,

$$\left( \frac{t_{AVB} - t_{TT}}{g_{TT,AVB}} - 1 \right) < \left\lfloor \frac{t_{AVB} - t_{TT}}{g_{TT,AVB}} \right\rfloor \le \frac{t_{AVB} - t_{TT}}{g_{TT,AVB}} \qquad (B.5)$$

Boundary of $q_{TT,AVB}$ is used in (B.5), we can get

$$\frac{-T_{TT} + L_{TT}}{g_{TT,AVB}} - 1 < q_{AVB,TT} \leq \frac{T_{AVB} - L_{AVB}}{g_{TT,AVB}} \tag{B.6}$$

Now, getting overall formulation for maximum of $\eta$. Values of $\eta < 1$ indicate timetables with reduced time limits, and as a result, they may be deemed unfeasibly tight. In other words, values of $\eta \geq 1$ imply that a particular issue is feasible to solve. Furthermore, higher values of guarantee that mistakes resulting from underestimate of gate time budgets, if any, are avoided, or even provide some flexibility for system experts in the event that gate time budgets need to be adjusted, as previously mentioned.

$$\sum_{k \in \mathcal{P}} a_{TT,k} = 1, \forall TT \in \Pi \tag{B.7}$$

$$a_{TT,k} \leq 1 - a_{AVB,k}, \forall k \in \mathcal{P}, \forall (TT, AVB) \in \mathcal{E} \tag{B.8}$$

$$\sum_{p \in c} a_{TT,p} \leq 1 - \sum_{p \in c} a_{AVB,p} , \forall c \in \mathcal{C}, \forall (TT, AVB) \in \mathcal{E}_c \tag{B.9}$$

$$a_{TT,k} \leq 1 - a_{AVB,k}, \forall k \in \mathcal{P}, \forall (TT, AVB) \in \mathcal{E}_a \tag{B.10}$$

$$(t_{AVB} - t_{TT}) - q_{AVB,TT} g_{AVB,TT} \tag{B.11}$$
$$\leq g_{TT,AVB} - \eta L_{AVB} + (2 - a_{TT,k} - a_{AVB,k})Z, \forall k$$
$$\in \mathcal{P}, \forall (TT, AVB) \in \Pi^2$$

$$(t_{AVB} - t_{TT}) - q_{AVB,TT} g_{TT,AVB} \geq \eta L_{TT} - (2 - a_{TT,k} - a_{AVB,k})Z, \tag{B.12}$$
$$\forall k \in \mathcal{P}, \forall (TT, AVB) \in \Pi^2$$

$$\frac{-T_{TT} + b_{TT}}{g_{TT,AVB}} - 1 < q_{AVB,TT} \leq \frac{T_{AVB} - b_{AVB}}{g_{TT,AVB}} , \forall (TT, AVB) \in \Pi^2$$
$$a_{TT,k} \in \{0,1\}, \forall k \in \mathcal{P}, \forall TT \in \Pi \tag{B.13}$$
$$t_{TT} \in \mathcal{T}_{TT}, \forall TT \in \Pi$$

***Theorem B3:*** *If Control Data Traffic (CDT), which is the highest priority traffic, are waiting*

*in queue for transferring and GCL gate are open for CDT traffic, then the service curve is given in following relationship*

$$\beta^h_{CBS\_TT} = I^x \left( \Delta t - \frac{V^x(t)}{I^x} - \frac{\alpha^h_{TT}}{C} \right) \tag{B.14}$$

**Proof:**

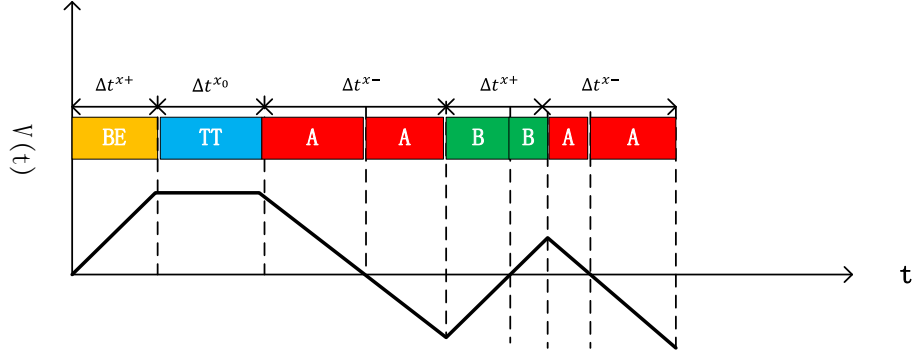

**Figure 61 CBS example for different traffic.**

Class A and B traffic both follow the CBS process; in addition, TT traffic has the greatest priority, while BE traffic has the lowest priority. To ensure proper data transmission, CBS credits are temporarily locked while the TT is in route, as shown in Fig.61. TT credits are not released until the TT completes the data transfer process.

Let Consider fig.61., Here is implied as the AVB class. And $V^x(t)$ is the credit of the class x over time 't'. According to the credit we get,

Credit of the any class = decreasing credit with time $\times$ decreasing rate of the slope $+$ iincreasing credit with time $\times$ increasing rate of the slope $+$ freezing credit with time $\times$ freezing credit rate

$$V^x(t) = (I^x)(\Delta t^{x+}) + (Slope\ of\ TT)(\Delta t^{TT}) + (S^x)(\Delta t^{x-}) \tag{B.15}$$

$$V^x(t) = I^x(-\Delta t^{x-} + \Delta t - \Delta t^{x_0}) + (\Delta t^{x-})(S^x) \tag{B.16}$$

$$(\Delta t^{x-})(I^x) - (\Delta t^{x-})(S^x) = (I^x)(\Delta t) - V^x(t) - (I^x)(\Delta t^{x_0}) \tag{B.17}$$

$$\Delta t^{x-} = \frac{(I^x)(\Delta t) - V^x(t) - (I^x)(\Delta t^{x_0})}{I^x - S^x} \tag{B.18}$$

Again,

$$O^x(t) - O^x(s) = C \times \Delta t^{x-} \tag{B.19}$$

$$O^x(t) - O^x(s) = C \times \frac{(I^x)(\Delta t) - V^x(t) - (I^x)(\Delta t^{x_0})}{I^x - S^x} \tag{B.20}$$

Again,

$$O^H(t) - O^H(s) = C \times \Delta t^{x-} \tag{B.21}$$

$$\Delta t^{x_0} = \frac{O^H(t) - O^H(s)}{C} \tag{B.22}$$

Again,

$$O^H(t) - O^H(s) = \max_{0 \leq i \leq N^h - 1}\{\alpha_{TT}^h(t)\} \tag{B.23}$$

$$\alpha_{TT}^h(t) = \max_{0 \leq i \leq N^h - 1} \sum_{j=i}^{i+N^h-1} L_{TT} \times \lceil \frac{t - O_{j,i}^h}{p_{GCL}^h} \rceil \tag{B.24}$$

from (B.23) we can get

$$O^x(t) - N^x(s) = C \times \frac{(I^x)(\Delta t) - V^x(t) - (I^x)(t^{x_0})}{I^x - S^x} \tag{B.25}$$

$$\beta_{CBS\_TT}^h = I^x(\Delta t - \frac{V^x(t)}{I^x} - \frac{\alpha_{TT}^h}{C}) \tag{B.26}$$

***Theorem B4:*** *Minimum service curve of Audio and video Bridge AVB class x which is written by in following relationship and relationship is found in output port h*

$$\beta^x(t) = \left\{ I^x \Delta t - V^x(t) - \frac{I^x \alpha_{TT}^h(t)}{C} - \frac{I^x \times \frac{\bar{l}}{C}}{C} \right\} \tag{B.27}$$

**Proof:** Let Consider fig.61., Here is implied as the AVB class. And $V^x(t)$ is the credit of the class x over time 't'.

Credit of the any class = decreasing credit with time $\times$ decreasing rate of the slope $+$ iincreasing credit with time $\times$ increasing rate of the slope $+$ freezing credit with time $\times$ freezing credit rate

$$V^x(t) = I^x \times \Delta t^{x+} + \Delta t^{x-} \times (S^x) + (\Delta t^{x_0})(I^{TT}) \tag{B.28}$$

$$V^x(t) = I^x(-\Delta t^{x-} + \Delta t - \Delta t^{x_0}) + (\Delta t^{x-})(S^x) \tag{B.29}$$

$$\Delta t^{x-} = (I^x)(\Delta t) - V^x(t) - \frac{I^x(\Delta t^{x_0})}{I^x - S^x} \tag{B.30}$$

Number of the bits are served to the strict priority which is $C\Delta t^{x-}$ .($\Delta t^{x-}$) that time bits are transfer to the TAS

$$O^x(t) - O^x(s) = C\Delta t^{x-} \quad [C = \text{Transmission Rate}] \tag{B.31}$$

$$O^x(t) - O^x(s) = C \times I^x \Delta t - V^x(t) - \frac{I^x \Delta t^{x_0}}{I^x - S^x} \tag{B.32}$$

Again, in the TT flows

$$O^H(t) - O^H(s) = C \times \Delta t^{x_0} \tag{B.33}$$

$$\Delta t^{x_0} = \{O^H(t) - O^H(s)\}/t \tag{B.34}$$

Again, Service curve of TT

$$\beta^H(t) = C[t - \bar{l}/C]^+ \quad [\bar{l} = \max(l_{ij}^A, l_{ij}^B, l_{ij}^E)] \tag{B.35}$$

$$\alpha_{TT}^h(t) = \max_{0 \le i \le N^h - 1} \sum_{j=i}^{i+N^h-1} \eta L_{TT} \times \lceil \frac{t - O_{j,i}^h}{p_{GCL}^h} \rceil \tag{B.36}$$

Using (B.33)

$$O^H(t) - O^H(s) \le \left(\frac{\alpha_{TT}^h}{\beta^H}\right)(\Delta t) = \alpha_{TT}^h(t) + \frac{b\bar{l}}{C} \tag{B.37}$$

$$\Delta t^{x_0} = \frac{\alpha_{TT}^h(t) + \frac{b\bar{l}}{C}}{C} \tag{B.38}$$

Again, $O^x(s) = N^x(p)$ and using (B.33) in (B.37)

$$O^x(t) - N^x(s) \ge \frac{CI^x \Delta t}{I^x - S^x} - \frac{CV^x(t)}{I^x - S^x} - \frac{CI^x}{I^x - S^x} \left\{ \frac{\alpha_{TT}^h(t) + \frac{b\bar{l}}{C}}{C} \right\} \tag{B.39}$$

$$\ge \frac{C}{I^x - S^x} \left\{ I^x \Delta t - V^x(t) - \frac{I^x \alpha_{TT}^h(t)}{C} - \frac{I^x \times \frac{\bar{l}}{C}}{C} \right\} \tag{B.40}$$

As we know $V^x(t) \le V^{x,max}$

$$\beta^x(t) = \left\{ I^x \Delta t - V^x(t) - \frac{I^x \alpha_{TT}^h(t)}{C} - \frac{I^x \times \overline{I}}{C} \right\} \qquad \text{(B.41)}$$

# Research Outcomes

**Awards:**

- Distinguished Foreign Student Scholarships of Beihang University 2021, First Prize

- Best Paper Award, The 6th International Conference on Intelligent, Interactive Systems and Application,2021

- Best Easy Competition Award, Beihang University,2020, First Prize

**Publications in Masters:**

[1] **Md Mehedi Hasan**, He Feng, Shahrukh Khan, Md Ibrahim Ullah, Md Tanvir Hasan, Bipro Gain," Improve Service Curve using Non-Overlapped Gate in Time Sensitive Network Switch", in 2021 IEEE 21$^{st}$ International Conference on Communication Technology (ICCT), Tianjin, China, Oct, 2021.DOI: [10.1109/ICCT52962.2021.9657974](10.1109/ICCT52962.2021.9657974)

[2] **Md Mehedi Hasan**, He Feng, Shahrukh Khan, Md Ibrahim Ullah, Md Tanvir Hasan, Bipro Gain, "Timing Analysis for Optimal Points in Credit-based Shaper of Time-Sensitive Network",2021 6$^{th}$ International Conference on Signal and Image Processing (ICSIP 2021), Nanjing, China, Oct, 2021 DOI:[10.1109/ICSIP52628.2021.9688723](10.1109/ICSIP52628.2021.9688723)

[3] **Md Mehedi Hasan,** He Feng, Md Tanvir Hasan, Bipro Gain, Md Ibrahim Ullah," Improved and Comparative End-to-End Delay Analysis in CBS and TAS using Data Compression for Time Sensitive Network", in IEEE The 3$^{rd}$ International Conference on Applied Machine Learning (ICAML 2021), in Changsha, Hunan, July,2021 DOI [10.1109/ICAML54311.2021.00049](10.1109/ICAML54311.2021.00049)

[4] **Md Mehedi Hasan,** Md Ibrahim Ullah, He Feng, Md Tanvir Hasan, Bipro Gain, "Convolutional Neural Network Based Smart System to Aid an Epileptic Patient", in IEEE The 3$^{rd}$ International Conference on Applied Machine Learning (ICAML 2021), in Changsha, Hunan, July,2021 DOI [10.1109/ICAML54311.2021.00087](10.1109/ICAML54311.2021.00087)

[5] **Md Mehedi Hasan,** Shahrukh Khan, He Feng, Qiao Li, Syed Mohammad Masum, Md Tanvir Hasan," Improved End-to-End Delay in CBS using Data Compression for Time Sensitive Network", in IEEE 2021 2$^{nd}$ Information Communication Technologies Conference (ICTC 2021), Nanjing,May2021,DOI:[10.1109/ICTC51749.2021.9441572](10.1109/ICTC51749.2021.9441572)

# Acknowledgement

Firstly, and most importantly, I express my appreciation to the Almighty "Allah" for providing me with the opportunity to complete my studies and get a degree. I am able to keep up the great work I have begun because of HIM's blessings.

I owe a debt of gratitude to **Prof. He Feng**, my supervisor. As a consequence of his efforts, he presented me with many opportunities to work on different projects. In the absence of his guidance, it's conceivable that my development would have stalled after just a few simulations. Many other international students were denied the opportunity to work on many projects, yet he gave me the opportunity to do just that. If it weren't for his continuous support and guidance throughout the process, this thesis would not be where it is now.

Thanks to everyone in the lab, but especially to **Dr Ershuai Li**, who have helped me much. They have been keeping checks on my progress from the beginning of the study. Their support has enabled me to take on new tasks while maintaining my focus on my academics. Please accept my gratitude for this **Prof Dr. Li** with Knowledge.

The Bangladeshi community in Beihang has been very kind, and I would want to use this opportunity to express my gratitude to everyone there, especially Sohag, Bipro Gain, Shahrukh Khan, Ibrahim Ullah, Rajib,Sagar for their generosity and compassion. I would not have survived the pandemic had it not been for them.

What's most essential is that I must express my gratitude to my parents and to my two beautiful sisters and brother, without whom I would be lost.

# Author's Profile

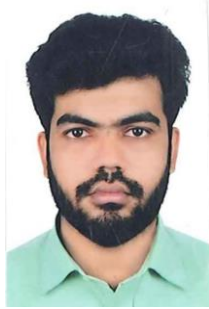

Md Mehedi Hasan was born in Chittagong, Sandwip to be exact, Bangladesh, 1993.He obtained his Bachelor of Engineering in Electrical and Electronic Engineering, University of Chittagong in December 2017, Chittagong, Bangladesh. He was a former lecturer in Physics at Presidency International School affiliated in University of Cambridge. He has joined Beihang University, Beijing, China in 2019 as a Master's degree scholar under major in Electronic and Communication Engineering, School of Electronic and Information Engineering. He pursued research under Prof. He Feng. During his Master's degree, he mainly worked on the Time Sensitive Network (TSN) and Scheduling mechanism. His Academic interest includes research on Credit-Based Shaper, and different scheduling mechanism of TSN network. His future decision to pursue a Doctor of Philosophy (PhD) in Time Sensitive Network from his personal experience right through school.